\documentclass{article}      

\usepackage{epsfig}
\usepackage{natbib}
\usepackage{amssymb}
\usepackage{amsmath}
\usepackage{color}

\usepackage{graphicx}
\begin{document}

\title{Velocity Derivatives in Turbulent Boundary Layers. Part II: Statistical Properties}
\author{William K. George$^1$ \and Michel Stanislas$^2$  \and  Jean-Marc Foucaut$^3$   \and Jean-Philippe Laval$^3$ \and
        Christophe Cuvier$^3$}
\maketitle

\noindent$^1$ Visiting Pr., Centrale Lille, F59651 Villeneuve d'Ascq, France\\
$^2$ Pr. Emeritus, Centrale Lille, F59651 Villeneuve d'Ascq, France\\
$^3$ Univ. Lille, CNRS, ONERA, Arts et Metiers Institute of Technology, Centrale Lille,
UMR 9014 - LMFL - Laboratoire de M\'{e}canique des Fluides de Lille - Kamp\'{e} de F\'{e}riet,
F-59000 Lille, France
\medskip

\begin{abstract}
An experiment was performed using Dual-plane-SPIV in the LMFL boundary layer facility to determine all of the derivative moments needed to estimate the average dissipation rate of the turbulent kinetic energy, $\varepsilon$, and its Reynolds stress counterpart the dissipation tensor, $\varepsilon_{ij}$. For this experiment, the Reynolds number was $Re_\theta = 7500$ or $Re_\tau = 2300$. Part I of this contribution \cite{stanislas20} presented in short the experiment and discussed in detail the dissipation profile and all twelve derivative moments required to compute it.  The data were compared to a channel flow DNS at approximately the same Reynolds number and to previous results. They were also used to evaluate recent theoretical results for the overlap region.

In this Part II the experimental and DNS results are used to evaluate the assumptions of `local isotropy', `local axisymmetry', and `local homogeneity'. They are extended to include the full dissipation tensor, $\varepsilon_{ij}$ and the `pseudo-dissipation tensor', $\mathcal{D}_{ij}$ and explain the strong anisotropy of the dissipation tensors observed. Two important results of the present  study 
are that {\it local isotropy}
is never valid inside the outer limit of the overlap region, $y/\delta_{99} \approx 0.1$; and that
the assumptions of {\it local axisymmetry} and {\it local homogeneity} fail  inside of $y^+ =100$. The implications of {\it homogeneity in planes parallel to the wall} is introduced to partially explain observations throughout the wall layer. The dissipation characteristics in this very near wall region show that $\varepsilon_{ij}$ is close to but different from $\mathcal{D}_{ij}$ .

\end{abstract} 

\section{Introduction}
\label{sec:introduction}

This contribution adresses the behaviour of the turbulence dissipation in the near-wall region of boundary layers and channels flows. A first part focussed only on the dissipation, $\varepsilon$, of the turbulence kinetic energy (TKE), $k$,  is presented in a separate Part I paper \cite{stanislas20}. Part I also included an extensive review of earlier experimental and DNS contributions, so that will not be repeated here. The details of the experimental technique and data processing tools used to obtain the present results are provided in Part III \cite{foucaut20} of this contribution as a separate paper targeted mainly toward experimentalists. In this paper, the different classical hypothesis --- {\it local isotropy}, {\it local axisymmetry} and {\it local homogeneity} --- are examined in detail for the whole flow before focussing on the very near wall region below $y^+ = 100$.  Before entering into the details of the results and their analysis, it is important to review the main equations governing the problem and to clarify the notations and denominations as these have  in the past been a significant source of misunderstanding.

\subsection{Review of Basic Equations \label{subsec-basic}}

First, if $u_i$ represents the fluctuating velocity components and $U_i$ the mean components, the Reynolds Stresses transport equations are:

\begin{eqnarray}
\begin{gathered}
\rho \frac{\partial \langle u_iu_j \rangle }{\partial t} + \rho U_l\frac{\partial \langle u_iu_j \rangle }{\partial x_l} = 
- \rho  \langle u_iu_l \rangle \frac{\partial U_j}{\partial x_l} - \rho  \langle u_ju_l \rangle\frac{\partial U_i}{\partial x_l}\\
+ \frac{\partial}{\partial x_l}\left(\langle u_i\tau_{jl} \rangle + \langle u_j\tau_{il} \rangle - \langle u_i p \rangle\delta _{jl} - \langle u_j p \rangle\delta _{il} - \rho\langle u_i u_ju_l \rangle\right)\\
+ \left(\langle p\frac{\partial u_i}{\partial x_j} \rangle + \langle p\frac{\partial u_j}{\partial x_i} \rangle\right)
- \langle \tau_{il}\frac{\partial u_j}{\partial x_l} \rangle
- \langle \tau_{jl}\frac{\partial u_i}{\partial x_l} \rangle
\end{gathered}
\label{rs_eq}
\end{eqnarray}
where $\tau_{ij} = 2\mu s_{ij} = \mu \left(\frac{\partial u_i}{\partial x_j} + \frac{\partial u_j}{\partial x_i}\right)$ is the viscous stress tensor (in a flow assumed to be incompressible of a Newtonian fluid) and $s_{ij} =  \frac{1}{2}\left(\frac{\partial u_i}{\partial x_j} + \frac{\partial u_j}{\partial x_i}\right)$ is the strain-rate tensor. 

From equation (\ref{rs_eq}), the dissipation tensor of the Reynolds stresses can be defined as:

\begin{equation}
\varepsilon _{ij} =  \langle \tau_{il}\frac{\partial u_j}{\partial x_l} \rangle +
\langle \tau_{jl}\frac{\partial u_i}{\partial x_l} \rangle	
\label{epsij}
\end{equation}
which, for an incompressible Newtonian fluid reduces to:

\begin{equation}
\varepsilon _{ij} =  2\nu \langle \frac{\partial u_i}{\partial x_l}\frac{\partial u_j}{\partial x_l}\rangle + \nu \langle \frac{\partial u_l}{\partial x_i}\frac{\partial u_j}{\partial x_l}\rangle + \nu \langle \frac{\partial u_l}{\partial x_j}\frac{\partial u_i}{\partial x_l}\rangle
\label{epsij2}
\end{equation}

By eliminating the cross-products terms of equation (\ref{epsij2})  between the dissipation and the viscous diffusion terms of equation (\ref{rs_eq}), one obtains the Reynolds stress transport equation used by most turbulence modellers:

\begin{eqnarray}
\begin{gathered}
\rho \frac{\partial \langle u_iu_j \rangle }{\partial t} + \rho U_l\frac{\partial \langle u_iu_j \rangle }{\partial x_l} = 
- \rho  \langle u_iu_l \rangle \frac{\partial U_j}{\partial x_l} - \rho  \langle u_ju_l \rangle\frac{\partial U_i}{\partial x_l}\\
+ \frac{\partial}{\partial x_l}\left(\mu\frac{\partial}{\partial x_l}[\langle u_iu_j \rangle] - \langle u_i p \rangle\delta _{jl} - \langle u_j p \rangle\delta _{il} - \rho\langle u_i u_ju_l \rangle\right)\\
+ \left(\langle p\frac{\partial u_i}{\partial x_j} \rangle + \langle p\frac{\partial u_j}{\partial x_i} \rangle\right)
- \rho \mathcal{D}_{ij}
\end{gathered}
\label{rs_dij}
\end{eqnarray}
where $\mathcal{D}_{ij}$ is properly called the pseudo-dissipation tensor (sometimes erronously called the homogeneous dissipation) defined as:

\begin{equation}
\mathcal{D}_{ij} = 2\nu \langle \frac{\partial u_i}{\partial x_l}\frac{\partial u_j}{\partial x_l}\rangle 
\label{dij}
\end{equation}
From equations (\ref{epsij2}) and (\ref{dij}) the dissipation tensor $\varepsilon _{ij}$ can be rewritten as:

\begin{equation}
\varepsilon_{ij} = \mathcal{D}_{ij} + \nu \langle \frac{\partial u_l}{\partial x_i}\frac{\partial u_j}{\partial x_l}\rangle + \nu \langle \frac{\partial u_l}{\partial x_j}\frac{\partial u_i}{\partial x_l}\rangle
\label{epsijdij}
\end{equation}

Taking half of the trace of equation (\ref{rs_eq}), gives the transport equation for the turbulent kinetic energy $k$ as:

\begin{equation}
\rho\frac{D k}{D t} = \rho\langle u_iu_j \rangle\frac{\partial U_j}{\partial x_j} +   \frac{\partial}{\partial x_j} \left[-\langle pu_j \rangle_- \frac{\rho}{2}\langle u_iu_iu_j \rangle  +   \langle u_i \tau _{ij} \rangle \right] ~- ~ \rho  \varepsilon 
\label{eq:k}
\end{equation}
with $k = \frac{1}{2}u_iu_i$, $D/D t$ the material derivative and 

\begin{equation}
\varepsilon = \frac{1}{2}\varepsilon_{ii} = 2\nu s_{ij}s_{ij} = \nu\langle\frac{\partial u_i}{\partial x_j}\frac{\partial u_i}{\partial x_j}\rangle + \nu \langle \frac{\partial u_i}{\partial x_j}\frac{\partial u_j}{\partial x_i}\rangle
\label{eq:eps}
\end{equation}
This can be unambiguously identified as the true dissipation since it occurs with opposite sign in the entropy transport equation.

By analogy, a pseudo-dissipation $\mathcal{D}$ can be defined for the TKE as:

\begin{equation}
\mathcal{D} = \frac{1}{2} D_{ii} = \nu\langle\frac{\partial u_i}{\partial x_j}\frac{\partial u_i}{\partial x_j}\rangle
\label{eq:d}
\end{equation}
This pseudo-dissipation, $\mathcal{D}$, and the true dissipation, $\varepsilon$, are related by:

\begin{equation}
\varepsilon = \mathcal{D} +\nu \langle \frac{\partial u_i}{\partial x_j}\frac{\partial u_j}{\partial x_i}\rangle
\label{epsd} 
\end{equation}

The transport equation for the turbulence kinetic energy based on $\mathcal{D}$ then reduces to:

\begin{equation}
\rho\frac{\overline{D} k}{\overline{D} t} = \rho\langle u_iu_j \rangle\frac{\partial U_j}{\partial x_j} +   \frac{\partial}{\partial x_j} \left[-\langle pu_j \rangle_- \frac{\rho}{2}\langle u_iu_iu_j \rangle  +  \mu \frac{\partial k}{\partial x_j} \right] ~- ~ \rho  \mathcal{D} 
\label{eq:k_d}
\end{equation}
Note that, as for the Reynolds stresses above, the difference in the viscous transport terms between equations (\ref{eq:k}) and (\ref{eq:k_d}) is exactly  the difference between $\varepsilon$ and $\mathcal{D}$ as written in equation (\ref{epsd}). That's why the two terms simplify. Thus both equations are exact, but only equation (\ref{eq:k}) involves the true dissipation. Equation (\ref{eq:k_d}) is, however, the equation mostly used in $k-\varepsilon$ turbulence models. 

As it will be at the heart of the following discussions, it is worth to mention already here that an important consequence of the hypothesis of {\it homogenous} turbulence is that the indices of  derivative moments such as those appearing in equations (\ref{epsij2}) can be permuted; i.e., 

\begin{equation}
\langle { \frac{\partial u_i}{\partial x_m}~\frac{\partial u_j}{\partial x_n} } \rangle = \langle { \frac{\partial u_i}{\partial x_n}~\frac{\partial u_j}{\partial x_m} } \rangle
\label{eq:homogderiv4}
\end{equation}  
If the flow is also incompressible, then it follows immediately that 
\begin{equation}
\langle { \frac{\partial u_i}{\partial x_j}~\frac{\partial u_j}{\partial x_i} } \rangle = \langle { \frac{\partial u_i}{\partial x_i}~\frac{\partial u_j}{\partial x_j} } \rangle = 0
\label{eq:homogderiv5}
\end{equation}
which cancels the cross products appearing at the right of equations (\ref{epsijdij}) and (\ref{epsd}
The immediate consequence is that in homogeneous incompressible turbulence both $\mathcal{D}_{ij} = \varepsilon_{ij}$ and $\mathcal{D} = \varepsilon$.
This is more general than the fact that the cross-products simplify themselves between the dissipation and the viscous diffusion in equations (\ref{rs_eq}) and (\ref{eq:k}). Here both are zero.

Since very few flows are statistically homogenous in all directions, additional assumptions are required to justify any assumed or approximate equality of $\mathcal{D}$ and $\varepsilon$.  The most popular assumption is to introduce the idea of {\it local} in which assumptions are believed to apply only to quantities most influenced by the smallest scales of motion, but not by the larger scales.  Examples include {\it local isotropy, local axisymmetry} and {\it local homogeneity}.  So in the context of the above, {\it local homogeneity} would imply that $\mathcal{D} \approx \varepsilon$, even if the transport terms could not be treated as homogeneous. Before the introduction of the idea of {\it local homogeneity} by \cite{george91}, it was wrongly believed that {\it local isotropy} was a necessary requirement \cite{reynolds76}.  This is fortunate, since many studies (\cite{honkan2001,antoniaetal86,antonia91,Fanetal2015})  have shown that {\it local isotropy} (for the derivative moments) is not satisfied in most flows (at least in those we can measure).  

This paper extends those earlier studies to wall-bounded flows at modestly high Reynolds numbers, and examines all of the usual assumptions about velocity derivative moments.  And also introduces some new ones.

\subsection{The experiment and DNS \label{subsec:expt}}

In Part I of this contribution \cite{stanislas20}, an experiment was described, using dual-plane SPIV to determine all of the velocity derivatives needed to calculate the true dissipation in a turbulent boundary layer. Table 1 below (taken from that paper)  summarizes the experimental parameters.

\begin{table}[h!]
	\caption{Table showing experimental flow parameters}
	\label{table-parameters}       
	\begin{center}
		\begin{tabular}{llllllll}
			\hline\noalign{\smallskip}
			$Ue$ & $\delta$ &  $\delta*$ & $\theta$ & $u_\tau$ & $Re_\theta$ & $Re_\tau$ & $C_f$  \\
			\noalign{\smallskip}\hline\noalign{\smallskip}
			3 m/s & 0.32 m & 48.2 mm & 36.2 mm & 0.113 m/s & 7634 & 2598 & 0.00275\\
			\noalign{\smallskip}\hline
		\end{tabular}
	\end{center}
\end{table}
The multi-plane SPIV allowed relatively noise-free measurements using techniques described briefly there, but in detail in Part III \cite{foucaut20}. Also utilized in Part I \cite{stanislas20} (and described therein) was a channel DNS at a value of  $H^+ = u_\tau H/2\nu$ where $H$ is the channel height and $u_\tau$ is the friction velocity, comparable to $\delta^+ = u_\tau \delta/\nu$.  For both sets of data, all of the derivative moments necessary to compute the dissipation directly were measured or computed, including the often-neglected cross-moments $\langle [\partial u_i/\partial x_j][\partial x_j/\partial x_i]\rangle$.

As reported in Part I \cite{stanislas20}, both sets of data were in excellent agreement with each other in  the overlap region, and compared favorably with the earlier hot-wire measurements of \cite{balint91,honkan97} at lower Reynolds numbers.  And both showed evidence of an overlap region from $100 < y^+ < 800$ approximately, in agreement with asymptotic analysis of \cite{george97b,wosnik00}, which argued that for the channel the dissipation varies as $\varepsilon^+ \propto 1/y^+$,
and for the boundary layer as $\varepsilon^+ \propto  1/{y^+}^{\gamma-1}$. Both theoretical prediction were well supported by the data but,  since $\gamma << 1$, the second result was indistinguishable from the first one.

\subsection{The goal of this paper \label{subsec:goal}}

What was  clear from the many derivative moments profiles presented in part I of this contribution \cite{stanislas20}, and will be discussed in more detail in the following, was that these derivative moments surely do not satisfy the conditions for {\it local isotropy}. 
The term {\it local} is used to refer to the idea that even though the flow as a whole does not satisfy the conditions, they may still apply locally in space for the smallest scales.  For example the high wavenumber spectra are often assumed to be {\it locally isotropic} e.g., \cite{saddoughi94}), in part because of the absence of other useful spectral relations.  
In this paper the term {\it local} is used only to apply to the derivative moments relations.
The defining relations will be reviewed below and used to evaluate the data.


This paper also reviews the dissipation rate tensor, $ \varepsilon_{ij}$ as defined by equation (\ref{epsij}), and the assumptions usually employed to simplify it.
Of particular interest is whether the velocity derivative moments in turbulent boundary layers can be assumed to be {\it locally isotropic}, {\it locally axisymmetric}, or even {\it locally homogeneous}.  It was noted in Part I \cite{stanislas20} (see appendix A) of this contribution that only homogeneity and incompressibility are required for $\varepsilon_{ij} = \mathcal{D}_{ij}$. This later is of particular importance since it is assumed by ALL turbulence models (e.g., \cite{pope00,leschziner2015}), with the single exception of \cite{jakirlic02}.

The following sections proceed from the least general (local isotropy) to the most general (local homogeneity). Note that  a flow can not be {\it locally} isotropic if it is not also {\it locally axisymmetric}.  And it cannot be {\it locally axisymmetric}  if it is not also {\it locally homogeneous}.  Finally, the very near wall region ($y^+ < 100$) is examined in detail.  

\section{Local isotropy \label{app-isotropic}}
By far the most common assumption by all experimenters has been the assumption of {\it local isotropy}, an idea originally introduced by Taylor\cite{Taylor1935}. (Note that the deductions from {\it local isotropy}  are often confused with implications from just continuity and {\it local homogeneity}  as noted by \cite{george91}.)  When applied to velocity derivatives, {\it local isotropy}  demands that all  mean square derivatives obey the isotropic relations; in particular:
\begin{eqnarray}
\langle { \left[ \frac{\partial u_1}{\partial x_1} \right]^2 } \rangle & = & \langle { \left[ \frac{\partial u_2}{\partial x_2} \right]^2 } \rangle = \langle { \left[ \frac{\partial u_3}{\partial x_3} \right]^2 } \rangle \label{iso1}\\
\langle { \left[ \frac{\partial u_1}{\partial x_2} \right]^2 } \rangle & =  & \langle { \left[ \frac{\partial u_2}{\partial x_1} \right]^2 } \rangle = \langle { \left[ \frac{\partial u_1}{\partial x_3} \right]^2 } \rangle = \langle { \left[ \frac{\partial u_3}{\partial x_1} \right]^2 } \rangle \\ & = & \langle { \left[ \frac{\partial u_2}{\partial x_3} \right]^2 } \rangle = \langle { \left[ \frac{\partial u_3}{\partial x_2} \right]^2 } \rangle = 2 \langle { \left[ \frac{\partial u_1}{\partial x_1} \right]^2 } \rangle \nonumber \\
\langle { \left[ \frac{\partial u_1}{\partial x_2}\frac{\partial u_2}{\partial x_1} \right] } \rangle &  = &\langle { \left[ \frac{\partial u_1}{\partial x_3}\frac{\partial u_3}{\partial x_1} \right] } \rangle = \langle { \left[ \frac{\partial u_2}{\partial x_3}\frac{\partial u_3}{\partial x_2} \right] } \rangle = -\frac{1}{2} \langle { \left[ \frac{\partial u_1}{\partial x_1} \right]^2 } \rangle \label{iso3}
\end{eqnarray}
Consequently, for local isotropy, there is only one-independent derivative moment so any one can be chosen.  The two most popular reduce the dissipation to: 
\begin{equation}
\varepsilon = 15 \nu \langle { \left[ \frac{\partial u_1}{\partial x_1} \right]^2 } \rangle  = \frac{15}{2} \langle { \left[ \frac{\partial u_2}{\partial x_1} \right]^2 } \rangle .
\label{iso_simp}
\end{equation}
Results of both (usually obtained using Taylor's hypothesis) are often cited side-by-side in the literature, even though they often yield very different answers.   \cite{antoniaetal86,antonia91} tabulate results from a large number of flows, almost none of which are consistent with the {\it local isotropy}  assumption (see also \cite{george91}). Nonetheless, this {\it local isotropy}  assumption is still widely used, mostly for the lack of easy alternatives.  A certain amount of luck would be required to obtain a reasonable estimate from the  measurement of a single derivative, especially when the Taylor hypothesis of frozen turbulence is used as well.  

\begin{figure}
	\resizebox{0.5\linewidth}{!}{\includegraphics[scale=1]{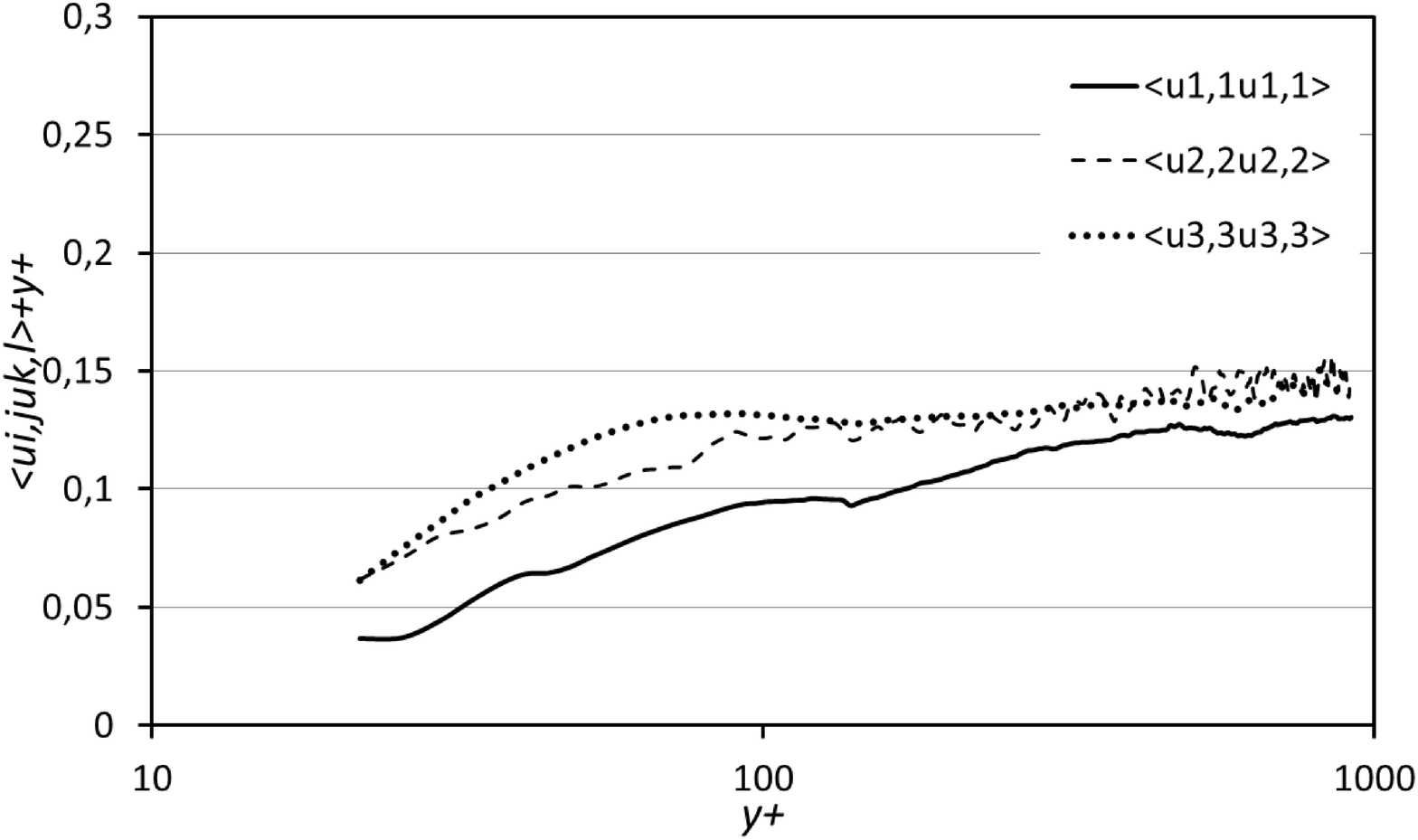}} 
	\resizebox{0.5\linewidth}{!}{\includegraphics[scale=1]{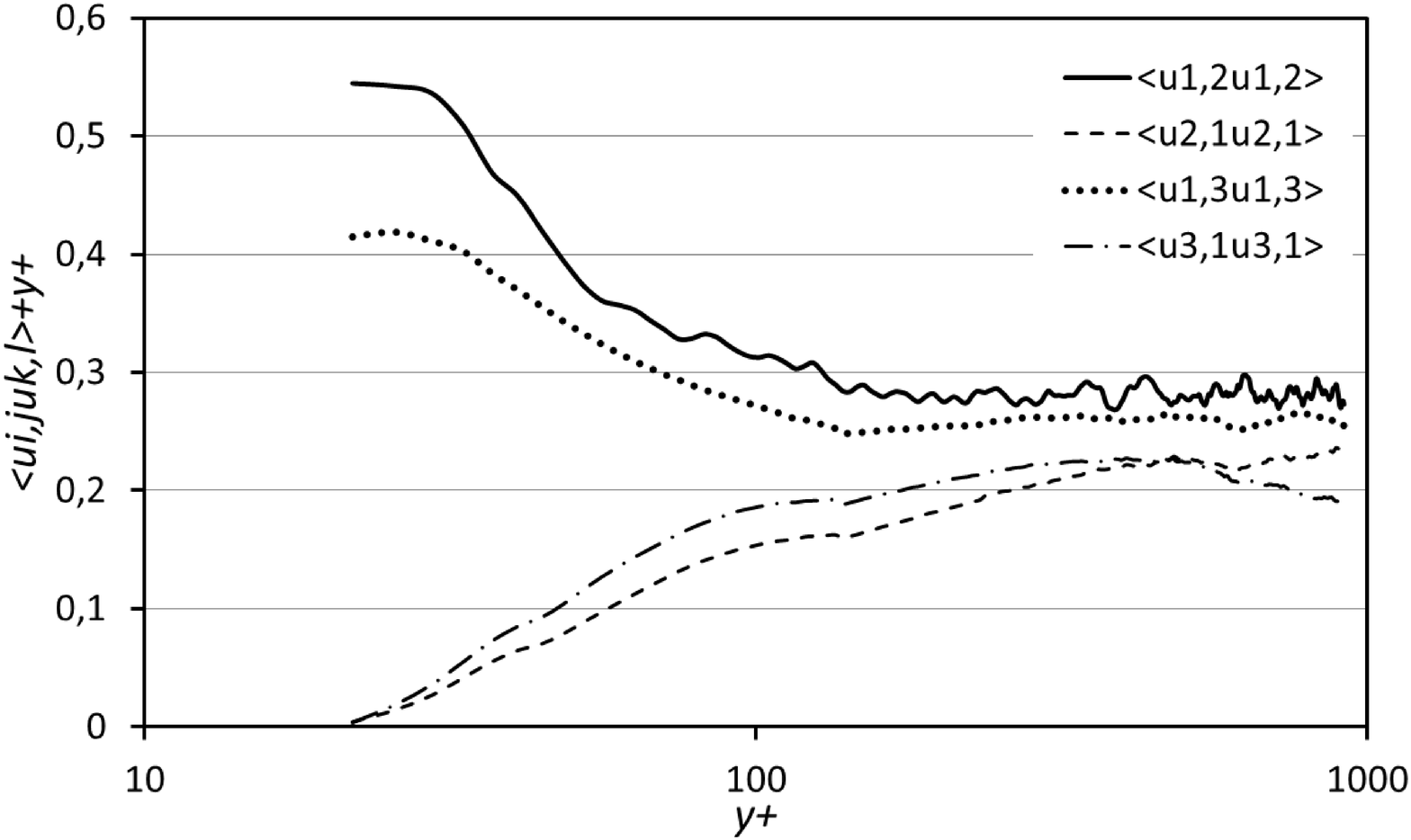}}\\
	\hspace*{3 cm}(a) \hspace{5 cm}(b)\\
	\resizebox{0.5\linewidth}{!}{\includegraphics[scale=1]{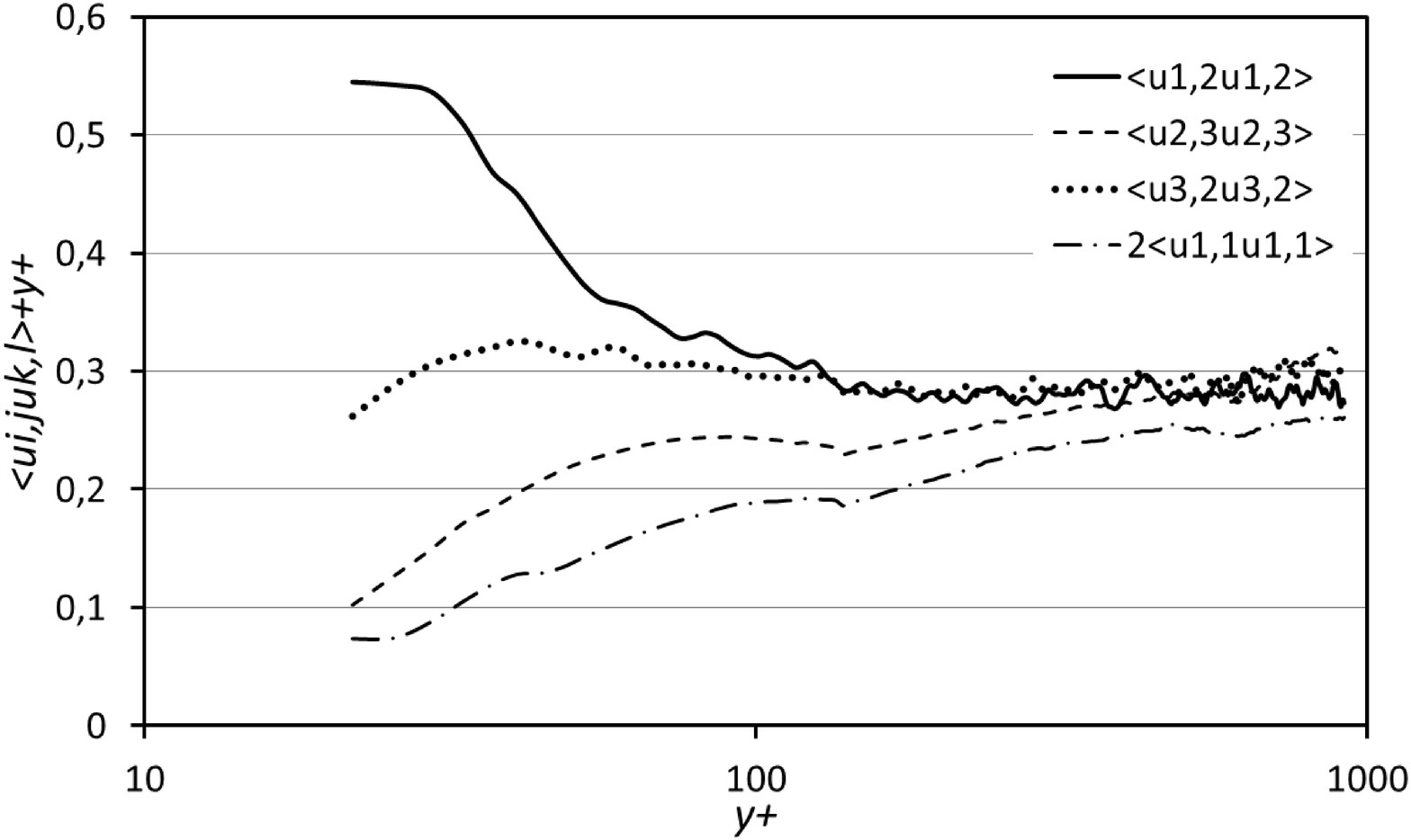}}
	\resizebox{0.5\linewidth}{!}{\includegraphics[scale=1]{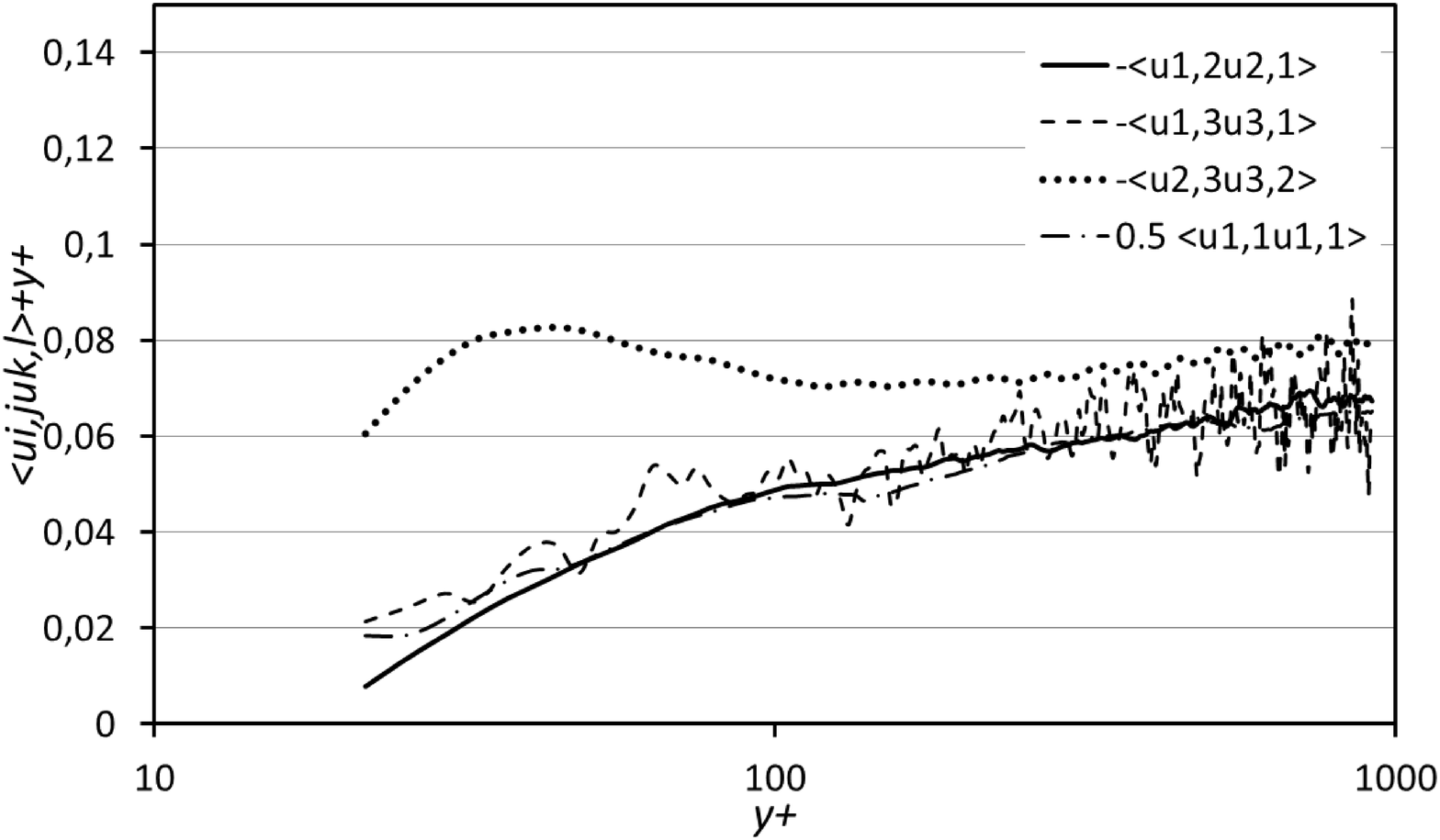}}\\
	\hspace*{3 cm}(c) \hspace{5 cm}(d)
	\caption{Isotropy test of experimental data showing lack of isotropy throughout wall and overlap regions (see equations: (\ref{iso1}) to (\ref{iso3}). Note that both ordinates have been multiplied by $y^+$.}
	\label{fig:spiviso}      
\end{figure}

\begin{figure}
	\resizebox{0.95\linewidth}{!}{\includegraphics[scale=1]{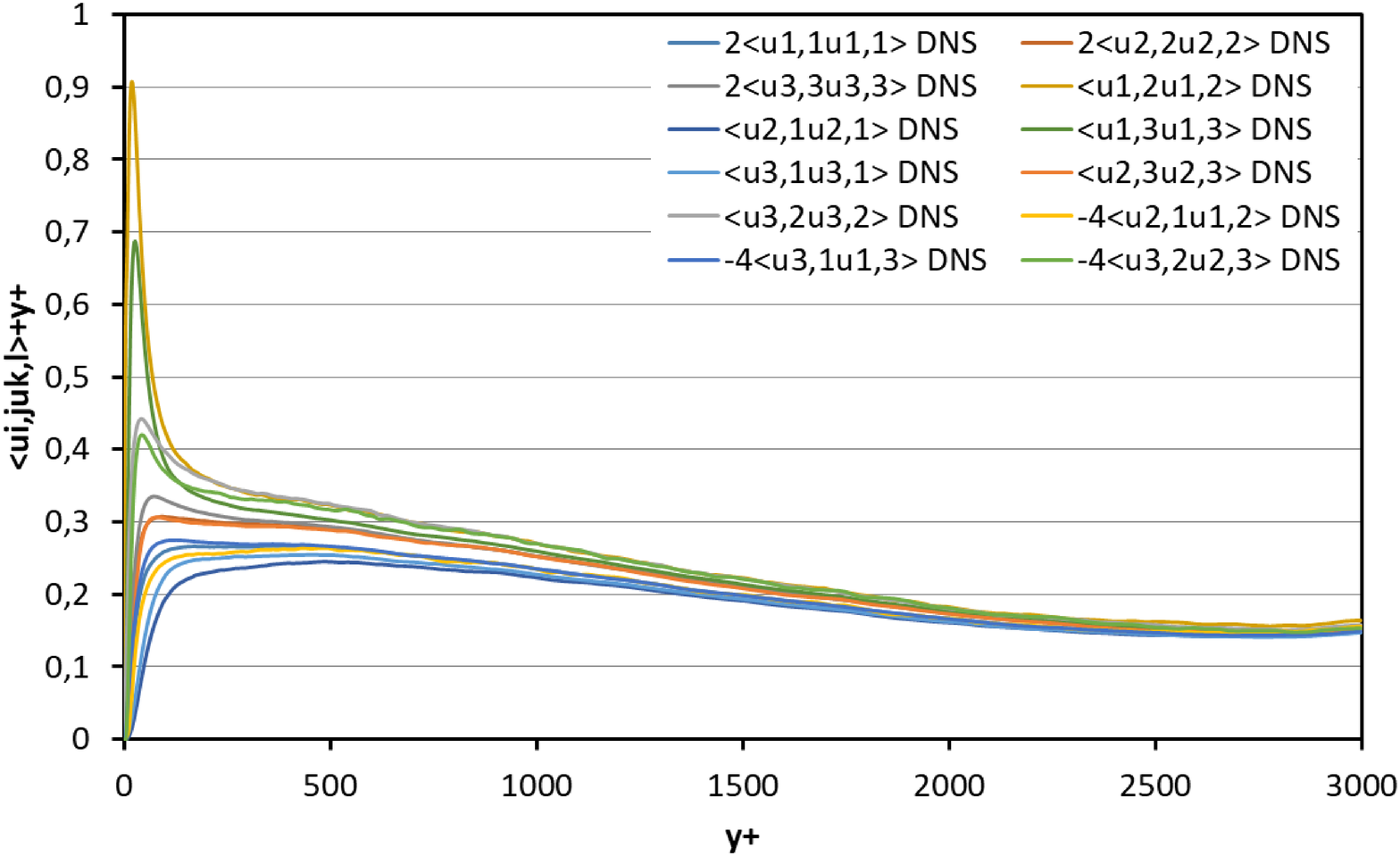}} \\
	\resizebox{0.95\linewidth}{!}{\includegraphics[scale=1]{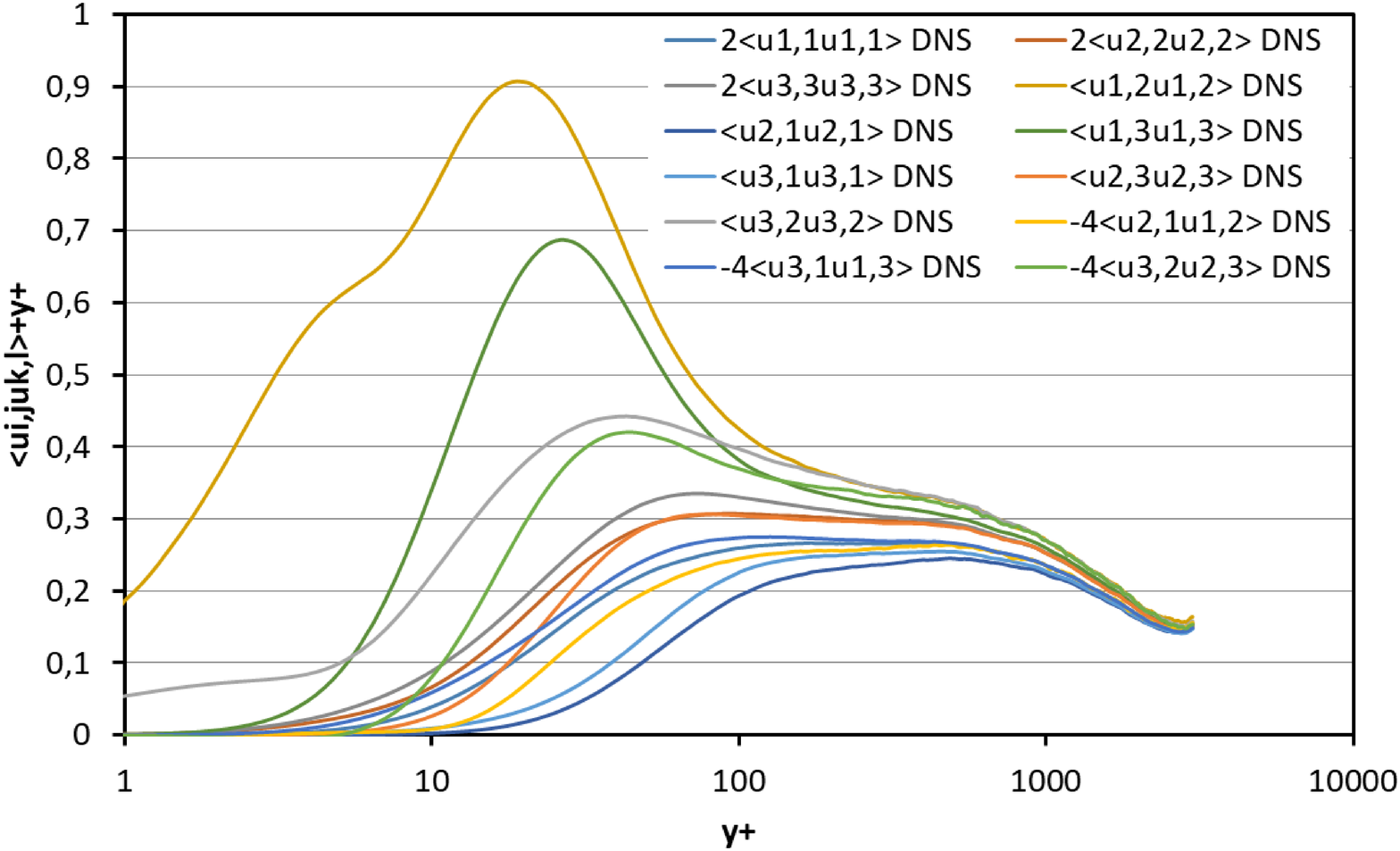}}
	\caption{Isotropy test of DNS data showing lack of isotropy for $y^+ < 2000$. Note that both ordinates have been multiplied by $y^+$.}
	\label{fig:dnsiso}     
\end{figure}

Figure~\ref{fig:spiviso} gives some of the SPIV derivative moments in inner variables  multiplied by $y^+$ in semi-log  plots.  The data have been plotted in groups (a) to (d)  to show  the expected equalities of equations (\ref{iso1}) to (\ref{iso3}), and  multiplied by the appropriate factors to make them equal if the flow were {\it locally isotropic}.  Figure~\ref{fig:dnsiso} shows a more global plot with all the moments from the DNS data. Note that inside of $y^+=30$ the experimental data are affected by the limited spatial resolution, but the DNS data are not.  And note also that the channel flow DNS data extend to the center of the channel, while the boundary layer data extend only till about $y/\delta_{0.99} = 0.1$, which is near the outer limit of the overlap (or log) region.

Inside of $y^+ = 100$, the departures from isotropy are catastrophic. Unfortunately it will be seen below in sections \ref{app-axisymmetric} and \ref{loc_homgen} that other assumed statistical hypotheses (e.g., local axisymmetry, local homogeneity) fail as well. The anisotropy very close the wall ($y^+ <5$) can be explained by a simple Taylor expansion of the variables there, since the continuity equation together with the boundary conditions dictates the behavior in that region.  This has already been discussed extensively by DNS and turbulence modellers \cite{manceau02a}, \cite{gerolymos16}.
Outside of $y^+ = 100$, the departures from isotropy for both sets of data are on the order of 50\% or more, at least throughout the overlap region.  The channel flow data shows some tendency toward isotropy in the core region  ($y^+ > 1000 $ or so), but no assessment can be made for the boundary layer outside of the overlap region.

{\it So {\it local isotropy} is clearly a very bad assumption for this boundary layer flow everywhere, at least within the overlap region and below.}  As they are relevant mostly below $y^+ = 100$, the anisotropy diagrams for the dissipation tensors will be presented and discussed in Section~\ref{sec-planehomogeneity} below, and will be seen to show the same characteristics. In any case, as can be seen in figures \ref{fig:spiviso} and \ref{fig:dnsiso}, the streamwise mean square derivative, $\langle [\partial u_1/\partial x_1]^2\rangle$, is a particularly bad representative of the rest.  This is unfortunate since it has been historically most commonly used  with time derivatives and Taylor's hypothesis to estimate dissipation in all flows.

Whether the smallest scales of motion themselves are isotropic is another matter which cannot be resolved with the current data. As noted above, this is a separate question from whether the isotropic derivative relations are approximately satisfied.\footnote{The measurements of \cite{saddoughi94} show that spectra from a very high Reynolds number at NASA/Ames agree reasonably well with the isotropic spectral relations at high wavenumbers. In the absence of anisotropic spectral relations though this does not constitute proof, since the anisotropic contributions to the spectral relations could be negligible.  This is especially problematical since the anisotropy of the intensities in the experiment is only about 20\% or less.} Perhaps at higher Reynolds numbers the core region of the channel will tend toward isotropy -- say outside of $y^+ > 0.1 H^+$. This would be consistent with the local spectral suggestions of \cite{george97b} and  the observations of ~\cite{georgetutkun2011} (using measurements at higher Reynolds number in the same facility) that the outer part of the boundary layer only reaches a true multi-point Reynolds number independence when $\delta^+ > 3000$ approximately, and even then only for $y^+ > 0.1 \delta^+$.  When these conditions are not satisfied, all scales of motion are affected by viscosity, so any K41 type arguments simply cannot apply -- including any tendency toward isotropy and an inertial range in the energy spectra (or structure functions).

\section{Local axisymmetry \label{app-axisymmetric}}
\begin{figure}
	\resizebox{0.5\linewidth}{!}{\includegraphics[scale=1]{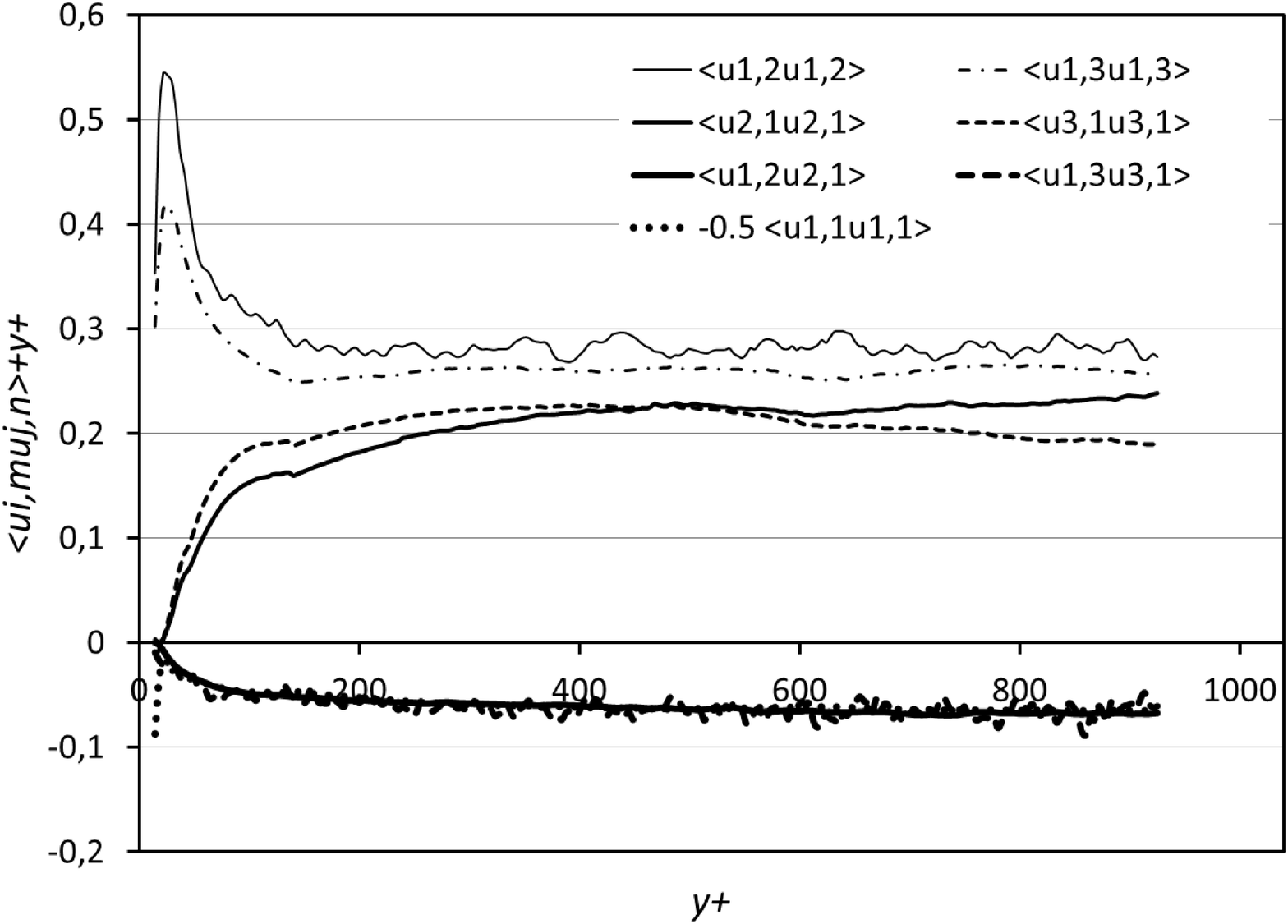}}
	\resizebox{0.5\linewidth}{!}{\includegraphics[scale=1]{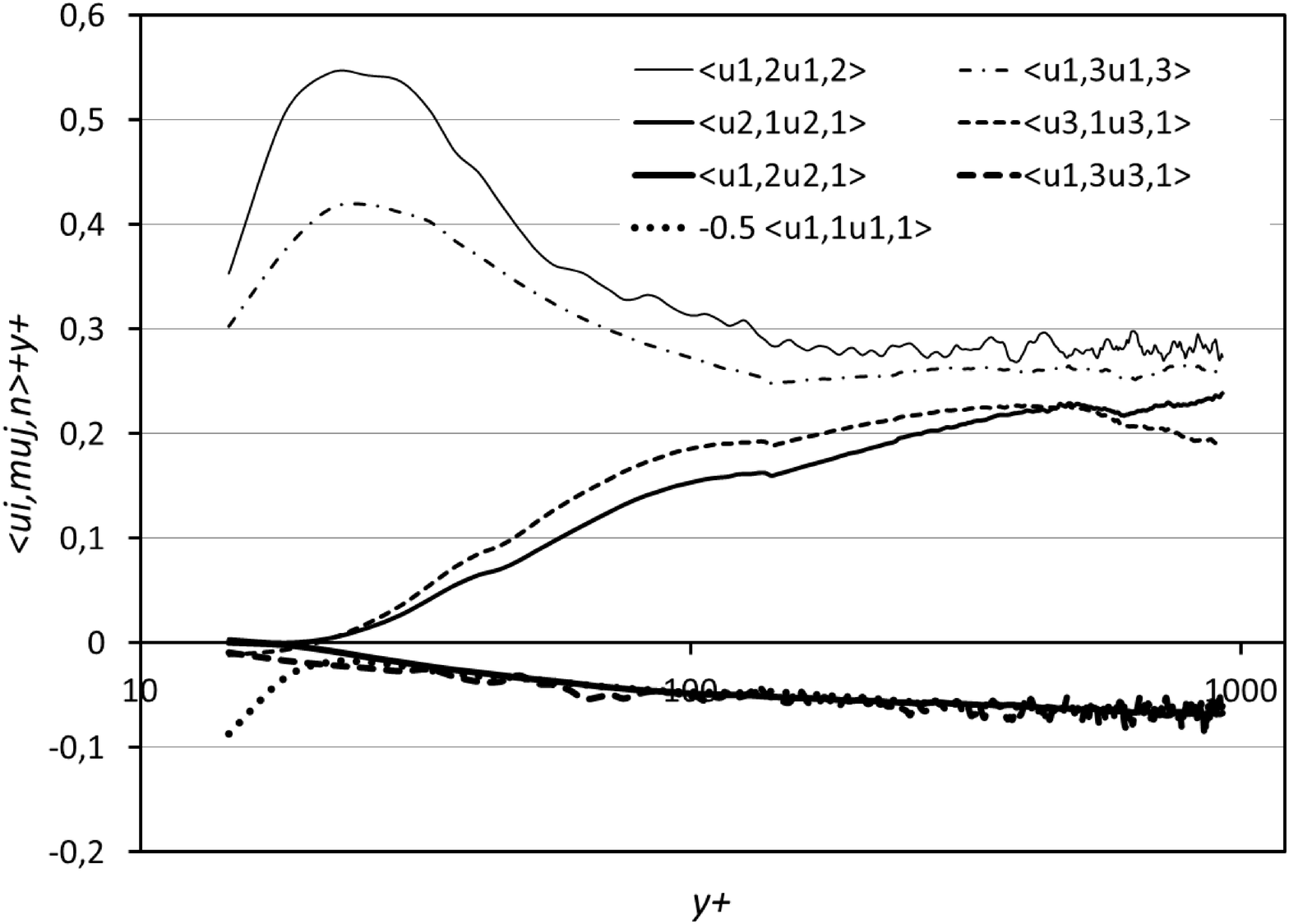}} \\
	\resizebox{0.5\linewidth}{!}{\includegraphics[scale=1]{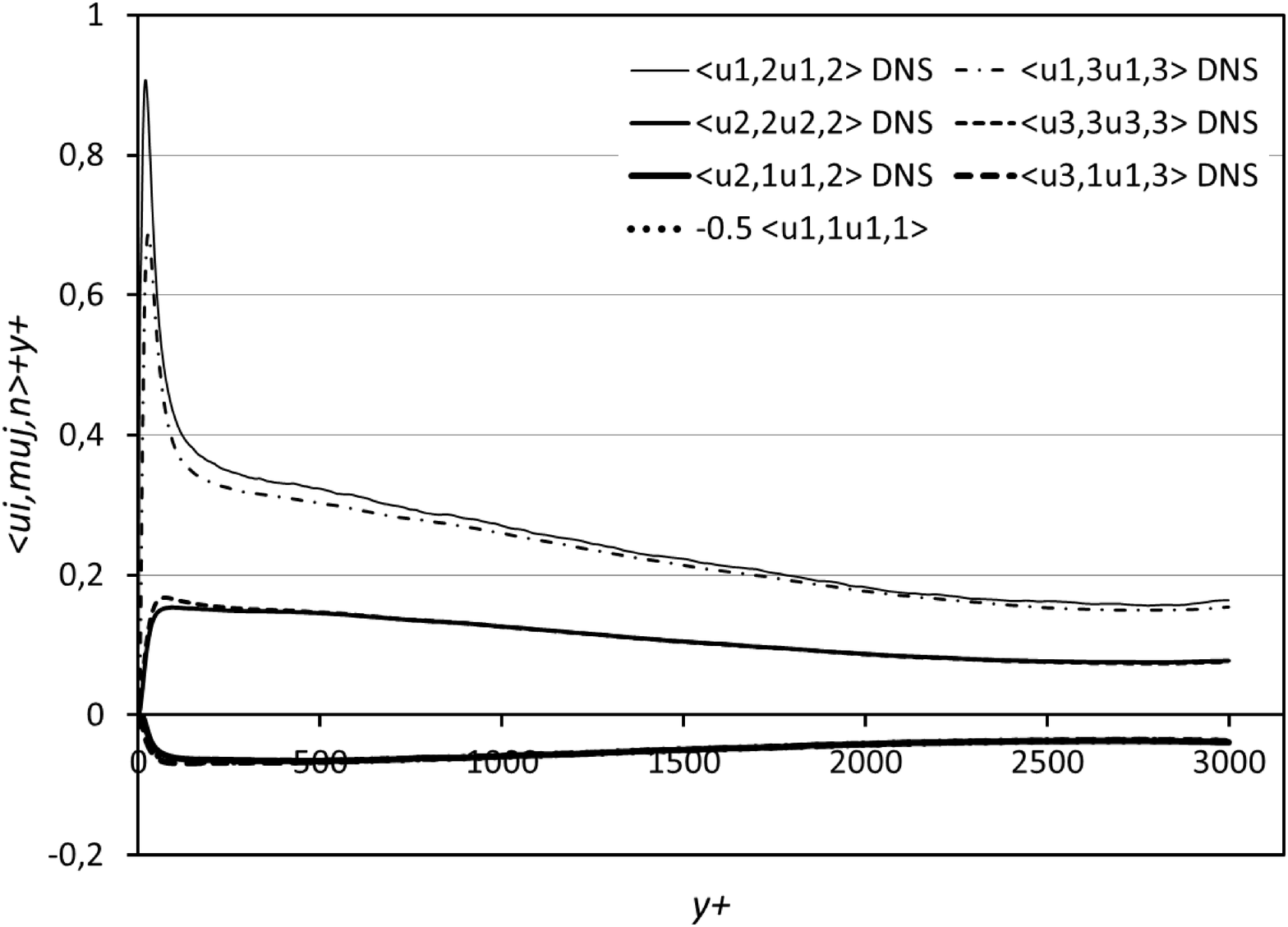}}
	\resizebox{0.5\linewidth}{!}{\includegraphics[scale=1]{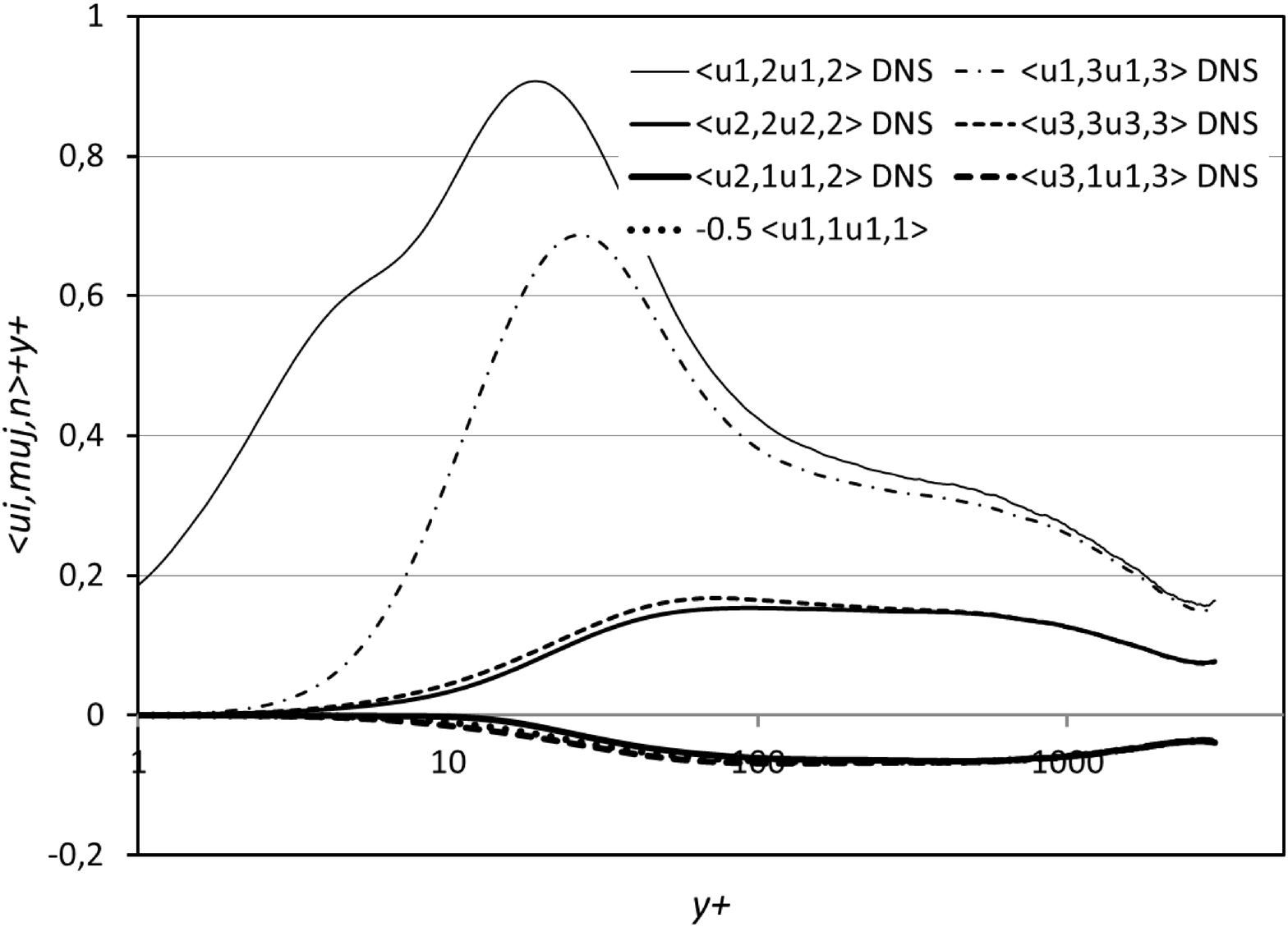}}
	\caption{Comparison of derivative moments from SPIV and DNS in both lin-lin and lin-log plots, showing support for local axisymmetry outside of $y^+=30$.}
	\label{fig:testaxi}      
\end{figure}

\begin{figure}
	\resizebox{0.5\linewidth}{!}{\includegraphics[scale=1]{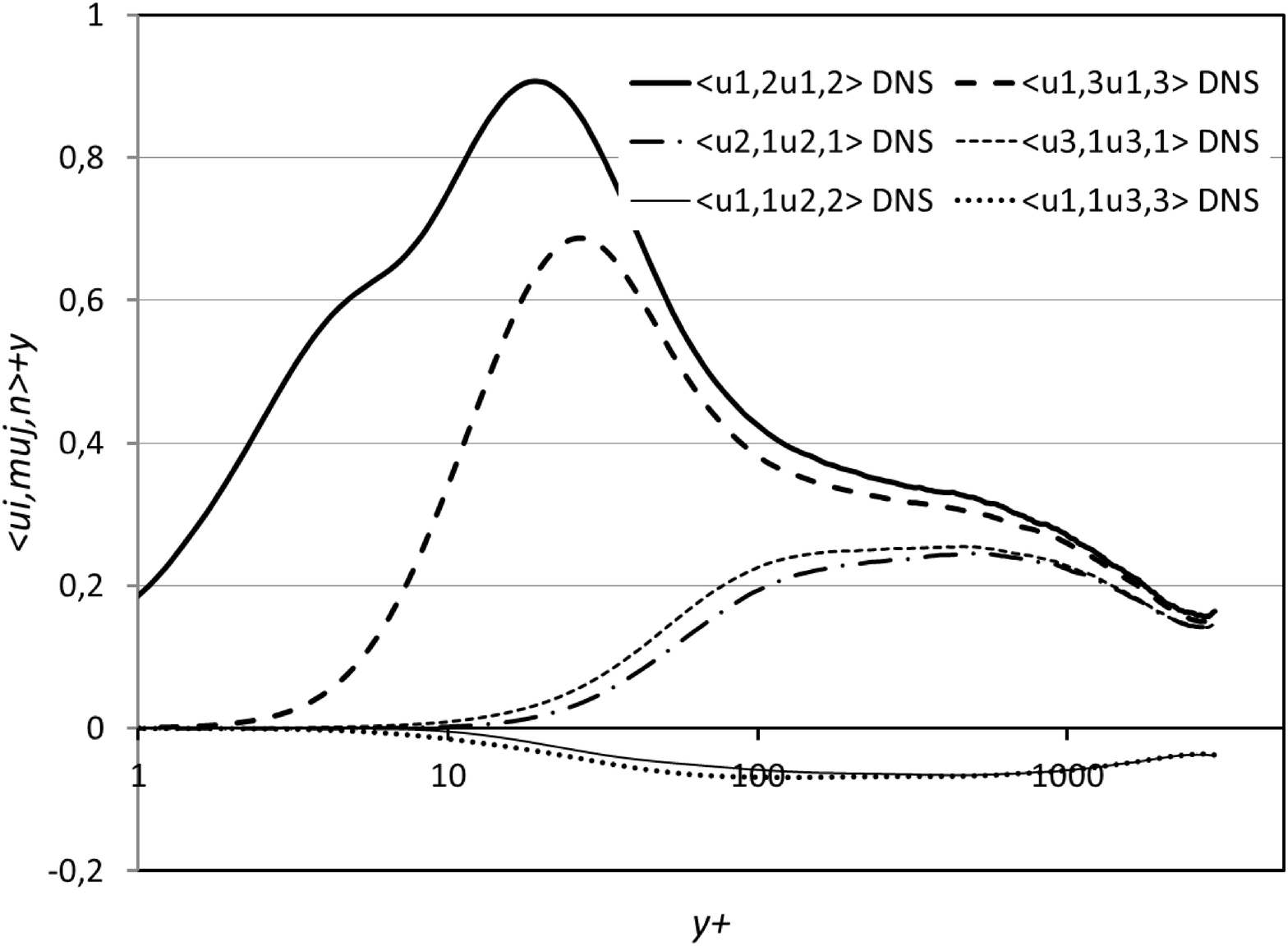}}
	\resizebox{0.5\linewidth}{!}{\includegraphics[scale=1]{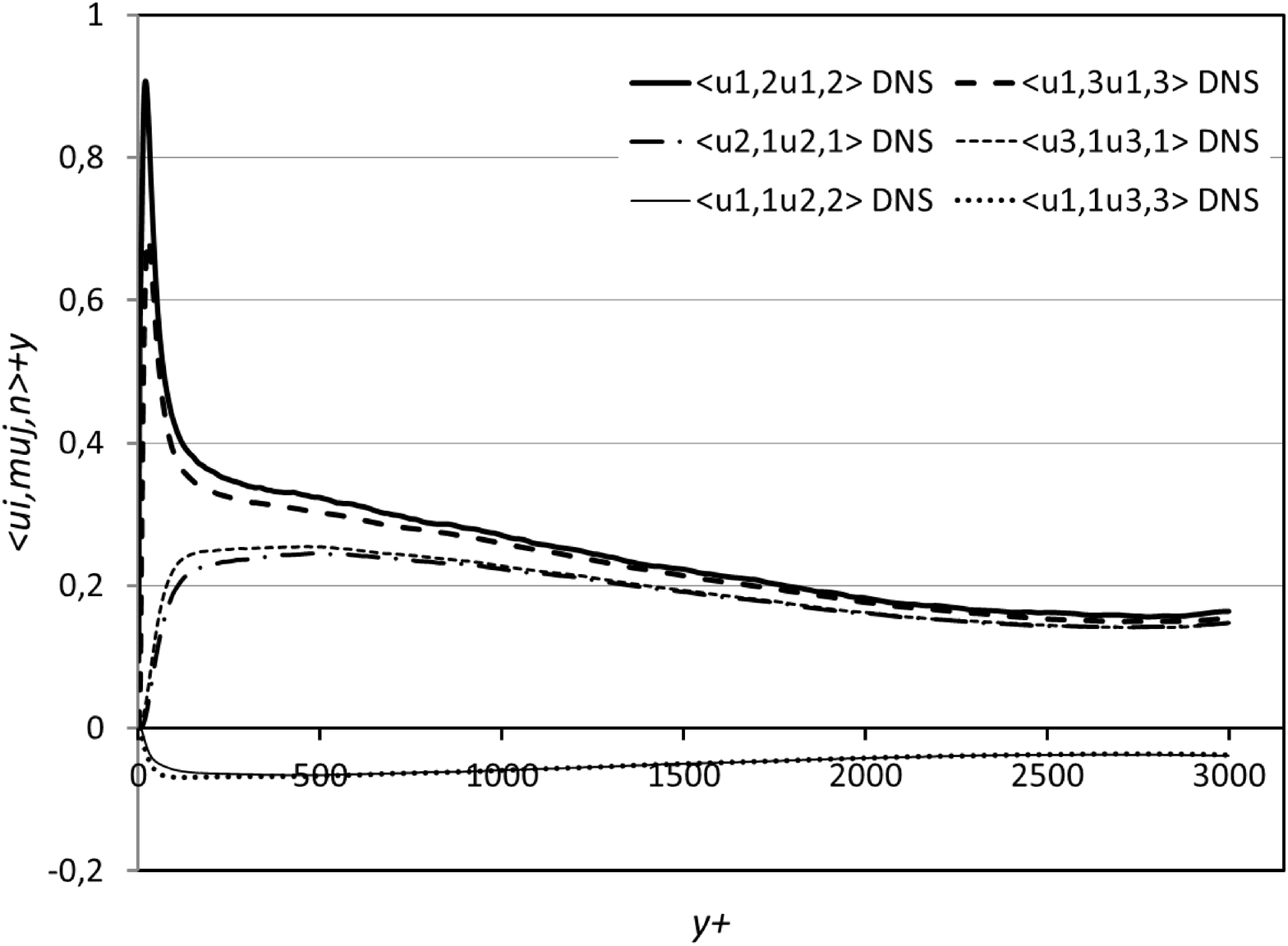}} \\
	\resizebox{0.5\linewidth}{!}{\includegraphics[scale=1]{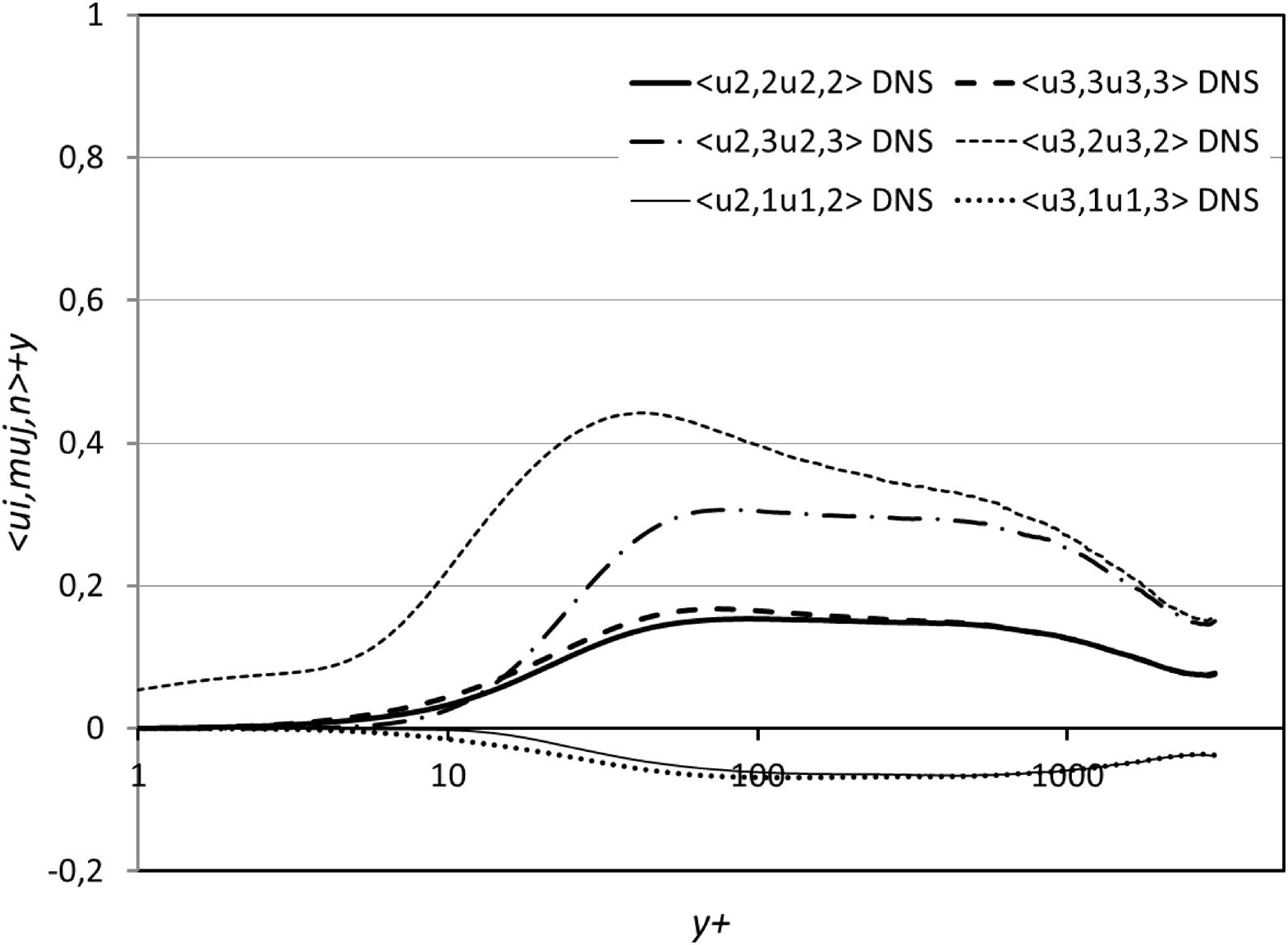}}
	\resizebox{0.5\linewidth}{!}{\includegraphics[scale=1]{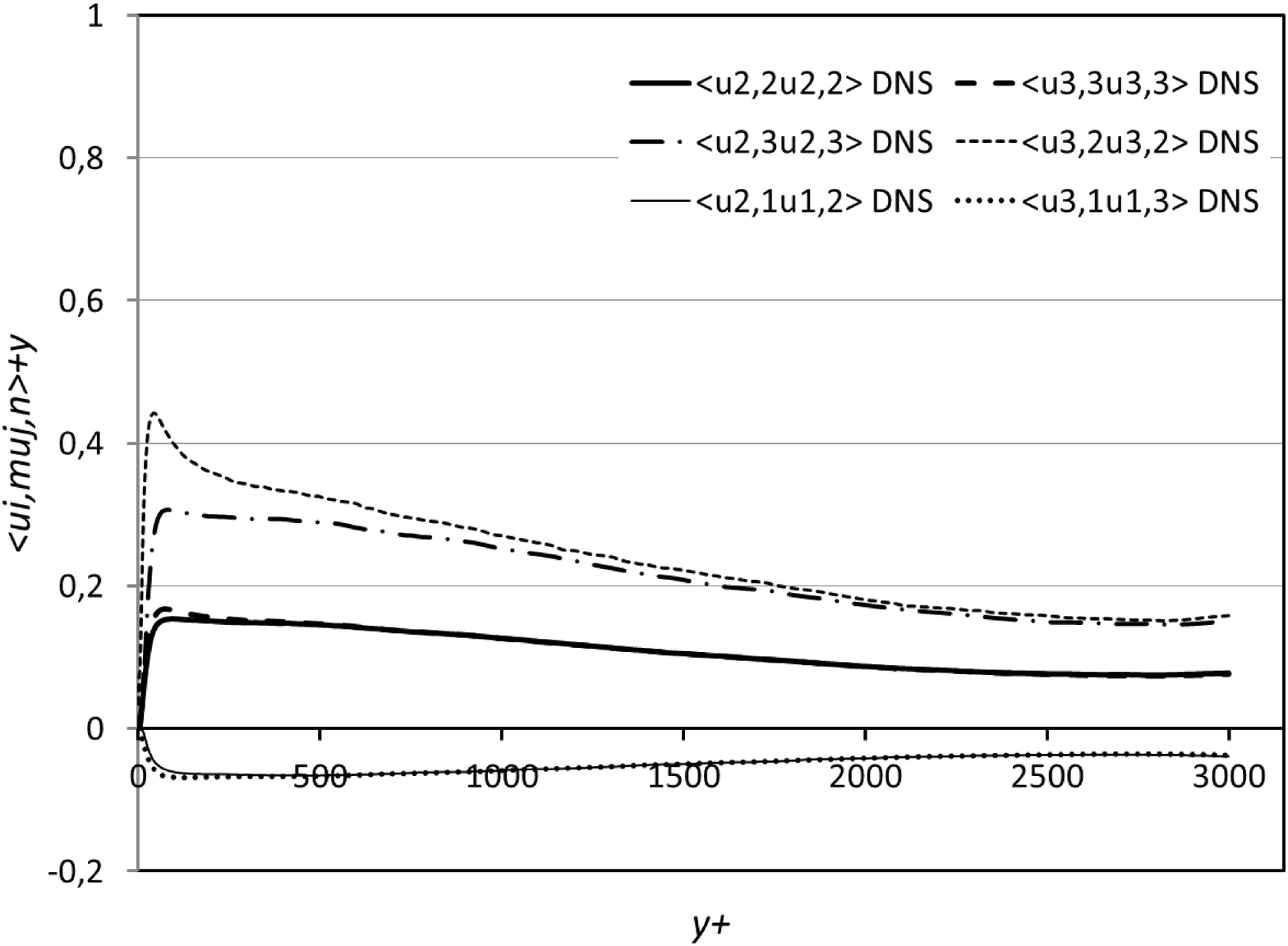}}
	\caption{Comparison of derivative moments from DNS showing support for local axisymmetry outside of $y^+=30$.}
	\label{fig:testaxidns}     
\end{figure}

The theory of axisymmetric turbulence was developed in parallel using different methodologies by \cite{batchelor46} and \cite{chandrasekar50}.  But it was \cite{george91} who realized its potential applicability to the turbulence derivative moments and coined the term {\it local axisymmetry}.  Note that the whole idea of {\it local} in this context is that the axisymmetric relations only apply to quantities dominated by the smallest scales of motion.  So the axisymmetric relations presented below only apply to the derivative moments, not to the velocity moments in general. 
\cite{george91} only developed the equations for an axis of symmetry which corresponded to the 1-axis.  (Other orthogonal orientations can be obtained by simply permuting the preferred axis-1 with a different axis.)  It was not obvious to them why the 1-axis should have been preferred, nor is it now.  But the data seemed to suggest strongly that the derivative moments arranged into pairs (with always one single exception).    {\it local axisymmetry}  (with the 1-axis as preferred) requires:

\begin{eqnarray}
\langle { \left[ \frac{\partial u_1}{\partial x_2} \right]^2 } \rangle & = &  \langle { \left[ \frac{\partial u_1}{\partial x_3} \right]^2 } \rangle \label{axi1}\\
\langle { \left[ \frac{\partial u_2}{\partial x_1} \right]^2 } \rangle & = &  \langle { \left[ \frac{\partial u_3}{\partial x_1} \right]^2 } \rangle \\
\langle { \left[ \frac{\partial u_2}{\partial x_2} \right]^2 } \rangle & = &  \langle { \left[ \frac{\partial u_3}{\partial x_3} \right]^2 } \rangle \\
\langle { \left[ \frac{\partial u_2}{\partial x_3} \right]^2 } \rangle & = &  \langle { \left[ \frac{\partial u_3}{\partial x_2} \right]^2 } \rangle \\
\langle { \left[ \frac{\partial u_2}{\partial x_2} \right]^2 } \rangle & = &  \frac{1}{3}\langle { \left[ \frac{\partial u_1}{\partial x_1} \right]^2 } \rangle  +  \frac{1}{3}\langle { \left[ \frac{\partial u_2}{\partial x_3} \right]^2 } \rangle\\
\langle { \left[ \frac{\partial u_2}{\partial x_3}  \frac{\partial u_3}{\partial x_2}\right] } \rangle & = &  \frac{1}{6}\langle { \left[ \frac{\partial u_1}{\partial x_1} \right]^2 } \rangle  - \frac{1}{3}\langle { \left[ \frac{\partial u_2}{\partial x_3} \right]^2 } \rangle\\
\langle { \left[ \frac{\partial u_1}{\partial x_2}\frac{\partial u_2}{\partial x_1} \right] } \rangle & = &  \langle { \left[ \frac{\partial u_1}{\partial x_3} \frac{\partial u_3}{\partial x_1}\right] } \rangle  = -  \frac{1}{2}\langle { \left[ \frac{\partial u_1}{\partial x_1} \right]^2 } \rangle\label{axi2}
\end{eqnarray}
As shown by figure \ref{fig:testaxi} for SPIV and figure \ref{fig:testaxidns} for the DNS data, all of the above relations can be checked across the boundary layer using the data presented herein.  Interestingly all appear to satisfy these relations outside of $y^+ > 100$ to an excellent approximation and some stay valid down to the wall. In fact they fail near the wall about the same place as {\it local homogeneity}  fails (as will be shown below). 

It is important to note that \cite{george91} pointed out that if the velocity derivative moments were strictly axisymmetric, then there could be no enstrophy (or dissipation) production direct from the mean shear, similar to the degeneracy of isotropy. So at least one derivative moment must not satisfy the axisymmetric relations -- as in fact observed in figures \ref{fig:testaxi} and \ref{fig:testaxidns} which show that two relations are more approximate in the outer part and clearly fail below $y^+ = 100$.  As we shall see in this paper, {\it local homogeneity}  breaks down about the same place, so the point appears to be moot. Interestingly it is the equality involving the 2-3 derivative which fails most near the wall. It is the same derivative which dominates the vorticity amplification or streamwise vorticity. Figures~\ref{fig:testaxi} and \ref{fig:testaxidns} show strong support for the hypothesis of {\it local axisymmetry}  outside of $y^+ =100$. Most strikingly, all the velocity derivatives moments except two pair up nicely.

Inside $y^+ =100$, is a different story.  {\it Both local axisymmetry} and {\it local isotropy} fail. As noted above, the anisotropy diagrams presented in Section~\ref{conclusion} below exhibit the same behavior. The reason for the failure of both inside $y^+ = 100$ will be obvious in the next section, and seen to be due to a complete failure of {\it local homogeneity} in this region, without which neither can be true.  Note, however, it will be seen below that in spite of the lack of symmetry of the derivative moments, for some combinations of derivatives the dissipation estimates near the wall will not be too bad.  This is only because some derivatives which are too large are compensated by those which are too small.

In any case, a clear advantage of {\it local axisymmetry}  is that only four derivative moments are independent, and they can be chosen for convenience.  Of the many combinations possible, \cite{george91} suggested and used two for the dissipation; namely,

\begin{eqnarray}
\varepsilon & \simeq & \nu \left\{  \langle { -\left[ \frac{\partial u_1}{\partial x_1} \right]^2 } \rangle + 8 
\langle { \left[ \frac{\partial u_2}{\partial x_2} \right]^2 } \rangle + 2 \langle { \left[ \frac{\partial u_1}{\partial x_2} \right]^2 } \rangle + 2 \langle { \left[ \frac{\partial u_2}{\partial x_1} \right]^2 } \rangle \right\}  \label{eq:axidissip1}
\end{eqnarray}
and
\begin{eqnarray}
\varepsilon & \simeq & \nu \left\{ \frac{5}{3} \langle { \left[ \frac{\partial u_1}{\partial x_1} \right]^2 } \rangle + 2 \langle { \left[ \frac{\partial u_1}{\partial x_2} \right]^2 } \rangle + 2 \langle { \left[ \frac{\partial u_2}{\partial x_1} \right]^2 } \rangle +  \frac{8}{3} \langle { \left[ \frac{\partial u_2}{\partial x_3} \right]^2 } \rangle \right\}.   \label{eq:axidissip2}
\end{eqnarray}
The first of these lends itself naturally to measurements in a plane (like planar PIV), but to our knowledge has not previously been used this way.  The second is most useful for the particular configuration  of parallel x-wires which can be rotated by 90 degrees. 
Another possibility, suited for SPIV measurements in a streamwise plane writes:

\begin{eqnarray}
\begin{gathered}
\varepsilon  \simeq  \nu \left\{ \frac{7}{3} \langle { \left[ \frac{\partial u_1}{\partial x_1} \right]^2 } \rangle + 4 \langle { \left[ \frac{\partial u_2}{\partial x_2} \right]^2 } \rangle + 2 \langle { \left[ \frac{\partial u_1}{\partial x_2} \right]^2 } \rangle + 2 \langle { \left[ \frac{\partial u_2}{\partial x_1} \right]^2 } \rangle \right\} \\
+  \nu \left\{ \frac{4}{3} \langle { \left[ \frac{\partial u_3}{\partial x_2} \right]^2 } \rangle   + 4 \langle   \frac{\partial u_1}{\partial x_2}\frac{\partial u_2}{\partial x_1} \rangle \right\}  \label{eq:axidissip3}
\end{gathered}
\end{eqnarray}

\begin{figure}
	\resizebox{0.5\linewidth}{!}{\includegraphics[scale=1]{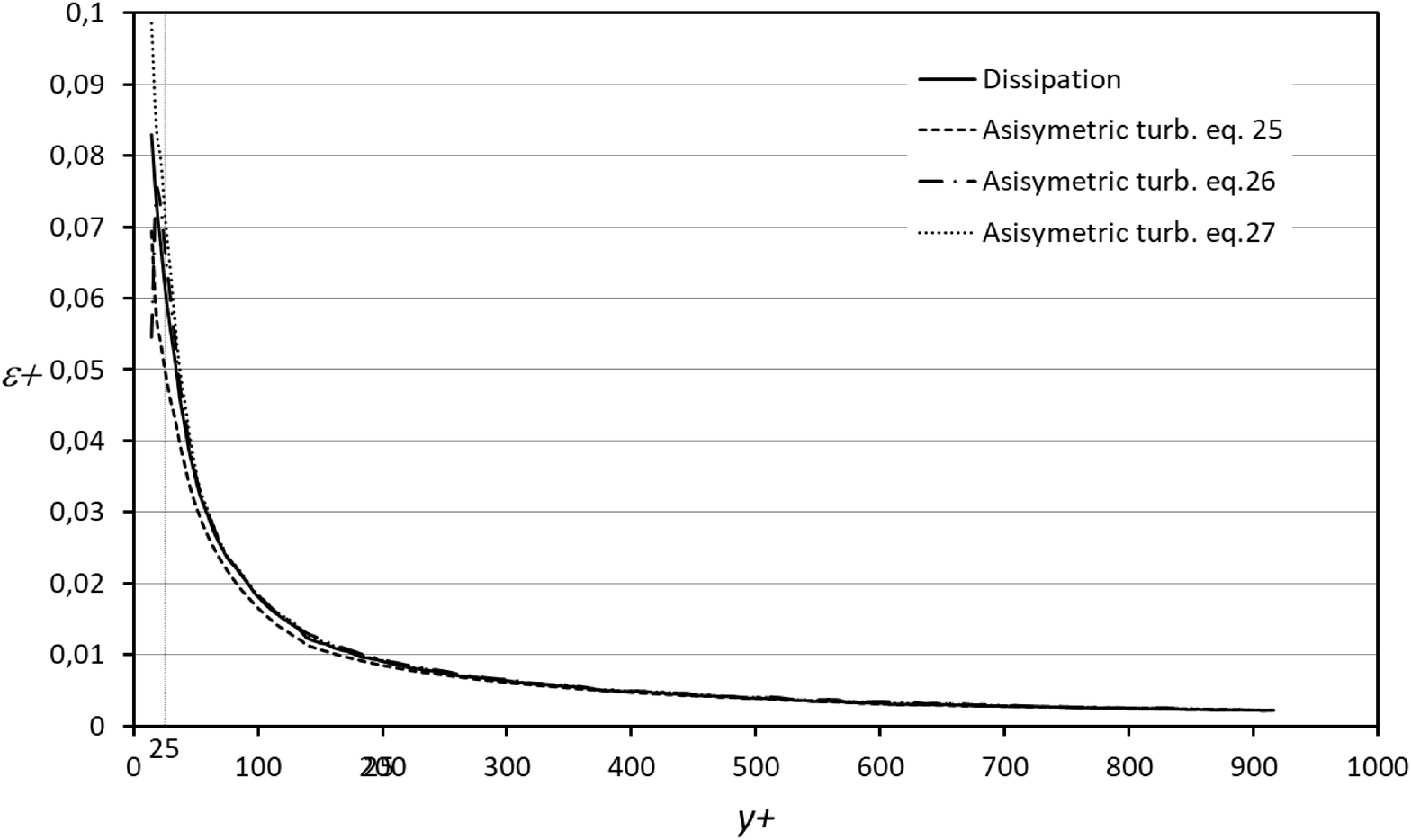}}
	\resizebox{0.5\linewidth}{!}{\includegraphics[scale=1]{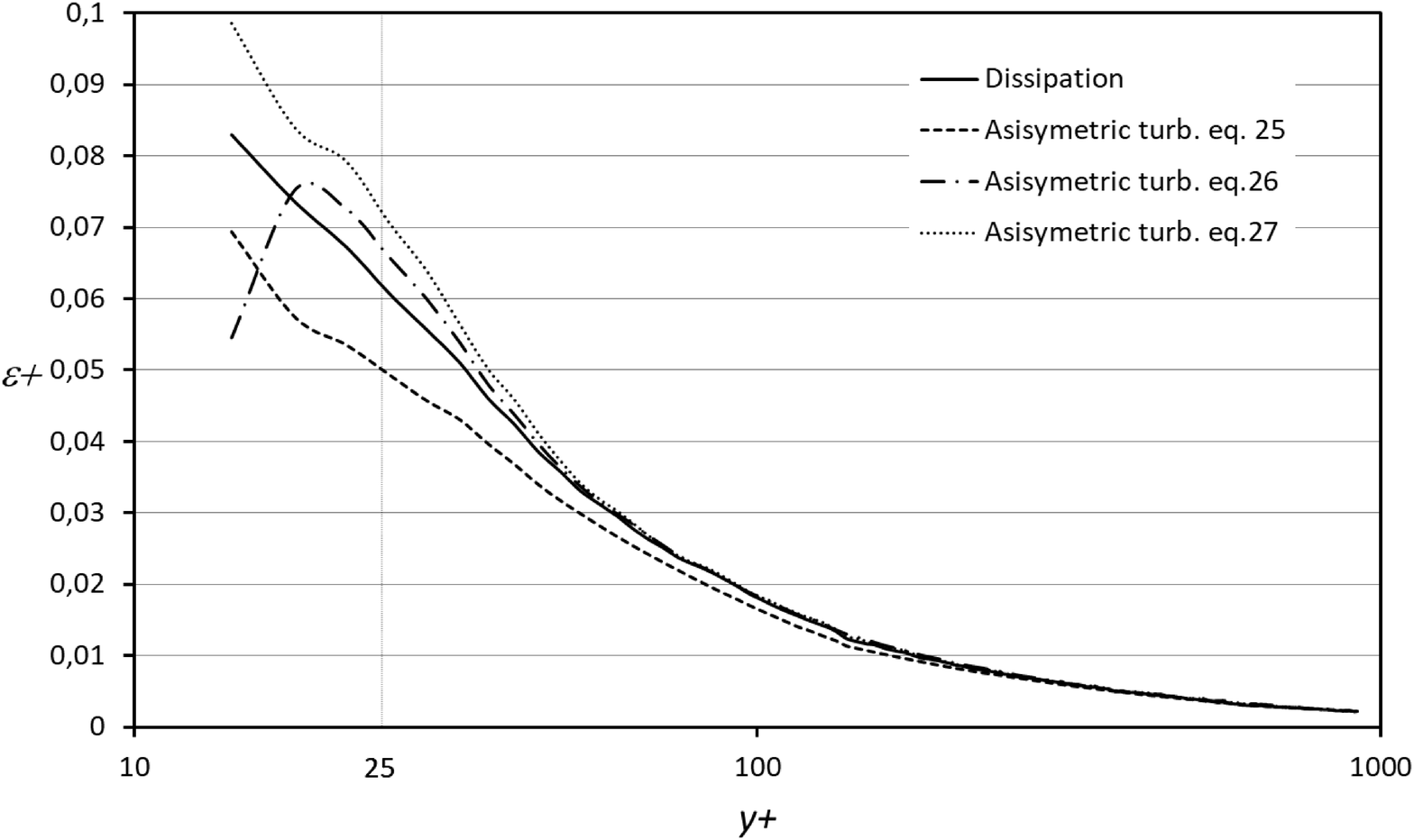}}  
	\caption{Validation of the  estimation of the full dissipation by the approximate equations (\ref{eq:axidissip1}), (\ref{eq:axidissip2}), (\ref{eq:axidissip3}) based on local axisymmetry.}
	\label{fig:testaxidiss}     
\end{figure}

Figure \ref{fig:testaxidiss} gives a comparison of the three above estimates with the full dissipation for the SPIV data. The first   estimate of equation (\ref{eq:axidissip1}) works well right to the wall because of cancellation of different terms in the sum.  As noted by ~\cite{antonia91} and  more recently  by \cite{Fanetal2015} as well, the second one (equation (\ref{eq:axidissip2})) fails inside of $y^+ =200$. The last proposal (equation (\ref{eq:axidissip3})) is fairly good down to $y^+ =50$, and then gives a slight overestimation. 

\section{Local homogeneity \label{loc_homgen}}
The basic ideas behind {\it local homogeneity} were first introduced by \cite{Taylor1935}, although he did not identify them as such.  So most subsequently confused these results with his conclusions from {\it local isotropy}, at least until  \cite{george91}.\footnote{Taylor's long forgotten results were independently rederived by George and Hussain \cite{george91}, and they coined the phrase {\it  locally homogeneous}, at least in the context of turbulence derivatives. Most turbulence models previously thought to depend on {\it local isotropy} or an isotropic dissipation assumption, in fact needed only local homogeneity.}

\subsection{What is local homogeneity}

The very idea of {\it local homogeneity}  as been at the core of turbulence thinking since the beginning of modern turbulence theory (c.f ~\cite{Taylor1935,Kolmogorov41a}), and was the rationale behind applying spectral and structure function relations to the smallest scales of inhomogeneous flows.  Curiously while the phrase {\it local isotropy}  is in common use since \cite{batchelor53},  to the best of our knowledge the actual term {\it local homogeneity}  appears to have first been introduced by \cite{george91}.  As with {\it local isotropy}  and local axisymmetry, the whole idea of {\it local homogeneity}  is that it only applies to statistical quantities dominated by the smallest scales of motion -- in this context the derivative moments.   Thus transport terms need not be zero, and only the derivative moment relations of equations (\ref{eq:homogderiv4})  need apply.  This leads to the following relevant equalities:

\begin{eqnarray}
\langle {\frac{\partial u_1}{\partial x_1} \frac{\partial u_2}{\partial x_2}  } \rangle & = & \langle {\frac{\partial u_1}{\partial x_2} \frac{\partial u_2}{\partial x_1}  } \rangle \label{lochom1}\\
\langle {\frac{\partial u_1}{\partial x_1} \frac{\partial u_2}{\partial x_3}  } \rangle & = & \langle {\frac{\partial u_1}{\partial x_3} \frac{\partial u_2}{\partial x_1}  } \rangle \\
\langle {\frac{\partial u_1}{\partial x_2} \frac{\partial u_2}{\partial x_3}  } \rangle & = & \langle {\frac{\partial u_1}{\partial x_3} \frac{\partial u_2}{\partial x_2}  } \rangle \\
\langle {\frac{\partial u_1}{\partial x_1} \frac{\partial u_3}{\partial x_2}  } \rangle & = & \langle {\frac{\partial u_1}{\partial x_2} \frac{\partial u_3}{\partial x_1}  } \rangle \\
\langle {\frac{\partial u_1}{\partial x_1} \frac{\partial u_3}{\partial x_3}  } \rangle & = & \langle {\frac{\partial u_1}{\partial x_3} \frac{\partial u_3}{\partial x_1}  } \rangle \\
\langle {\frac{\partial u_1}{\partial x_2} \frac{\partial u_3}{\partial x_3}  } \rangle & = & \langle {\frac{\partial u_1}{\partial x_3} \frac{\partial u_3}{\partial x_2}  } \rangle \\
\langle {\frac{\partial u_2}{\partial x_1} \frac{\partial u_3}{\partial x_2}  } \rangle & = & \langle {\frac{\partial u_2}{\partial x_2} \frac{\partial u_3}{\partial x_1}  } \rangle\\
\langle {\frac{\partial u_2}{\partial x_1} \frac{\partial u_3}{\partial x_3}  } \rangle & = & \langle {\frac{\partial u_2}{\partial x_3} \frac{\partial u_3}{\partial x_1}  } \rangle\\
\langle {\frac{\partial u_2}{\partial x_2} \frac{\partial u_3}{\partial x_3}  } \rangle & = & \langle {\frac{\partial u_2}{\partial x_3} \frac{\partial u_3}{\partial x_2}  } \rangle 
\label{lochom2}
\end{eqnarray}

It has been noted already that, thanks to equation (\ref{eq:homogderiv5}), an immediate consequence of {\it local homogeneity}  and incompressibility is that $\mathcal{D}$ as defined by equation (\ref{eq:d}) is equal to the true dissipation, $\varepsilon$ as given by equation (\ref{eq:eps}).  For this reason  $\mathcal{D}$ has been referred to by \cite{george91} as the {\it pseudo-dissipation} or by many as the homogeneous dissipation.  Some have erroneously construed this to mean that $\mathcal{D}$  is somehow more fundamental than $\varepsilon$. Nothing could be further from the truth:  $\varepsilon$ is {\it always} the true dissipation.  $\mathcal{D}$ {\bf only} equals it in a homogeneous flow, and contains other terms involving rotation when the flow is not homogenous. Another consequence of homogeneity is that the mean square strain-rate and mean square rotation rates are equal; i.e.,  $\langle \omega_{ij} \omega_{ij} \rangle=  2\langle s_{ij} s_{ij} \rangle$. Appendix A of Part I of this contribution \cite{stanislas20} discusses the relation between $\mathcal{D}$ and $\varepsilon$ in detail (summarized in the present introduction), along with the forms of the Reynolds averaged kinetic energy equation containing them.

\subsection{Tests of local homogeniety from experiment and DNS}

The large number of derivative moment combinations available in the present experiments and DNS makes it possible to exhaustively test the {\it local homogeneity}  hypothesis provided by equations (\ref{lochom1}) to (\ref{lochom2}), even for derivative combinations which do not appear in either the dissipation or the enstrophy. 

\begin{figure}
	\resizebox{0.5\linewidth}{!}{\includegraphics[scale=1]{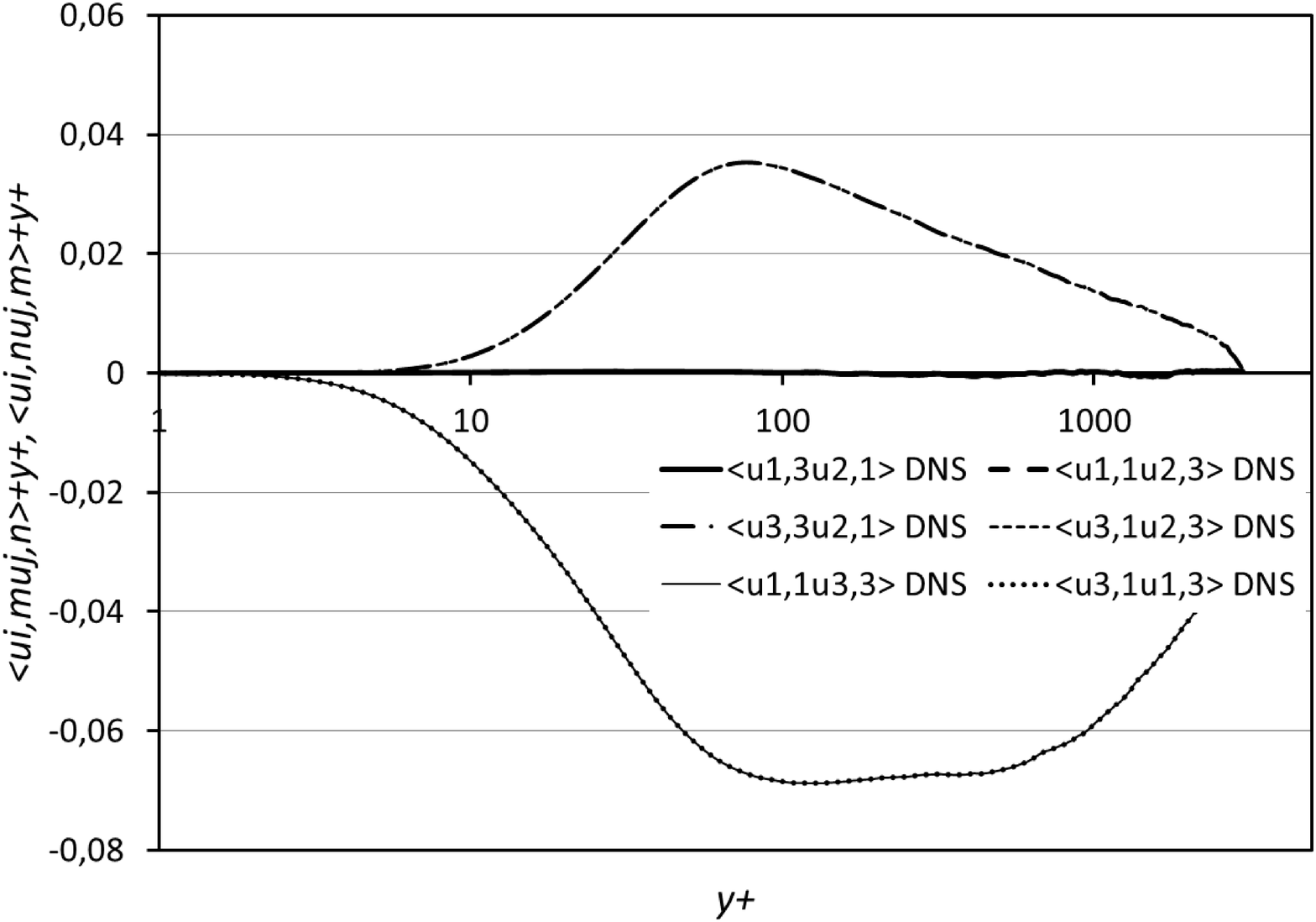}} 
	\resizebox{0.5\linewidth}{!}{\includegraphics[scale=1]{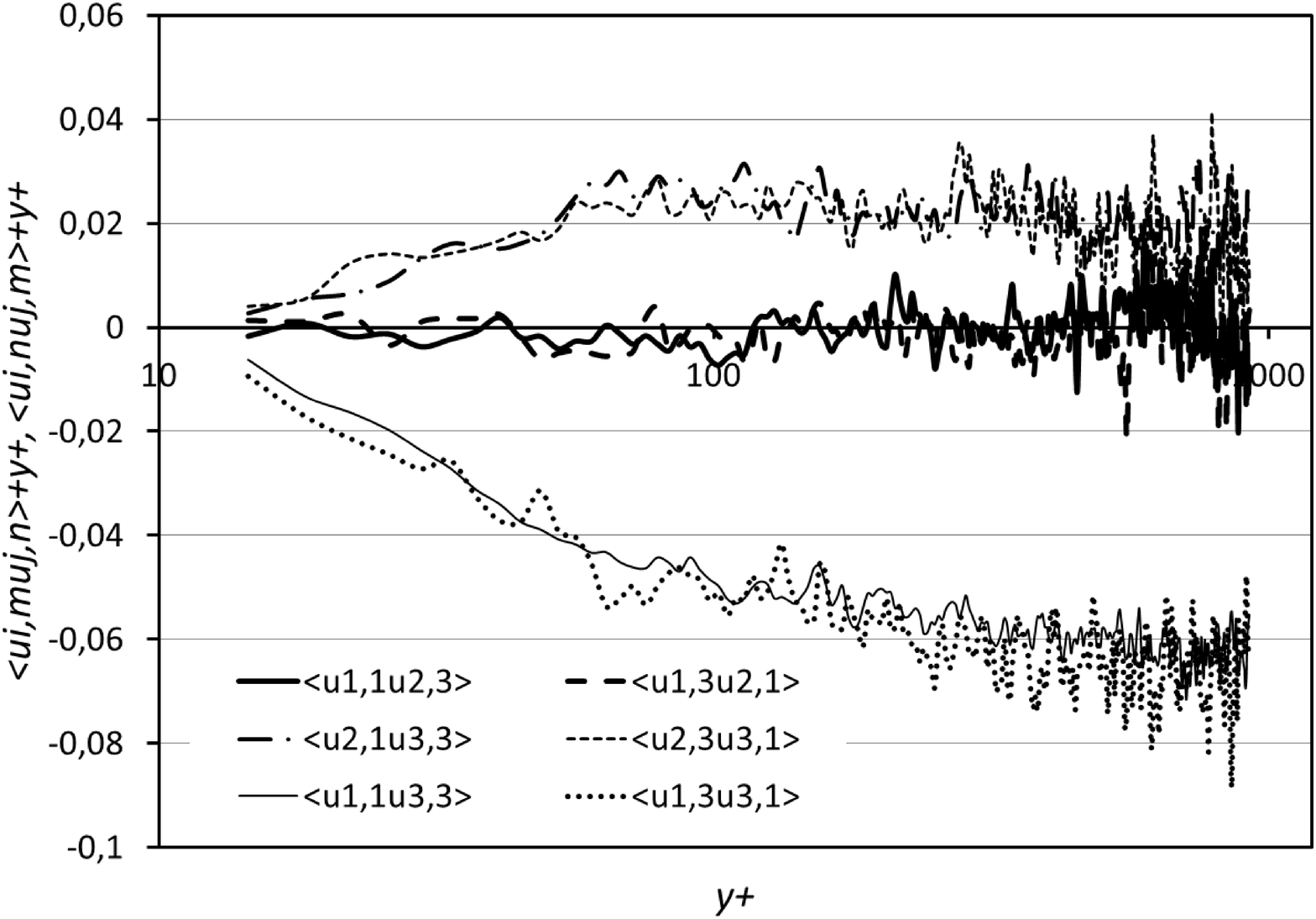}}\\
	\hspace*{3 cm}(a) \hspace{5 cm}(b)\\
	\resizebox{0.5\linewidth}{!}{\includegraphics[scale=1]{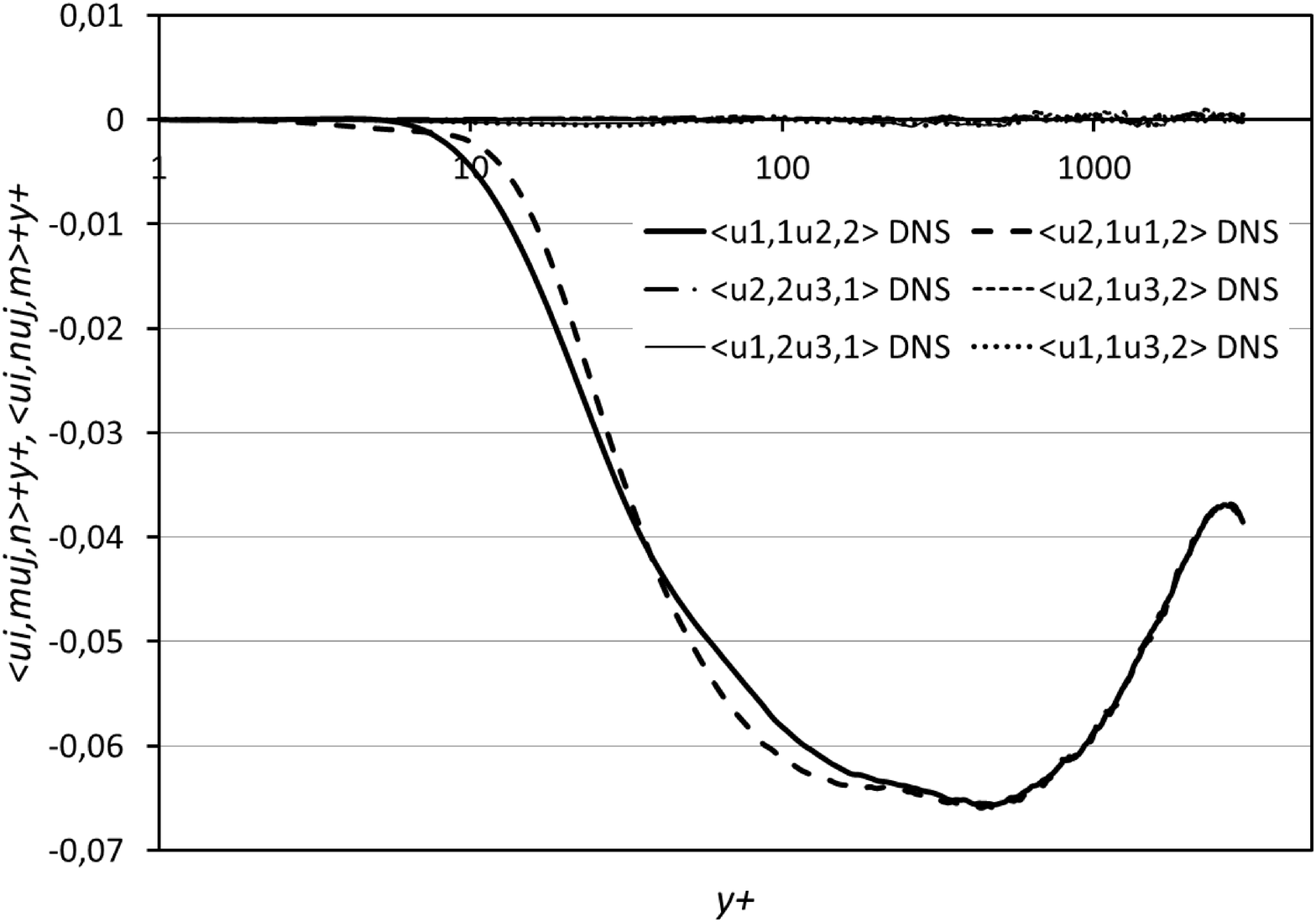}}
	\resizebox{0.5\linewidth}{!}{\includegraphics[scale=1]{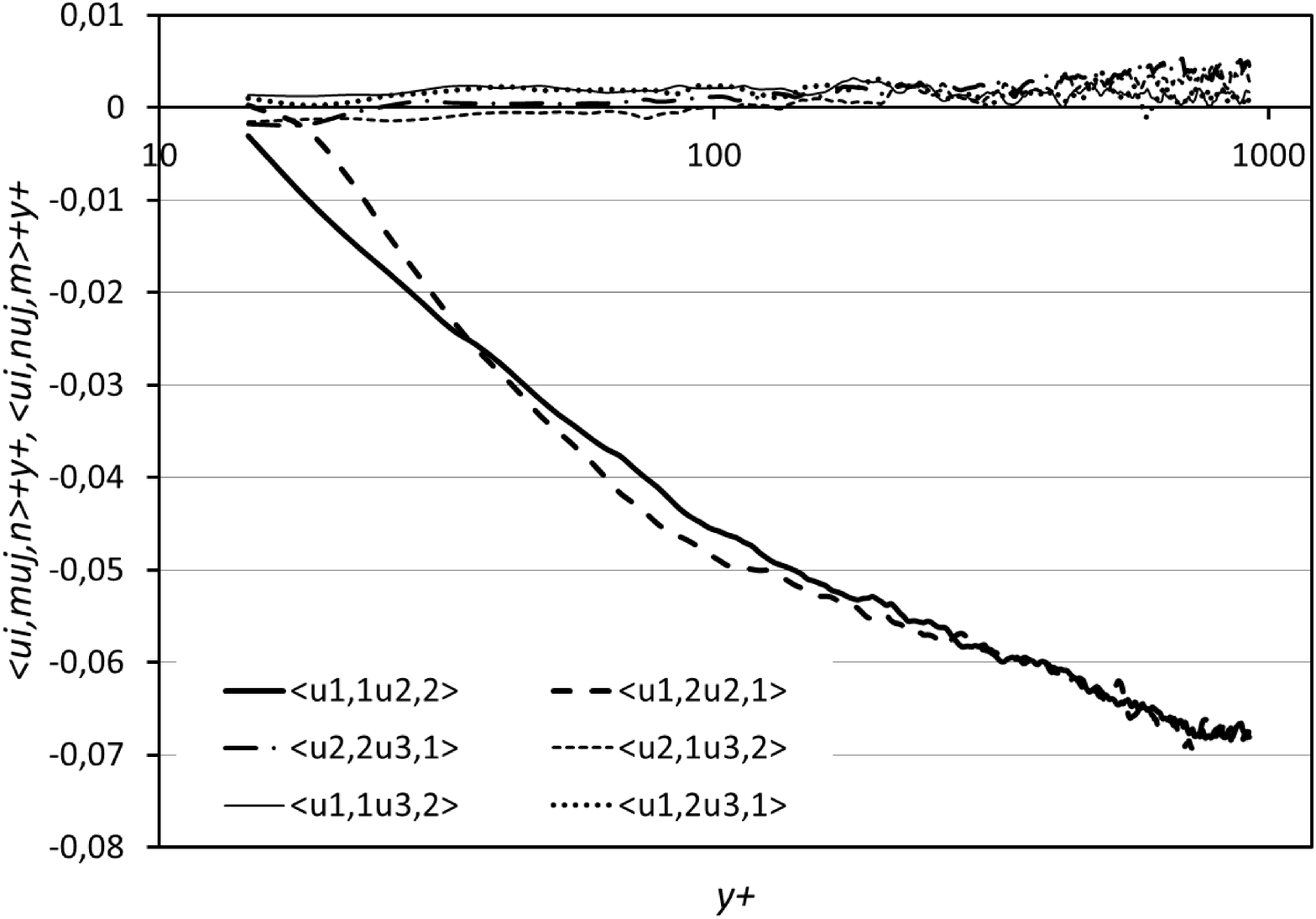}}\\
	\hspace*{3 cm}(c) \hspace{5 cm}(d)\\
	\resizebox{0.5\linewidth}{!}{\includegraphics[scale=1]{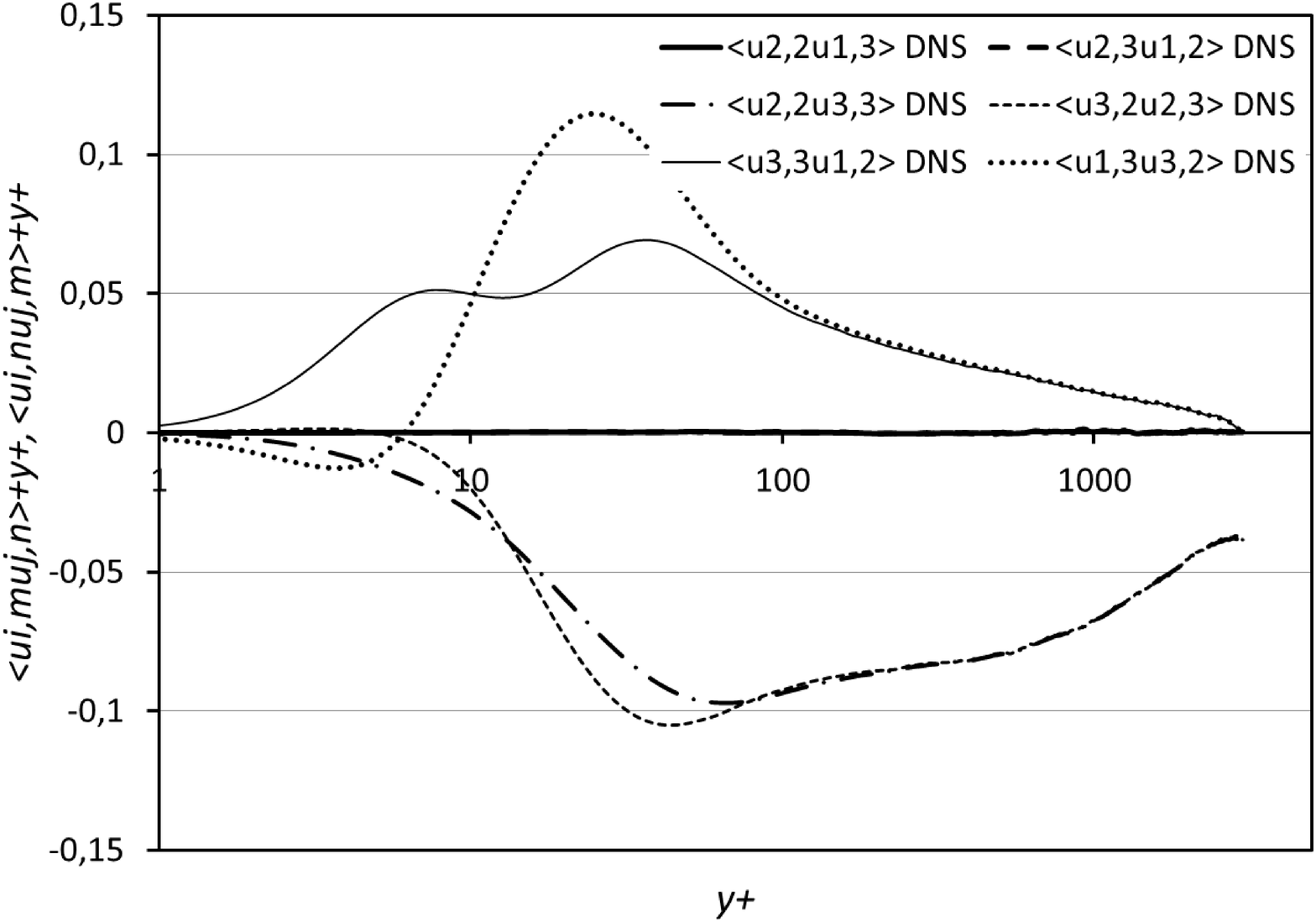}}
	\resizebox{0.5\linewidth}{!}{\includegraphics[scale=1]{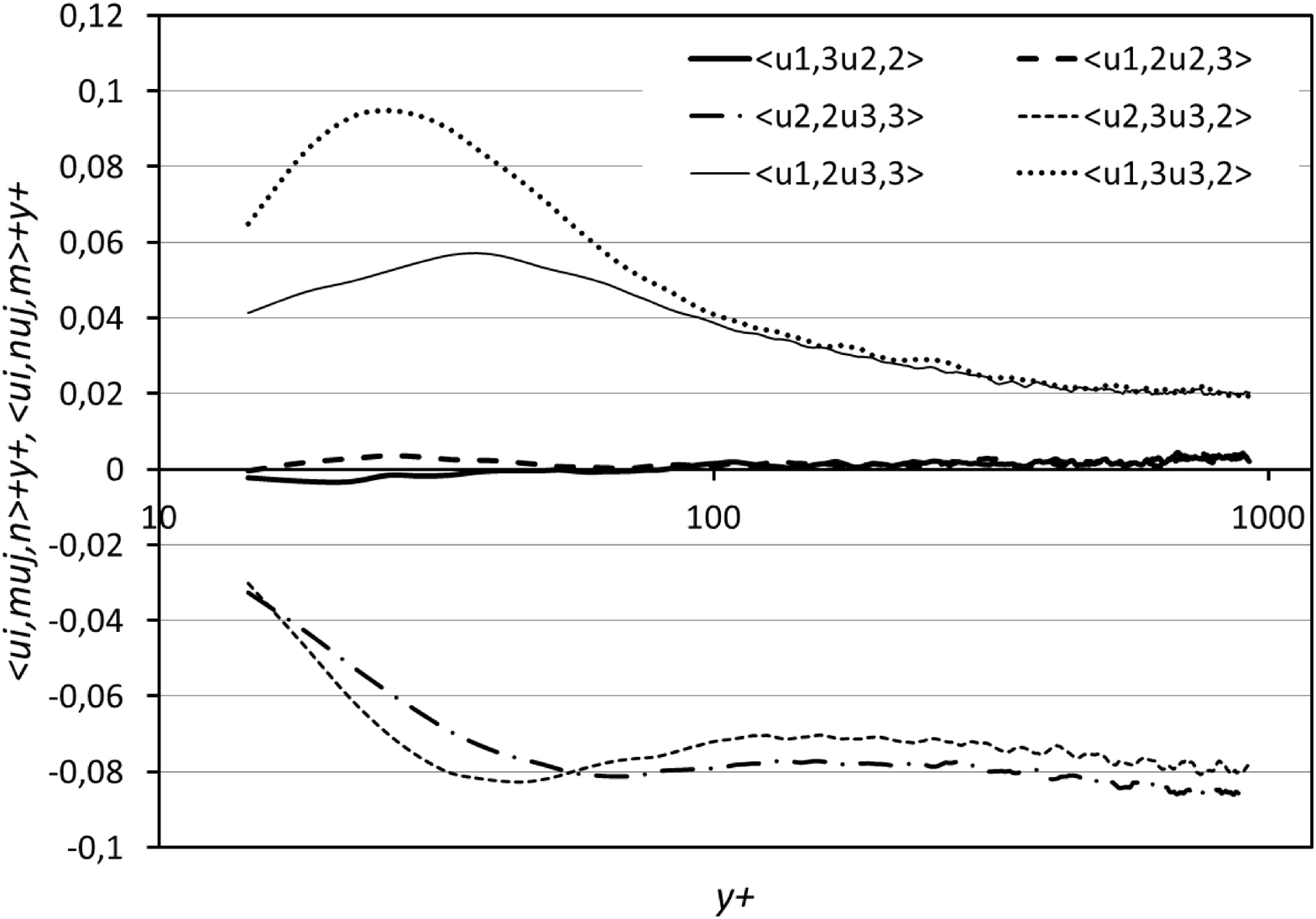}}\\
	\hspace*{3 cm}(e) \hspace{5 cm}(f)
	\caption{Comparison of derivative moments from DNS (left column) and SPIV (right column) showing support for local homogeneity outside of $y^+=100$.}
	\label{fig:homodiss}      
\end{figure}

{Figures~\ref{fig:homodiss} and \ref{fig:dnshomodiss} show plots of the various pairs of cross-derivative moments, from the BL experiment and  channel DNS respectively, which should fulfill  equations (\ref{lochom1}) to (\ref{lochom2}) if the turbulence can be considered {\it locally homogeneous}.  The close correspondence of the derivative pairs {\bf outside of $y^+ =100$} is truly remarkable. Figure~\ref{fig:dnshomonondiss} makes it clear that the derivative combinations which do not appear in either the dissipation or the enstrophy are in a range  more than an order of magnitude smaller. A careful analysis shows that they also fullfill the {\it local homogeneity}  condition in the overlap region.  
The general conclusion is then that outside of $y^+ =100$, {\it local homogeneity}  is an excellent approximation. 

\begin{figure}
	\resizebox{0.5\linewidth}{!}{\includegraphics[scale=1]{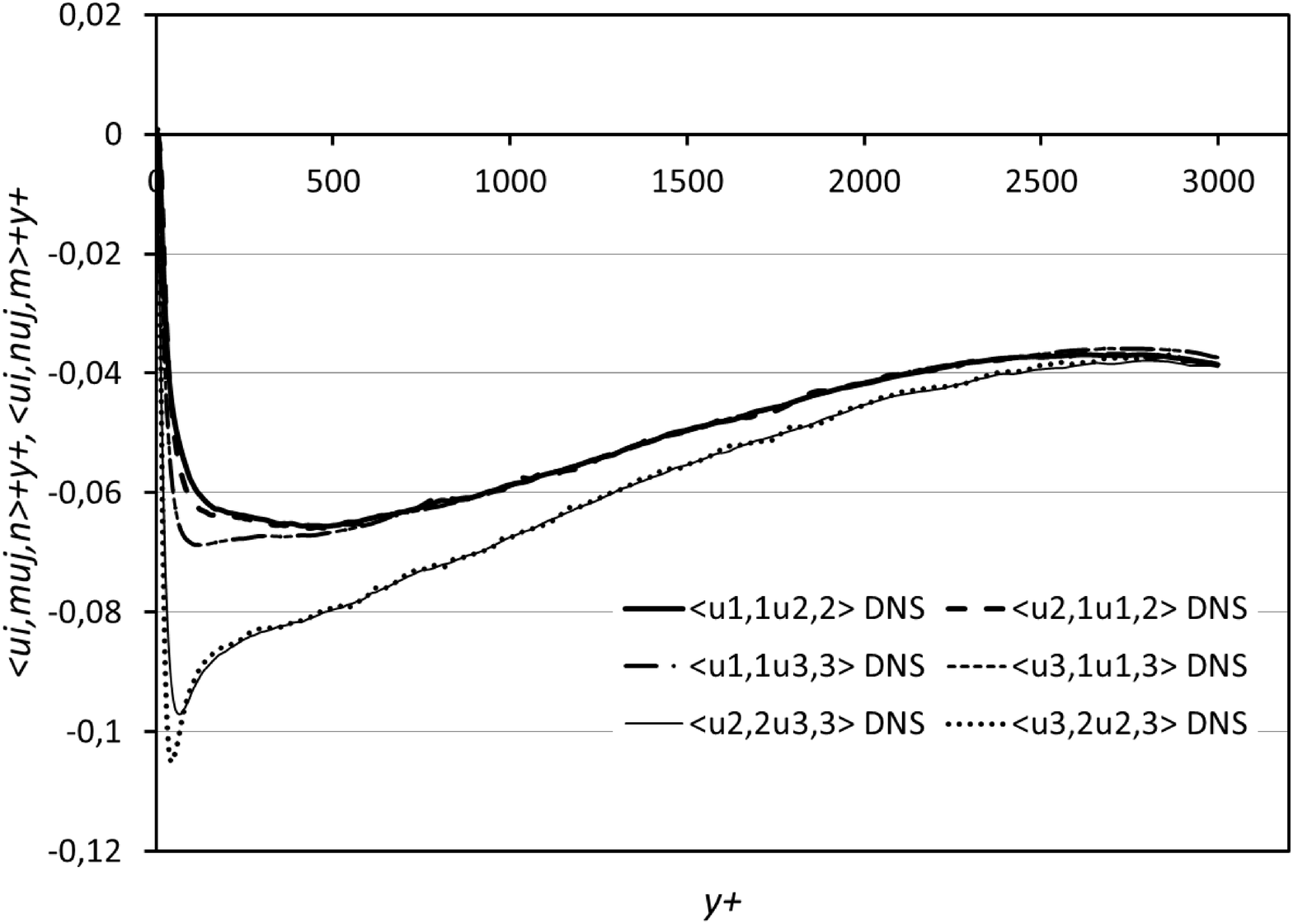}}
	\resizebox{0.5\linewidth}{!}{\includegraphics[scale=1]{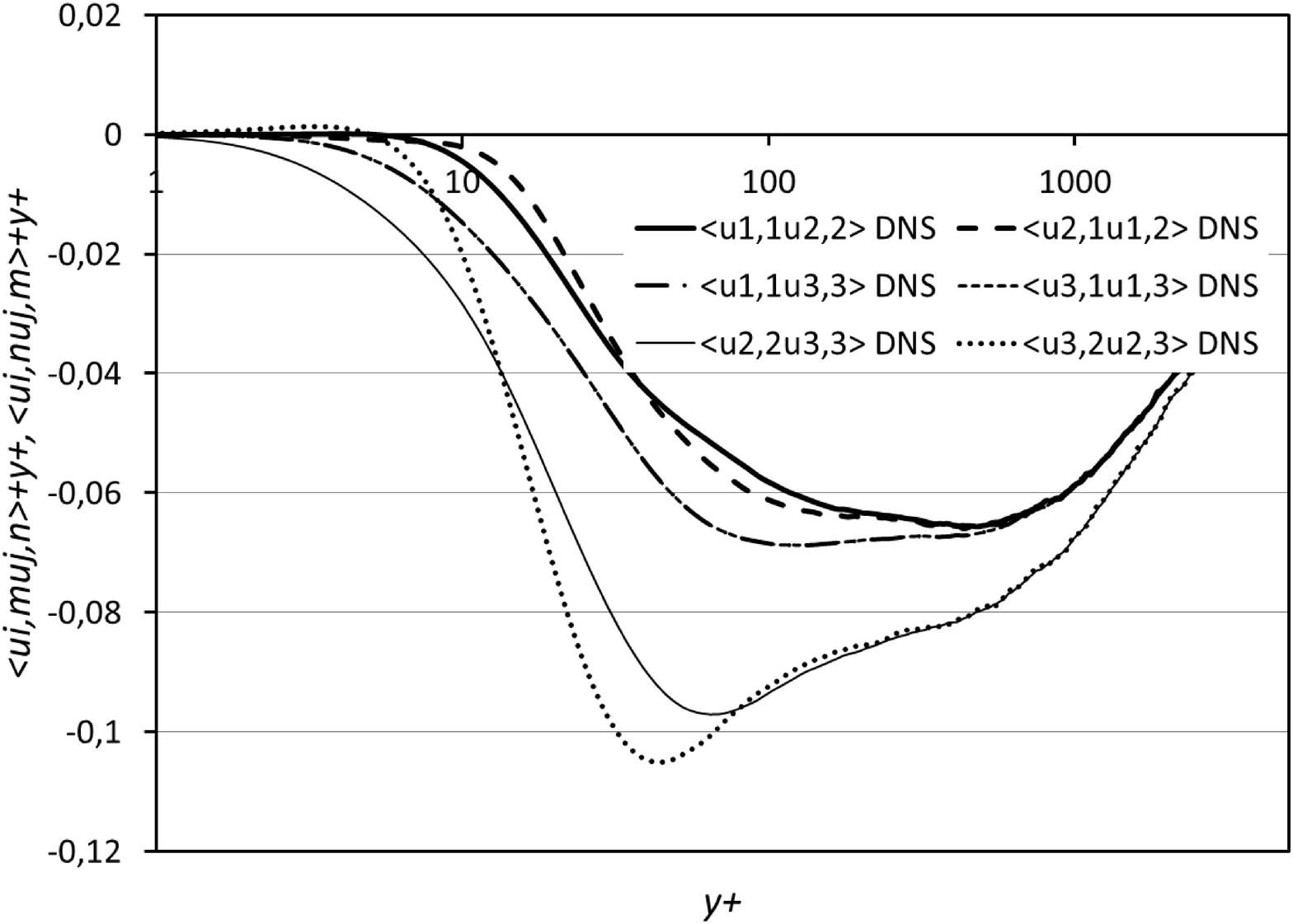}} \\
	\hspace*{3 cm}(a) \hspace{5 cm}(b)\\
	\resizebox{0.5\linewidth}{!}{\includegraphics[scale=1]{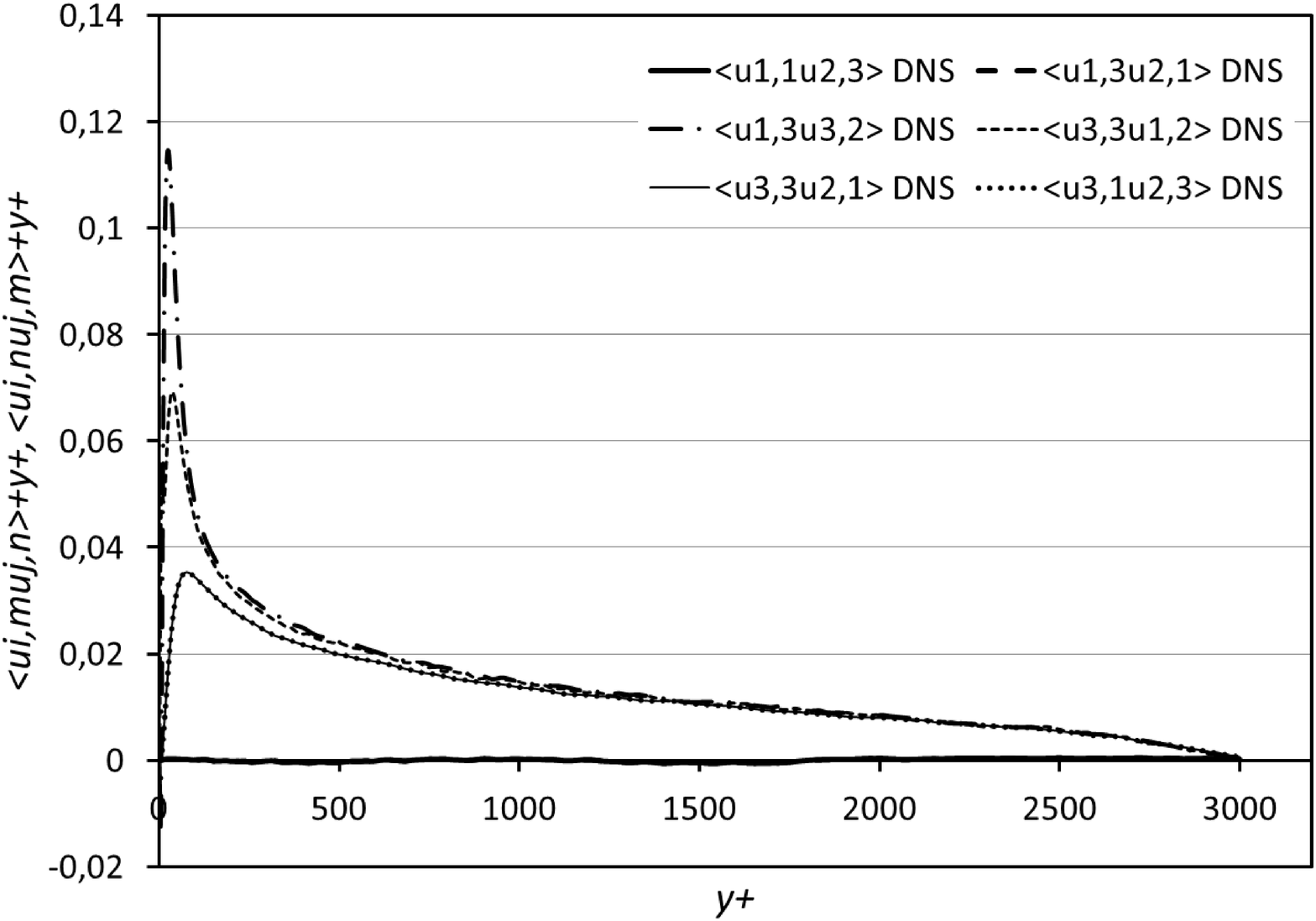}}
	\resizebox{0.5\linewidth}{!}{\includegraphics[scale=1]{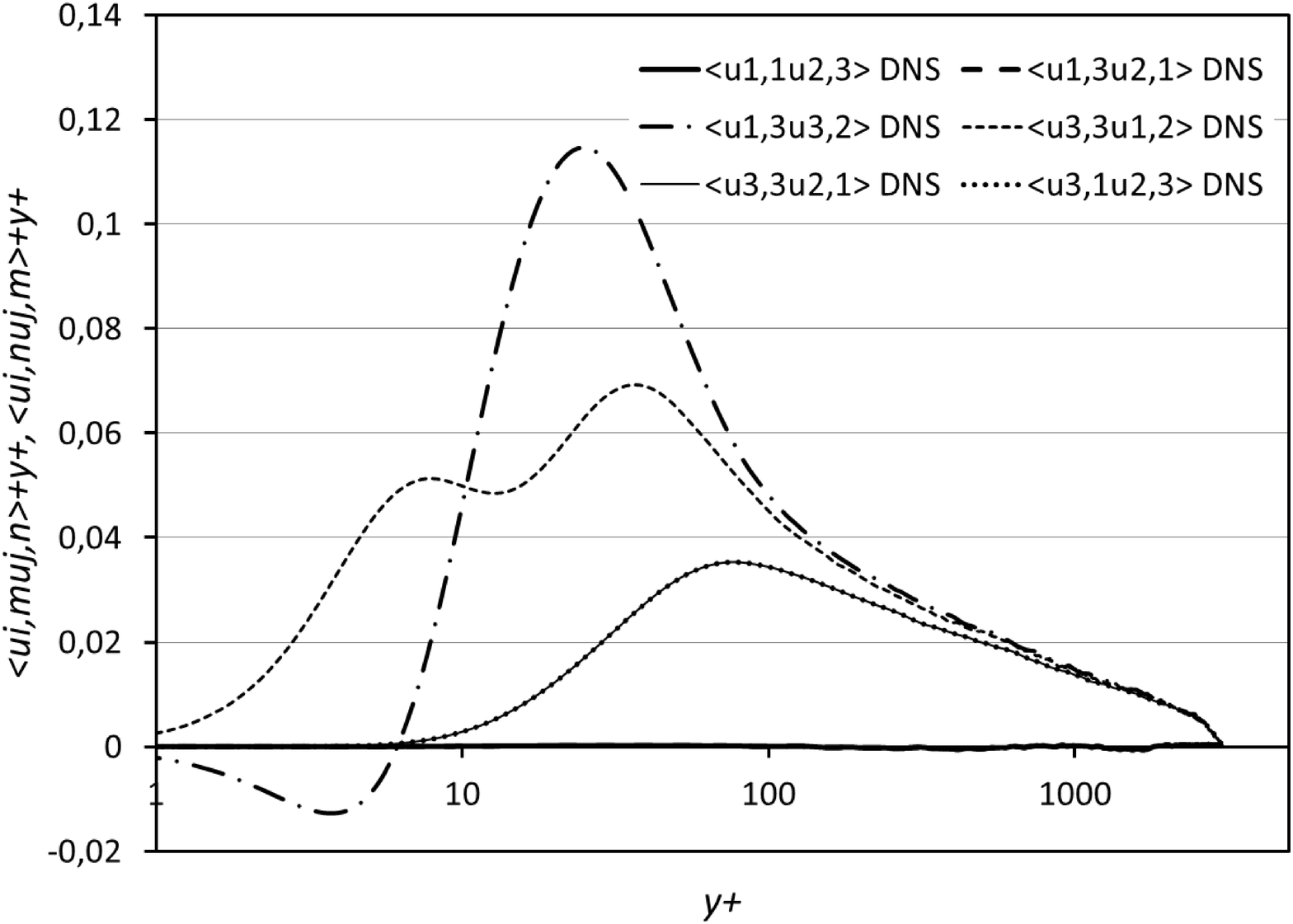}}
	\hspace*{3 cm}(c) \hspace{5 cm}(d)\\
	\caption{Comparison of derivative moments for DNS showing support for local homogeneity outside of $y^+=100$.}
	\label{fig:dnshomodiss}      
\end{figure}
 
\begin{figure}
	\resizebox{0.5\linewidth}{!}{\includegraphics[scale=1]{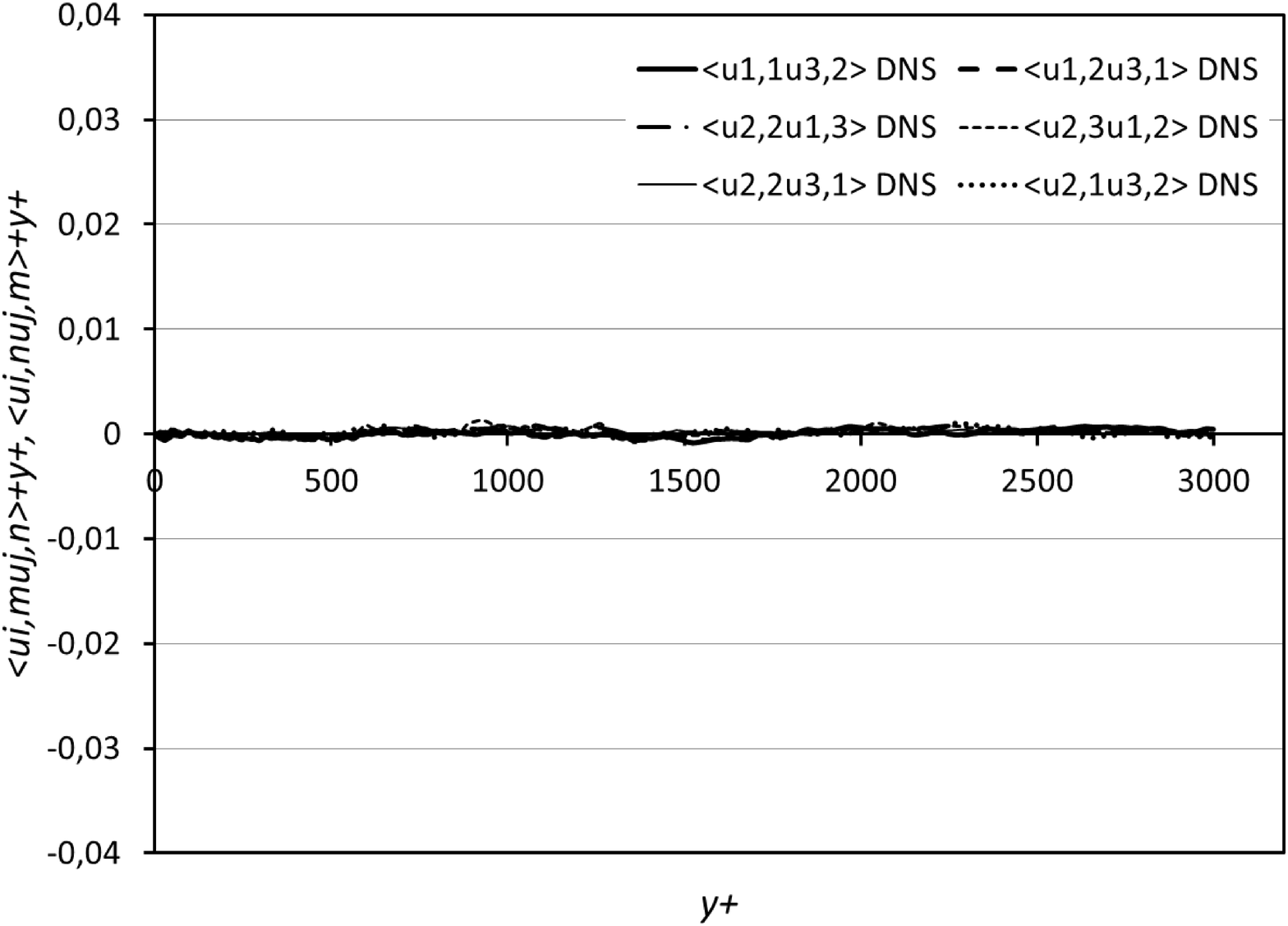}} 
	\resizebox{0.5\linewidth}{!}{\includegraphics[scale=1]{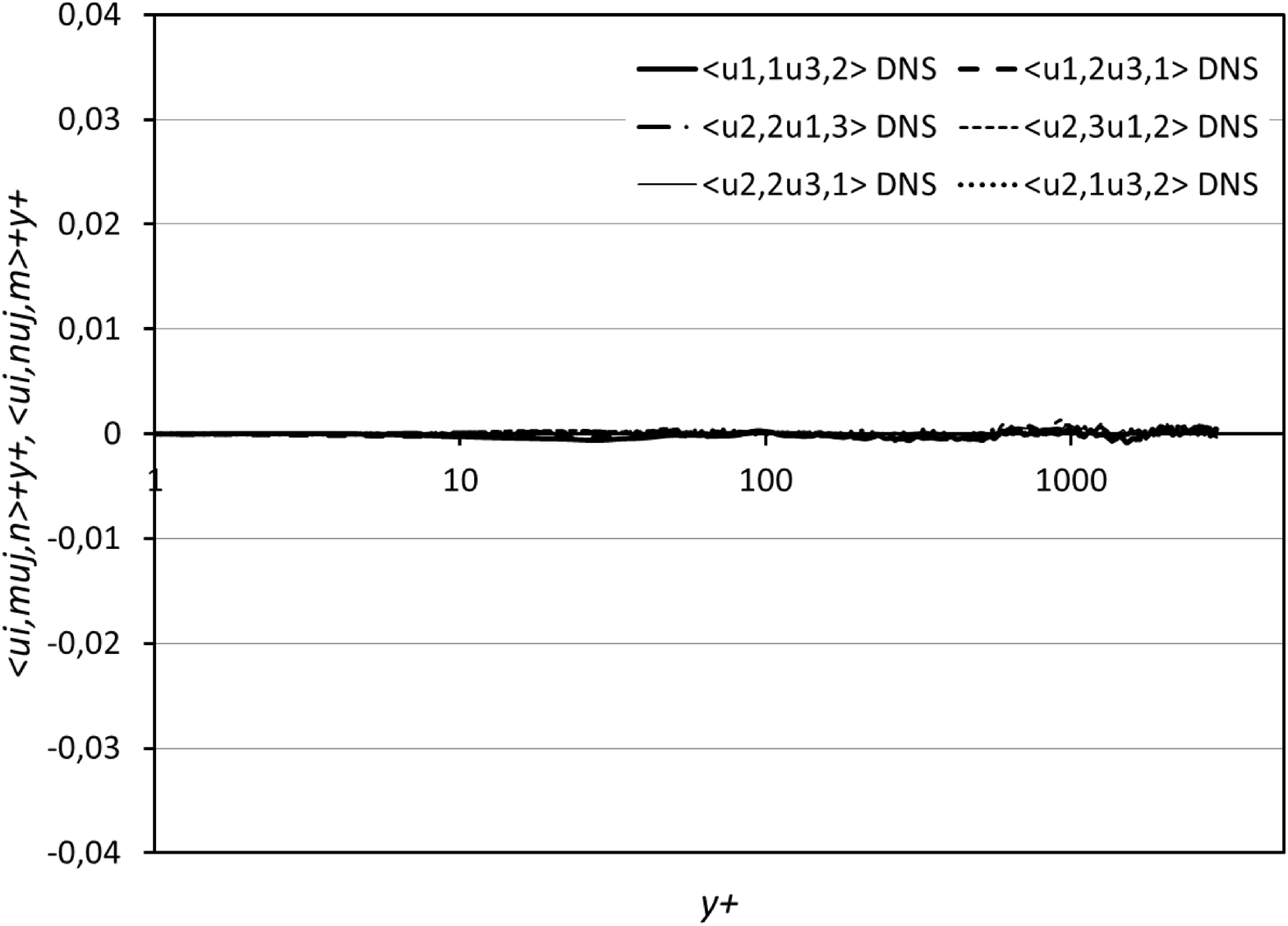}}
	\hspace*{3 cm}(a) \hspace{5 cm}(b)\\
	\caption{Comparison of DNS non-dissipative and non-vortical derivative moments.}
	\label{fig:dnshomonondiss}      
\end{figure}
	
Inside of $y^+=100$ is another story altogether. Note that only the DNS channel data are reliable in this region. There is no evidence of any {\it local homogeneity}  at all.  In fact only three derivatives completely dominate, and there is no apparent relation among them.  (Note that a fourth only appears to be important because it has been multiplied by 4.)
	
	As mentionned earlier, the absence of {\it local homogeneity}  inside of $y^+=100$ eliminates any possibility of eventual {\it local axisymmetry}  or local isotropy. This is  a matter of  concern as under such conditions, we cannot expect to have $\varepsilon_{ij} = \mathcal{D}_{ij}$. This fact is adressed in more detail in the next section.

\section{Dissipation in the near wall region $y^+ < 100$ \label{sec-planehomogeneity}}

All of our experimental and DNS provide overwhelming evidence that the conditions for {\it local homogeneity} are not  met even approximately inside of $y^+ = 100$. Yet, even as noted already in Part I \cite{stanislas20} (and assumed by most of the turbulence community), $\mathcal{D} \approx \varepsilon$, even close to the wall. The question which rises immediately is whether any other hypothesis exists which can explain what is happening in this very near wall region. 

\subsection{The near wall data}

\begin{figure}
	\resizebox{0.95\linewidth}{!}{\includegraphics[scale=1]{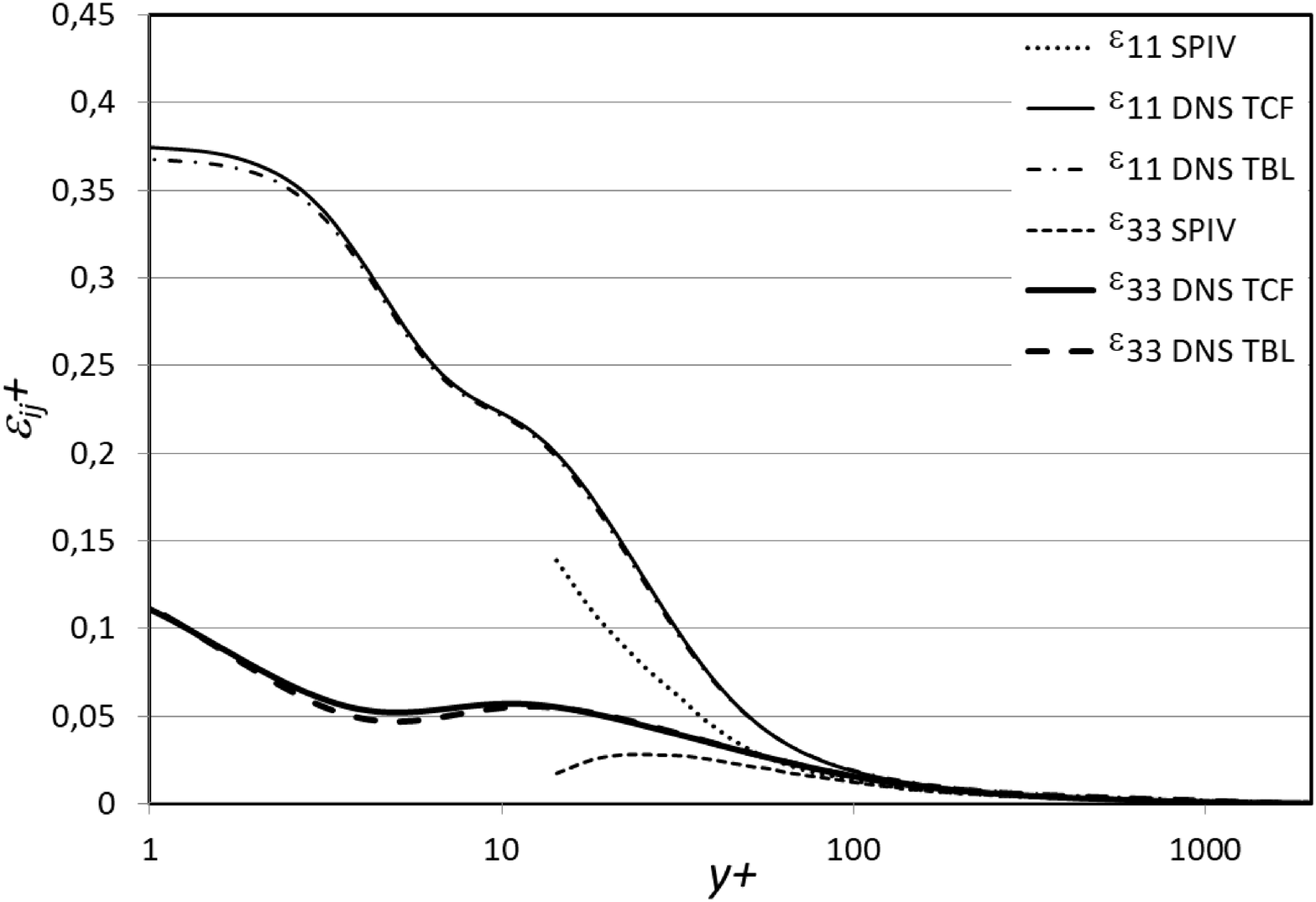}} \\
	\resizebox{0.95\linewidth}{!}{\includegraphics[scale=1]{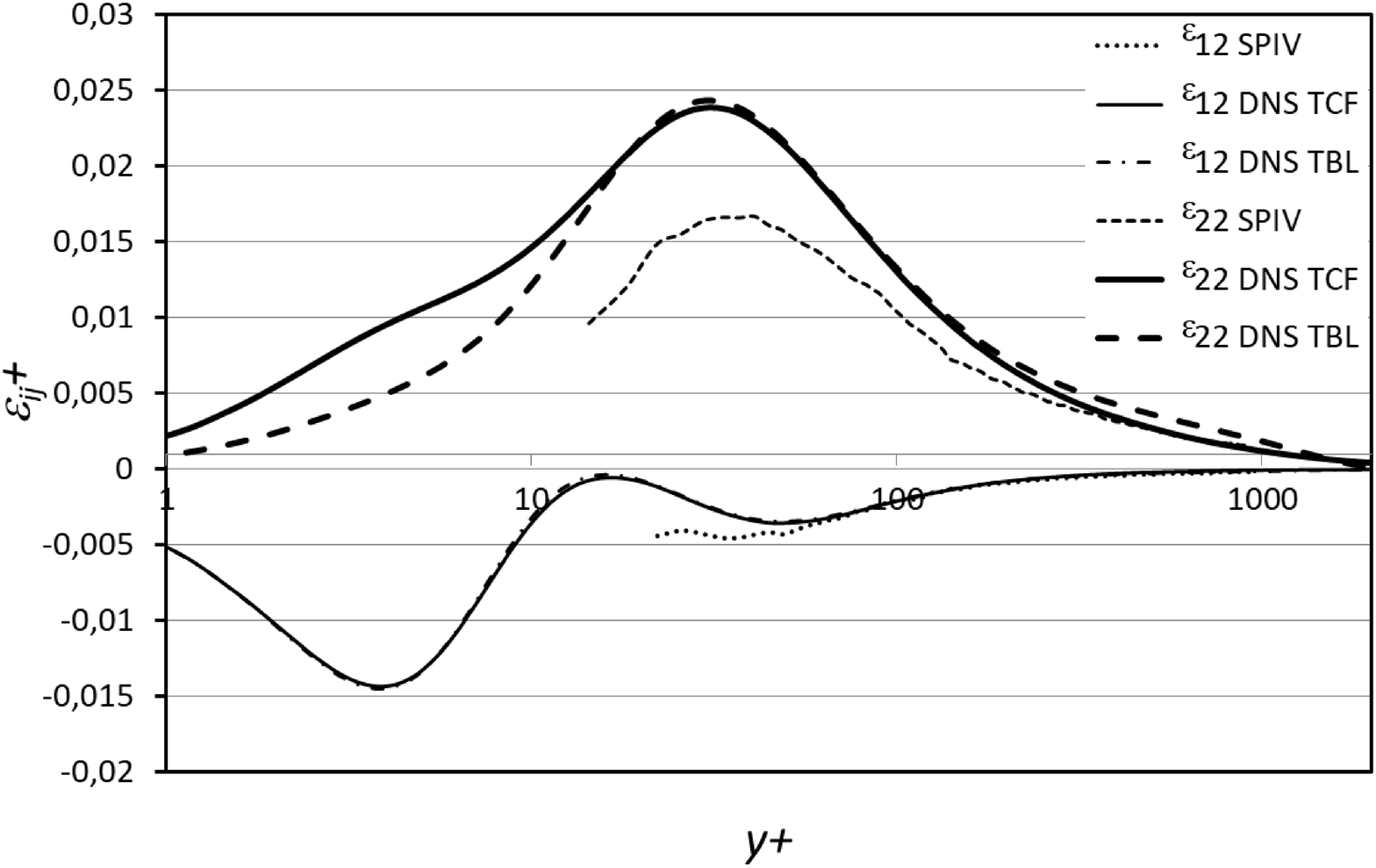}}
	\caption{Diagonal terms, $\varepsilon_{11}$, $\varepsilon_{22}$, and $\varepsilon_{33}$, of the dissipation tensor, together with the shear stress component dissipation, $\varepsilon_{12}$ from the SPIV boundary layer experiment, the channel DNS and the boundary layer DNS of \cite{sillero13}}. 
	\label{fig:componentdissipations}      
\end{figure}

To answer this question we examine first  the shape of the dissipation tensor $\varepsilon_{ij}$. Figure \ref{fig:componentdissipations} gives a lin-log plot of the main components of this tensor in the near wall region. Data are compared from the SPIV boundary layer experiment, the present channel DNS of \cite{thais11} and the Boundary Layer DNS of \cite{sillero13}. The top figure gives the diagonal terms and the bottom one the main cross-term. The first thing to notice is that the cross-term $\varepsilon_{12}$ is negative (consistent with the fact that $\langle u_1u_2\rangle$ is also negative) and an order of magnitude smaller than the diagonal terms, except  for $\varepsilon_{33}$. A second point to notice is that the SPIV data, although partly affected by spatial filtering, give the same tendencies as the DNS. An other very remarkable point is the nearly perfect superposition of the channel and boundary layer DNS down to the wall, which means that the dissipation physics is very similar between the two flows\footnote{This is in spite of the fact that their Taylor expansions from the wall are different, namely  the non-zero second-order term arising from the streamwise pressure gradient in the channel mean velocity.}. Finally, one should point out the strong anisotropy between the three diagonal terms, with a clear domination of $\varepsilon_{11}$ over the two others.

\begin{figure}
	\resizebox{0.95\linewidth}{!}{\includegraphics[scale=1]{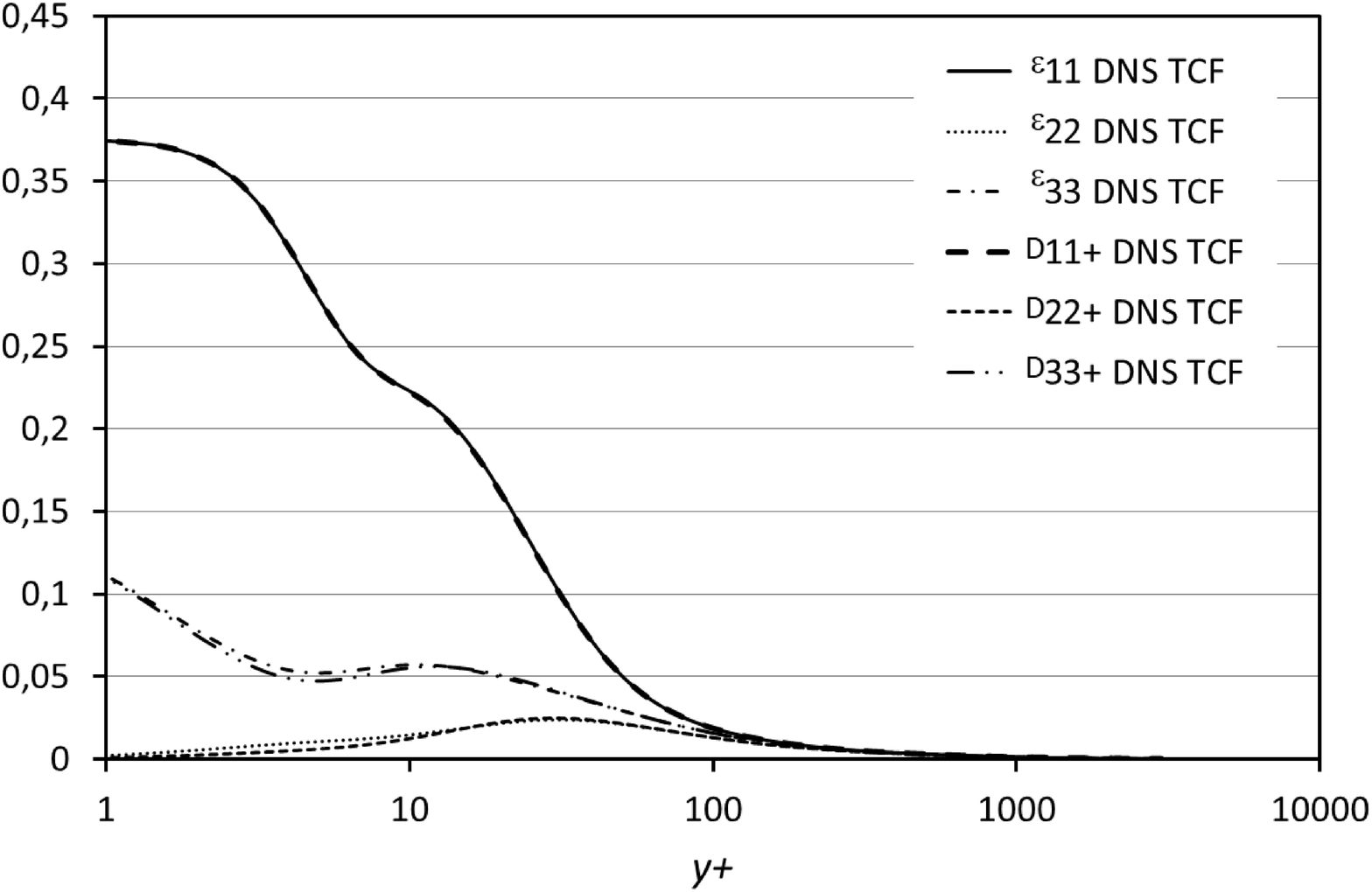}} 
	\caption{Comparison between the true dissipation tensor diagonal terms $\varepsilon _{ii}$ and the pseudo dissipation ones $\mathcal{D}_{ii}$ using data from the channel DNS.}
	\label{fig:compardissipation}     
\end{figure}

Now the key question is the behaviour of the pseudo-dissipation $\mathcal{D}_{ij}$ compared to the true dissipation tensor $\varepsilon_{ij}$. Figure \ref{fig:compardissipation} gives a lin-log plot of the three diagonal components of $\varepsilon_{ij}$ (which, as shown above, are the dominating ones) compared to the corresponding terms of $\mathcal{D}_{ij}$. The data are from the channel DNS. As can be seen, there is little difference between the corresponding components. 
It is not obvious why this is true given our conclusions about the departures from {\it local homogeneity} based on figures \ref{fig:homodiss}  and \ref{fig:dnshomodiss}.


\subsection{Cross-derivative relations if the flow is only homogeneous in a plane. \label{subsec-planehomog}}
Clearly none of the above analysis explains the near equality (if not exactly equality) of $\varepsilon_{ij}$ and $\mathcal{D}_{ij}$ near the wall. One possibility that does not seem to have been previously considered is that this behavior might be a direct consequence of homogeneity only in the ($x_1,x_3$) plane.  Certainly all experiments and DNS considered herein (and elsewhere) have one feature in common: they are homogeneous in the plane parallel to the wall.  Note that this is exactly true for the channel, but approximately true for the boundary layers (and most free shear flows as well) due to their slow spreading rate.

What we would need to be true for $\mathcal{D}_{ij}$ and $\varepsilon_{ij}$ to be equal is for the cross-derivative terms, $\langle \partial u_i/\partial x_j \partial x_j /\partial x_i\rangle$ to sum to zero, without being able to use the fully-homogeneous condition of equation~(\ref{eq:homogderiv4}) to permute the indices.  But this can be deduced from the continuity equation only if the following equalities hold:

\begin{eqnarray}
\langle {\frac{\partial u_1}{\partial x_1} \frac{\partial u_2}{\partial x_2}  } \rangle & = & \langle {\frac{\partial u_1}{\partial x_2} \frac{\partial u_2}{\partial x_1}  } \rangle \label{plhom1} \\
\langle {\frac{\partial u_1}{\partial x_1} \frac{\partial u_3}{\partial x_3}  } \rangle & = & \langle {\frac{\partial u_1}{\partial x_3} \frac{\partial u_3}{\partial x_1}  } \rangle \label{plhom2} \\
\langle {\frac{\partial u_2}{\partial x_2} \frac{\partial u_3}{\partial x_3}  } \rangle & = & \langle {\frac{\partial u_2}{\partial x_3} \frac{\partial u_3}{\partial x_2}  } \rangle \label{plhom3}
\end{eqnarray}

In planar homogeneity parallel to the wall, equation~(\ref{eq:homogderiv4}) is restricted to permutation of $m$ and $n$ only equal to $1$ and $3$. The development of all the possibilities leads to the conclusion that only three equalities are meaningful and that only equation (\ref{plhom2}) is valid in plane homogeneous turbulence. This is confirmed by figure \ref{fig:homodiss}a which also shows that the two other equalites are exactly true in plane homogeneous turbulence, but  neither of these appears in the dissipation. Figures \ref{fig:homodiss}e and \ref{fig:homodiss}f clearly show that the two other above equalities (\ref{plhom1}) and (\ref{plhom3}) are violated below $y^+ = 100$.
So we have exactly what we need for only the 1-3 derivative moments.  Unfortunately without assuming homogeneity in the 2-direction as well, it is impossible to derive a similar result for the other two mixed moments. 

\begin{figure}
	\resizebox{0.95\linewidth}{!}{\includegraphics[scale=1]{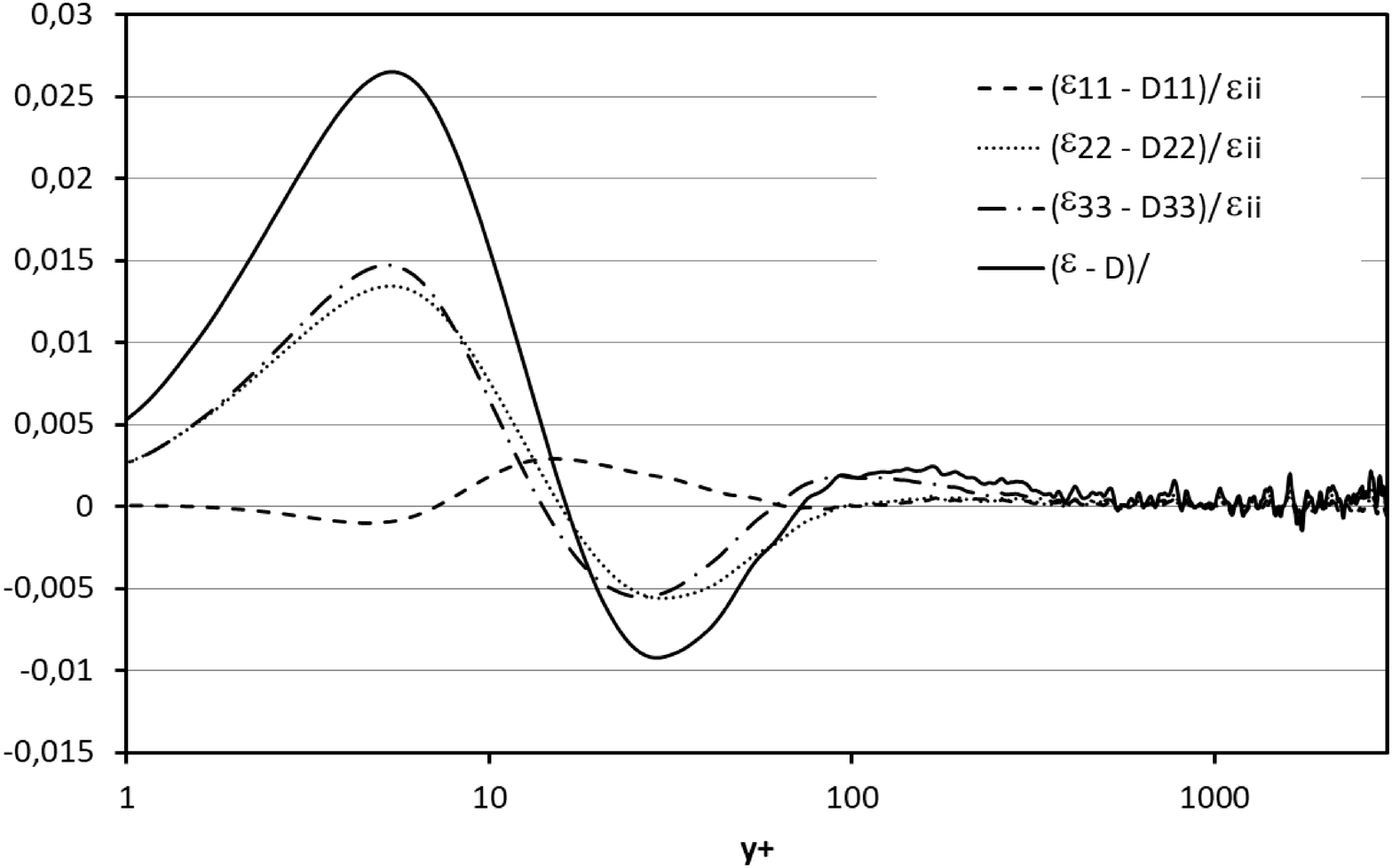}} 
	\caption{Difference between the normal components  $\varepsilon _{ii}$ and $\mathcal{D}_{ii}$ of the full and pseudo dissipation tensors and between $\varepsilon$ and $\mathcal{D}$, data from the channel DNS.}
	\label{fig:eps-D}     
\end{figure}

Figures \ref{fig:compardissipation} and \ref{fig:eps-D} respectively plot the component dissipations and their differences. The clear and obvious conclusion is that $\mathcal{D}_{ij}$ and $\varepsilon_{ij}$ (as well as $\mathcal{D}$ and $\varepsilon$) must be fundamentally unequal, no matter how close they may appear to be in practice. Whether these differences are significant enough for turbulence modellers to worry about is beyond the scope of this paper.  But if it is significant, then it means that modelling should be done at the level of equation (\ref{rs_eq}), before the simplification of the cross moments which leads to equation (\ref{rs_dij}). This should probably affect also the near-wall models for the viscous transport and pressure-strain-rate terms.

\subsection{The dissipation anisotropy very near the wall}

Figure \ref{fig:compardissipation} provides interesting information about the dissipation of the normal stresses. As can be seen, below $y^+ = 100$ this dissipation becomes very anisotropic, with $\mathcal{D}_{11}$ dominating the two other terms. 
A now classical way to examine the same data in terms of anisotropy is to look at the so-called `Lumley-triangle' formed by plotting the second and third invariants of the dissipation anisotropy tensor. As in the region of interest ($y^+ < 100$) $\varepsilon_{ij}$ and $\mathcal{D}_{ij}$ differ slightly, it is of interest to look at the anisotropy of both. The two invariants are obtained in the following manner for the full dissipation tensor $\varepsilon_{ij}$:

\begin{eqnarray}
II_\varepsilon & = & \varepsilon_{ij} \varepsilon_{ij} \\
III_\varepsilon & = & \varepsilon_{ij}\varepsilon_{jk}\varepsilon_{ki}
\end{eqnarray} 
with similar relations for $II_\mathcal{D}$ and $III_\mathcal{D}$. 

Figure~ \ref{fig:lumley}  plots $II$ versus $III$ for both $\varepsilon_{ij}$ and $\mathcal{D}_{ij}$ in the case of the channel DNS.  Also shown are the usual limiting lines. Two inserts provide an enlarged view of two regions of interest.
To the best of our knowledge this is the first time the diagrams for $\varepsilon_{ij}$ and $\mathcal{D}_{ij}$ have been plotted together.  One point of interest is that the curves begin to diverge only below $y^+=10$, and significantly around $y^+ = 5$. This is far below $y^+ = 100$ where local homogeneity was put in question.  

\begin{figure}
	\resizebox{0.95\linewidth}{!}{\includegraphics[scale=1]{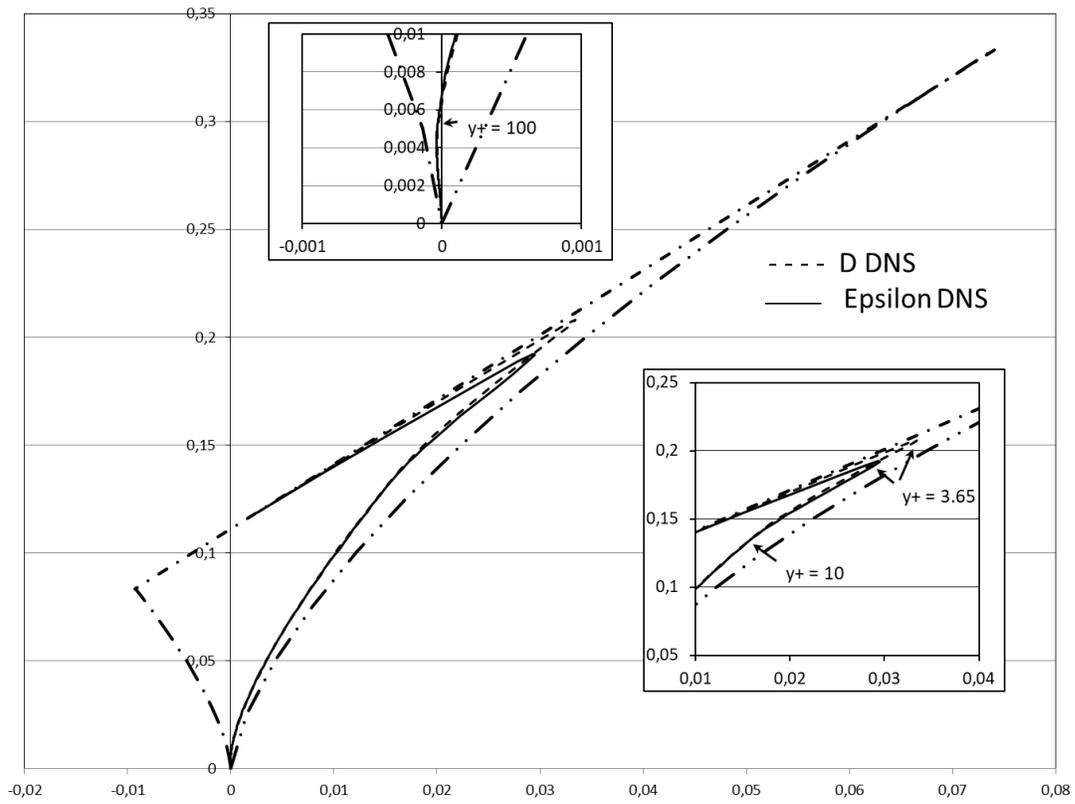}}
	\caption{Lumley triangle for both $\varepsilon_{ij}$ and $\mathcal{D}_{ij}$ computed from the present DNS data, with a zoom in the near wall region. }
	\label{fig:lumley}      
\end{figure}

Very similar results have been presented and discussed in detail for DNS channel flow for $\mathcal{D}_{ij}$ in the comprehensive paper by ~\cite{gerolymos16} (Note that they incorrectly label $\mathcal{D}_{ij}$ the dissipation).  Therefore we note that our results are virtually identical to theirs, and refer the reader to their paper.  We do note three things, however. First the very near wall region ($y^+ < 5$) is nearly two-dimensional, not surprising given the suppression of the normal velocity component by the wall.  Second, the diagram rapidly tends toward the axisymmetric asymptote, but never quite arrives. This is consistent with the idea of `local' (but not complete) axisymmetry outside of $y^+ = 100$.  And finally, we note that what might be interpreted as a trend toward isotropy appears to contradict the detailed analysis of Section~\ref{app-isotropic} and of the above conclusions. Note that it is arguable whether there is a trend toward isotropy at all since all curves in figure \ref{fig:spiviso} and \ref{fig:dnsiso} are maintaining the same relative distance from each other.  Regardless, any tendency toward isotropy is well into the core region or outer boundary layer which was not considered in this paper.  

\subsection{Potential link to the near wall turbulence structure}

It is of interest at this stage to try to understand better the physical origin of the anisotropy detailed above and the mechanisms at the origin of this dissipation. In the last 50 years, a large amount of research has been devoted to the study of the specific organization of turbulence near the wall (e.g.,\cite{theodorsen52,Kline67, Zhou1999,panton97}). Even if the outer part organisation is still a subject of intensive research with much to understand especially at high Reynolds number \cite{smits11}, the inner layer organization is fairly well characterized and understood (e.g.,\cite{jimenez99b,Kline1997}) A global picture including indications of scales is provided by \cite{lin08} and reproduced in figure \ref{fig:lin08}. The classical low and high speed streaks are sketched together with the quasi-streamwise vortices and sweeps and ejections. What is interesting is that the characteristic dimensions  provided in this picture range from about $y^+ = 100$ to about $y^+ = 20$ (50 to 10 Kolmogorov units). It is clear from this sketch that this near wall organization is very far from the classical picture of homogeneous isotropic turbulence and of any cascade. The main difference is that production is now  at scales which are the same as the dissipative scales. 

\begin{figure}
	\resizebox{0.95\linewidth}{!}{\includegraphics[scale=1]{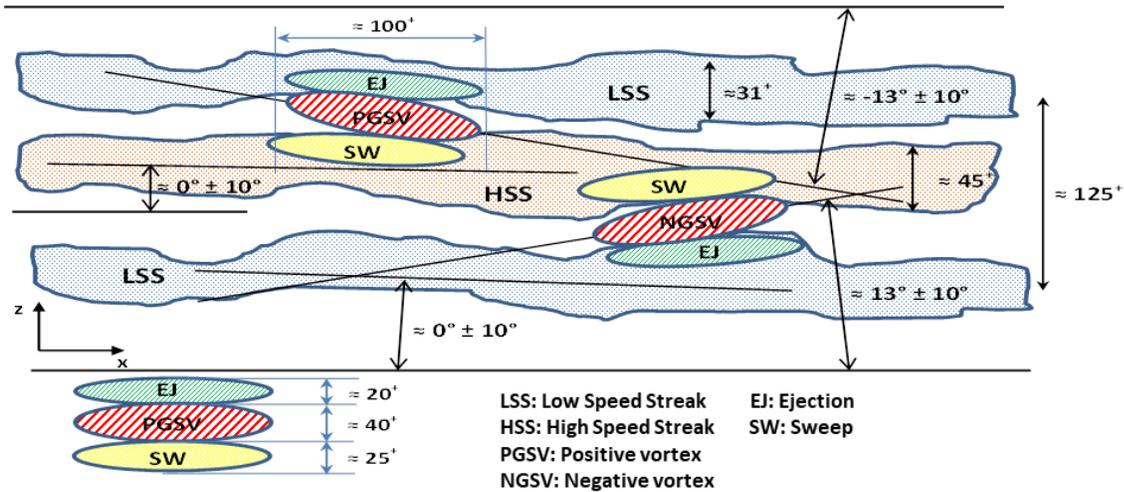}}
	\caption{Sketch of the very near wall turbulence organization proposed by \cite{lin08}. }
	\label{fig:lin08}      
\end{figure}

It is interesting to note that although this `turbulence organization' research has been very active around the world, it has stayed relatively disconnected from the modelling community and has not yet stimulated significant progress on the modelling side  leading to a universal near wall model \footnote{The WALLTURB European research project, which is at the origin of the present contribution, was aimed at that.}. Having in-hand detailed information on a key unknown of turbulence, the dissipation tensor, and knowing now that most of its anisotropy is occuring in the region where the turbulence organization is reliably characterized, it is worth trying to see if this organization has any relation to the dissipation behaviour very near the wall. The pseudo-dissipation $\mathcal{D}_{ij}$ given by equation (\ref{dij}) can now be considered as the relevant tensor to represent dissipation and can be developed into the following for the diagonal terms and the main off-diagonal term $\mathcal{D}_{12}$:

\begin{eqnarray}
\mathcal{D}_{11} &=&  2\nu \langle \left[\frac{\partial u_1}{\partial x_1}\right]^2\rangle + 2\nu \langle \left[\frac{\partial u_1}{\partial x_2}\right]^2\rangle + 2\nu \langle \left[\frac{\partial u_1}{\partial x_3}\right]^2\rangle
\label{equ:d11}\\
\mathcal{D}_{22} &=&  2\nu \langle \left[\frac{\partial u_2}{\partial x_1}\right]^2\rangle + 2\nu \langle \left[\frac{\partial u_2}{\partial x_2}\right]^2\rangle + 2\nu \langle \left[\frac{\partial u_2}{\partial x_3}\right]^2\rangle
\label{equ:d22}\\
\mathcal{D}_{33} &=&  2\nu \langle \left[\frac{\partial u_3}{\partial x_1}\right]^2\rangle + 2\nu \langle \left[\frac{\partial u_3}{\partial x_2}\right]^2\rangle + 2\nu \langle \left[\frac{\partial u_3}{\partial x_3}\right]^2\rangle
\label{equ:d33}\\
\mathcal{D}_{12} &=&  2\nu \langle \frac{\partial u_1}{\partial x_1}\frac{\partial u_2}{\partial x_1}\rangle + 2\nu \langle \frac{\partial u_1}{\partial x_2}\frac{\partial u_2}{\partial x_2}\rangle + 2\nu \langle \frac{\partial u_1}{\partial x_3}\frac{\partial u_2}{\partial x_3}\rangle.
\label{equ:d12}
\end{eqnarray}
These show clearly the different velocity derivative moments contributing to each pseudo-dissipation component.

\begin{figure}
	\resizebox{0.5\linewidth}{!}{\includegraphics[scale=1]{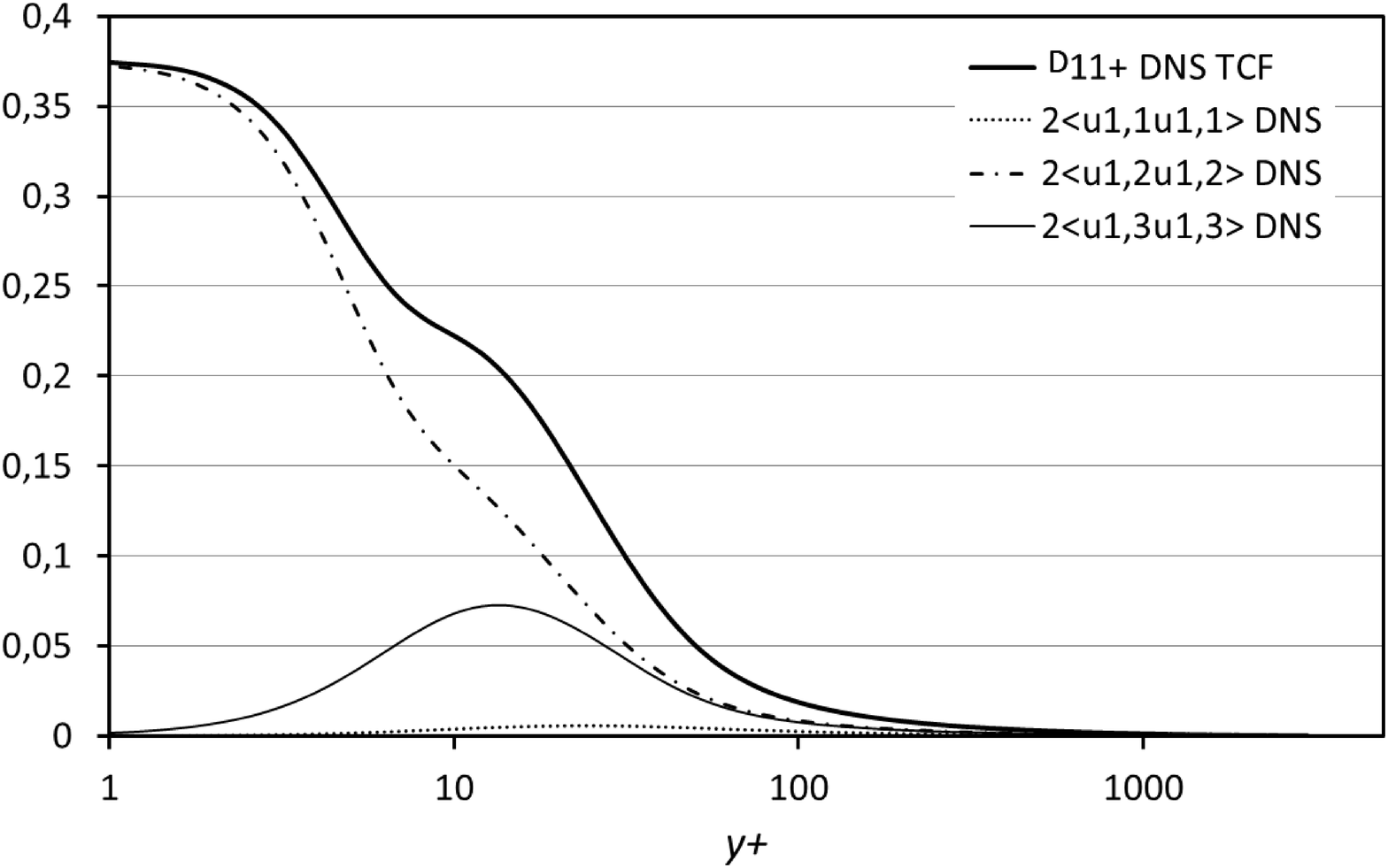}} 
	\resizebox{0.5\linewidth}{!}{\includegraphics[scale=1]{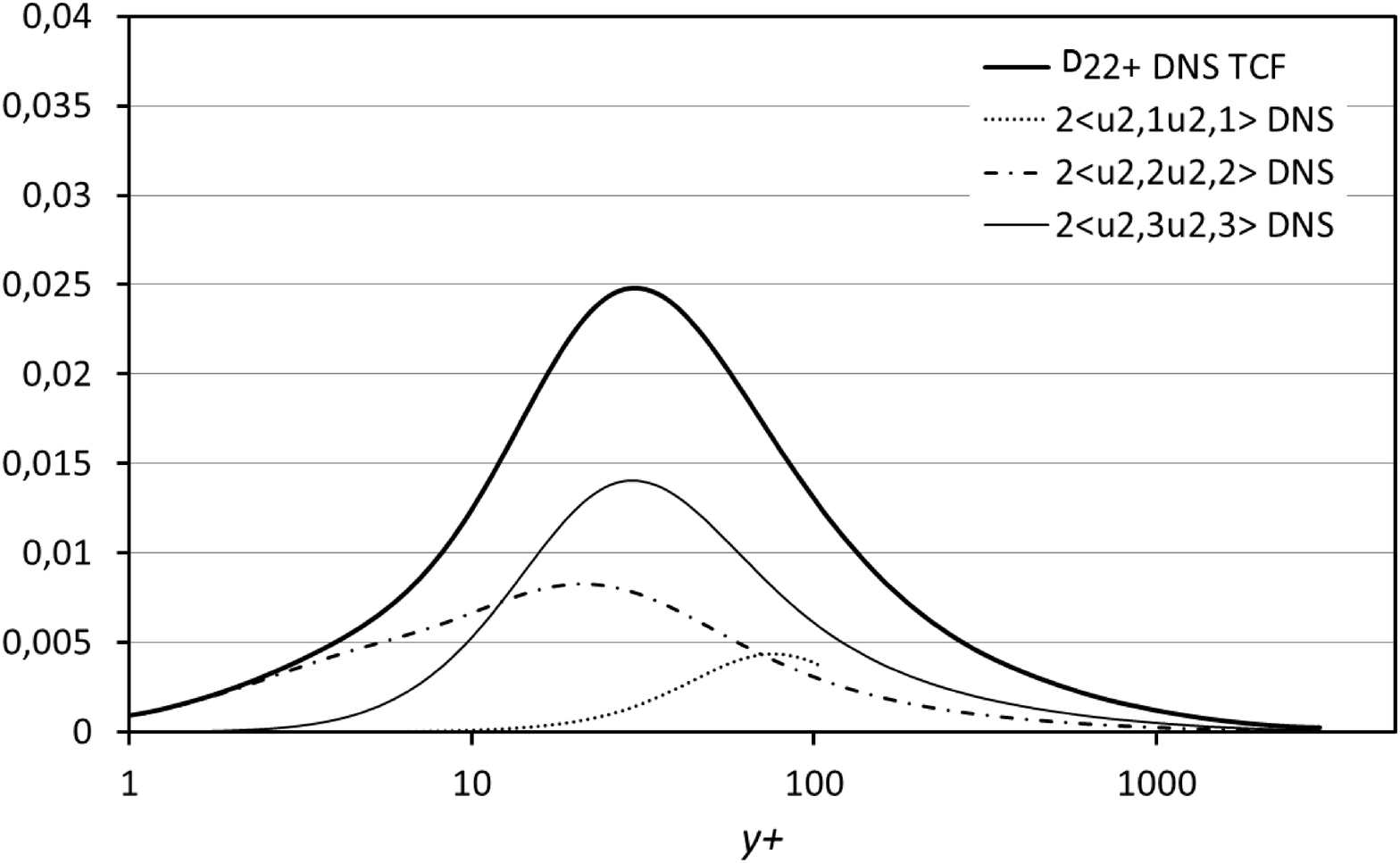}}\\
	\hspace*{3 cm}(a) \hspace{5 cm}(b)\\
	\resizebox{0.5\linewidth}{!}{\includegraphics[scale=1]{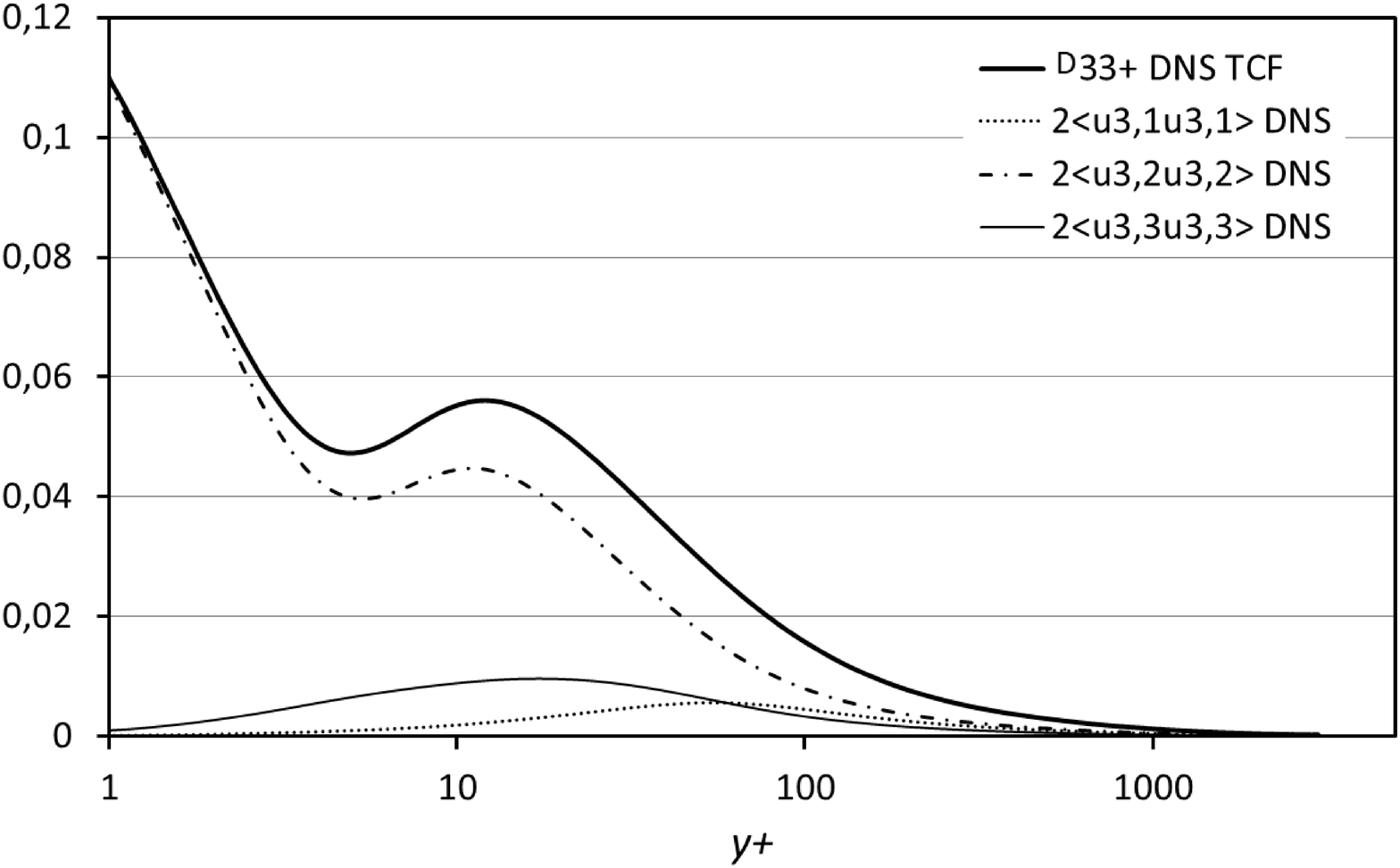}}
	\resizebox{0.5\linewidth}{!}{\includegraphics[scale=1]{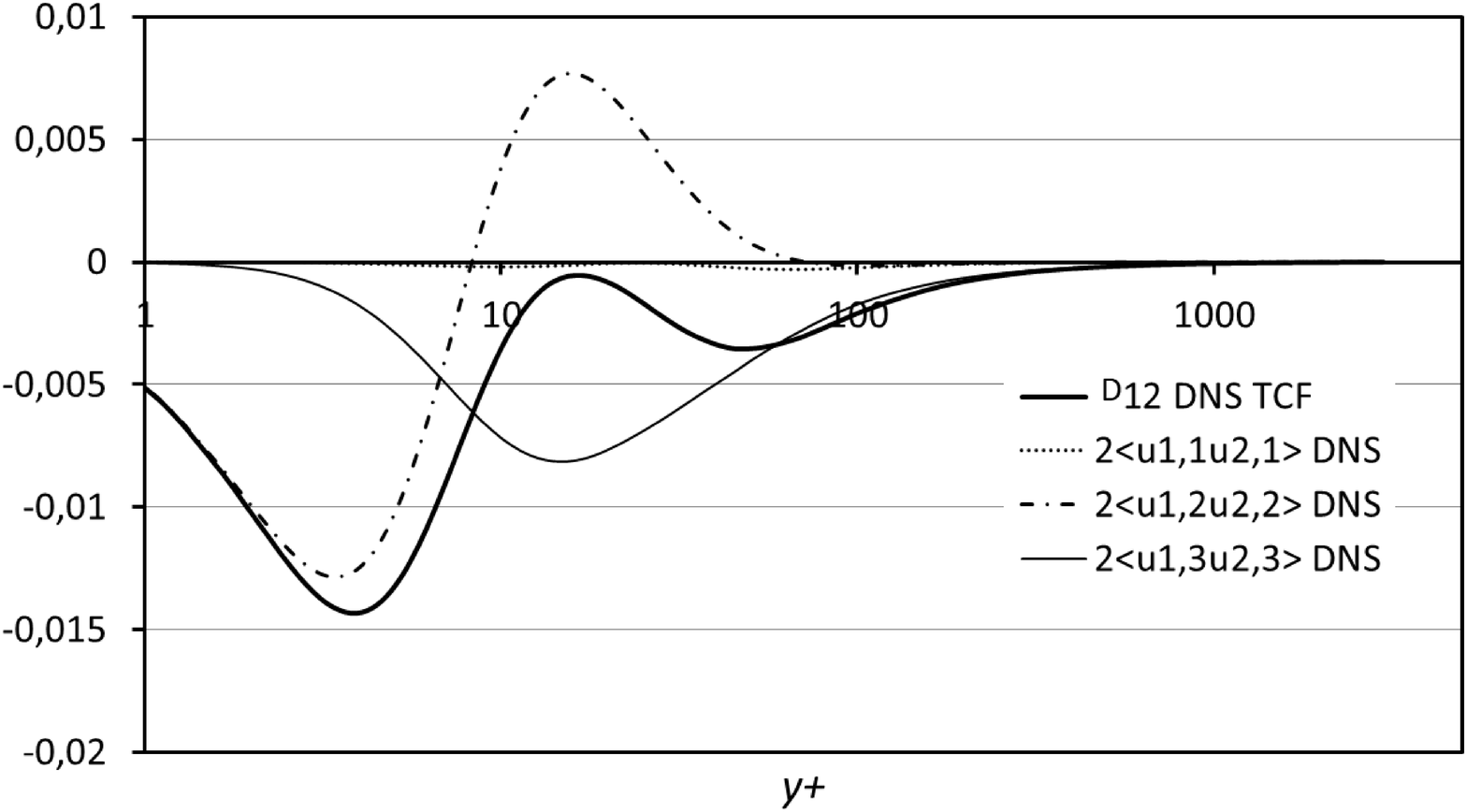}}\\
	\hspace*{3 cm}(c) \hspace{5 cm}(d)\\
	\caption{Derivative moments contributing to the main terms of the pseudo-dissipation tensor $D_{ij}$.}
	\label{fig:dnsdii}      
\end{figure}

Figure \ref{fig:dnsdii} gives a plot of each of these moments for each of the pseudo-dissipation tensor terms of equations \ref{equ:d11} to \ref{equ:d12}. For the first $\mathcal{D}_{11}$ term, the dominating contribution is clearly $\langle\left[\frac{\partial u_1}{\partial x_2}\right]^2\rangle$ which is maximum at the wall.  The second pseudo-dissipation component $\mathcal{D}_{22}$ is an order of magnitude smaller than the first one (as is the corresponding Reynolds stress). The peak of the two main contributing moments,  $\langle\left[\frac{\partial u_2}{\partial x_2}\right]^2\rangle$ and $\langle\left[\frac{\partial u_2}{\partial x_3}\right]^2\rangle$, is located  around the peak of TKE ($y^+ \simeq 15-20$), which is also the location of the center of the quasi-streamwise vortices as found by \cite{lin08}.  The last diagonal $\mathcal{D}_{33}$ term  is, like the first one, largely dominated by one moment, $\langle\left[\frac{\partial u_3}{\partial x_2}\right]^2\rangle$, which is again maximum at the wall. Both $\mathcal{D}_{11}$ and $\mathcal{D}_{33}$ show  a intriguing kink around $y^+ \simeq 15-20$. As could be expected, these three diagonal terms are mainly dominated by the moments corresponding to gradients normal to the wall.

\begin{figure}
	\resizebox{0.5\linewidth}{!}{\includegraphics[scale=1]{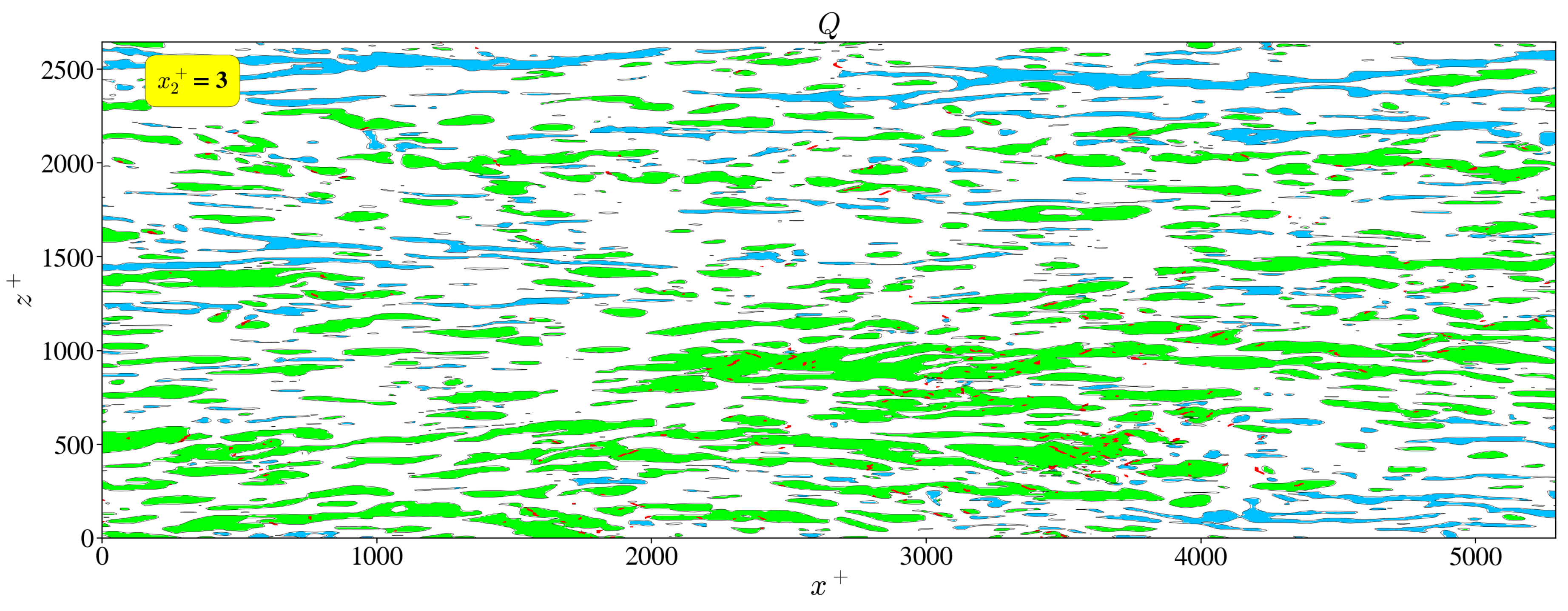}} 
	\resizebox{0.5\linewidth}{!}{\includegraphics[scale=1]{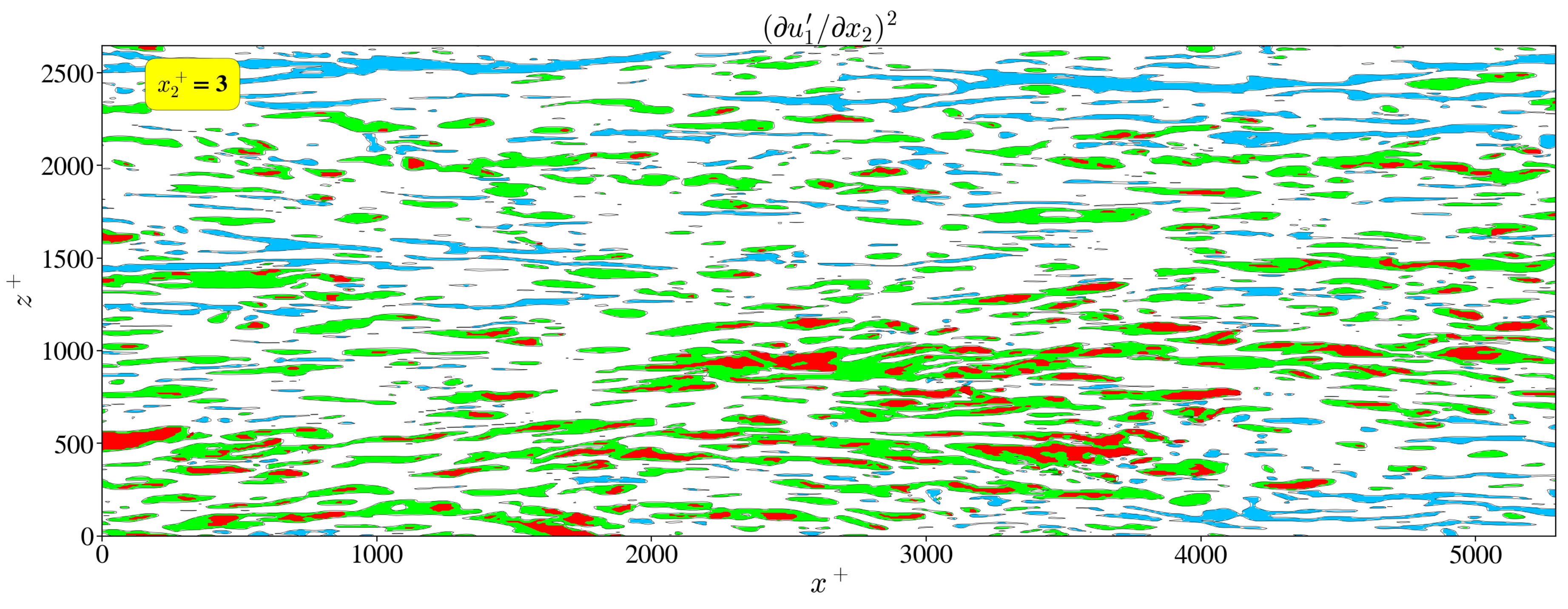}}\\
	\hspace*{3 cm}(a) \hspace{5 cm}(b)\\
	\resizebox{0.5\linewidth}{!}{\includegraphics[scale=1]{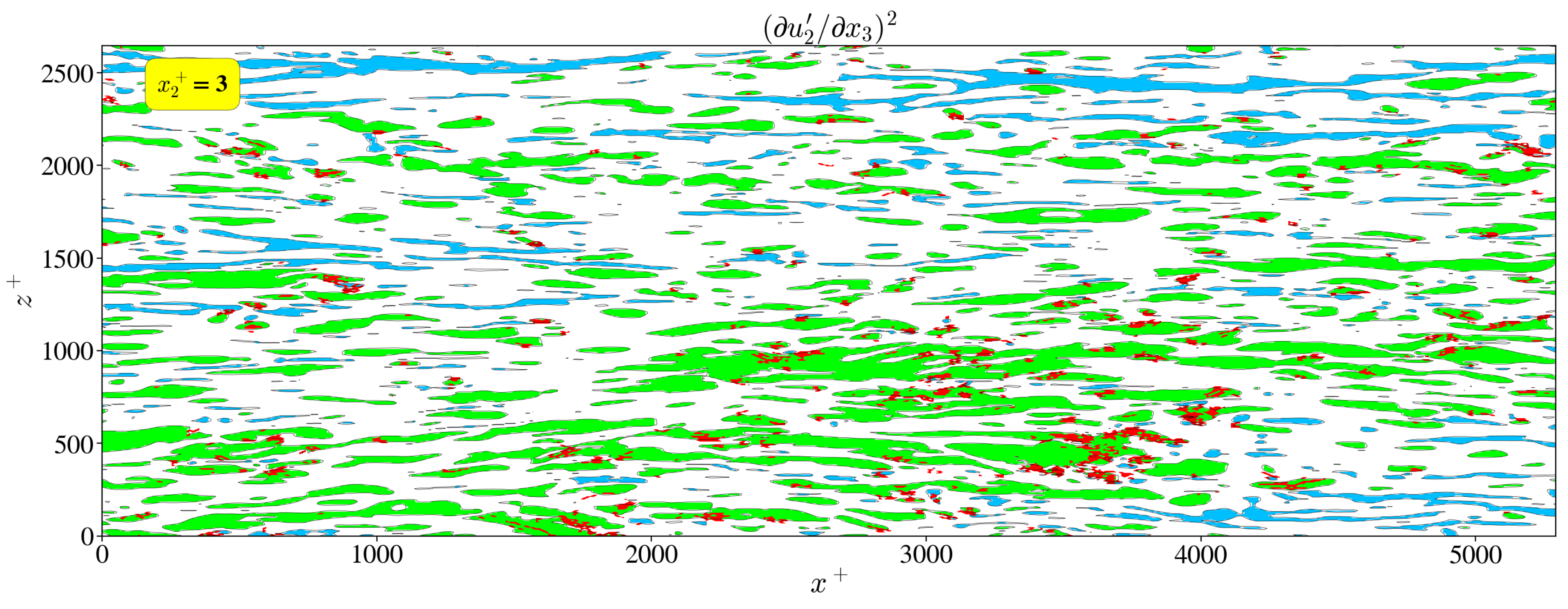}}
	\resizebox{0.5\linewidth}{!}{\includegraphics[scale=1]{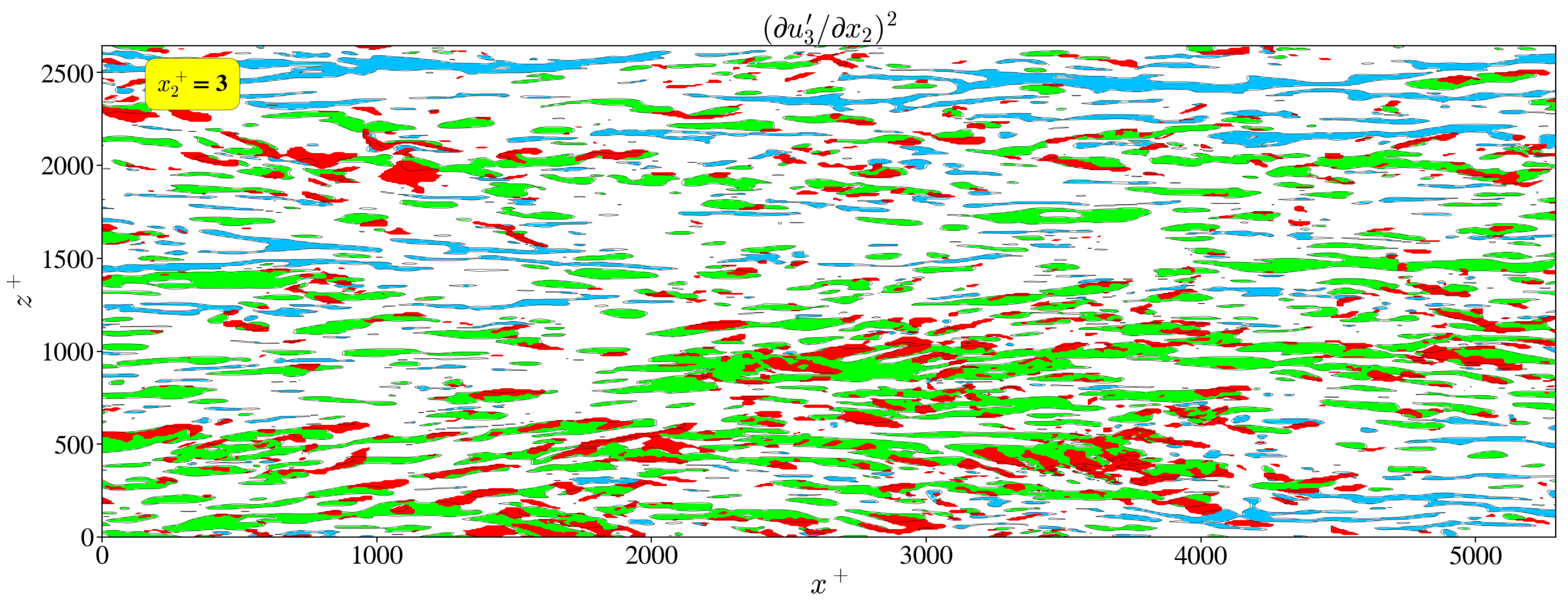}}\\
	\hspace*{3 cm}(c) \hspace{5 cm}(d)\\
	\caption{Instantaneous view of the derivative moments contributing to the main terms of the pseudo-dissipation tensor $D_{ij}$ at $x_2^+ = 3$. High speed streaks: green, low speed streaks: blue. In red: (a) Q criterion, (b) main term of $\mathcal{D}_{11}$. (c) main term of $\mathcal{D}_{22}$. (d) main term of $\mathcal{D}_{33}$. }
	\label{fig:instant_3}      
\end{figure}

\begin{figure}
	\resizebox{0.5\linewidth}{!}{\includegraphics[scale=1]{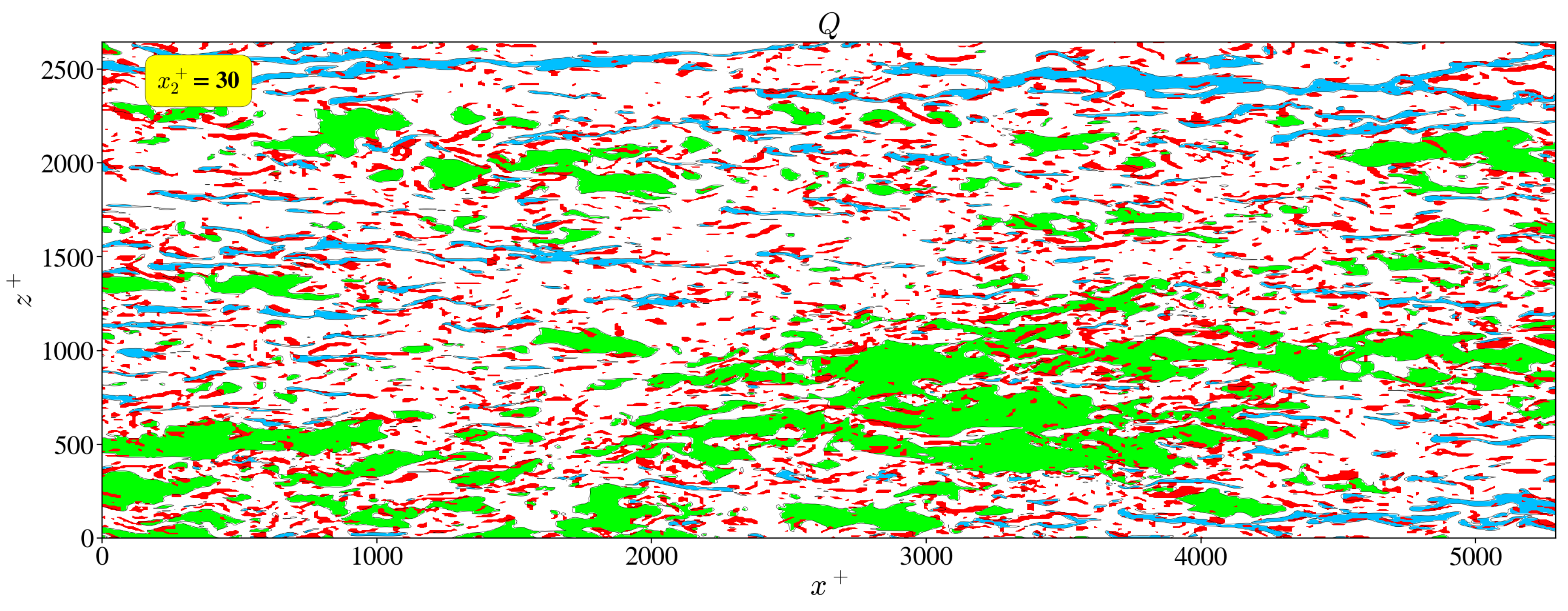}} 
	\resizebox{0.5\linewidth}{!}{\includegraphics[scale=1]{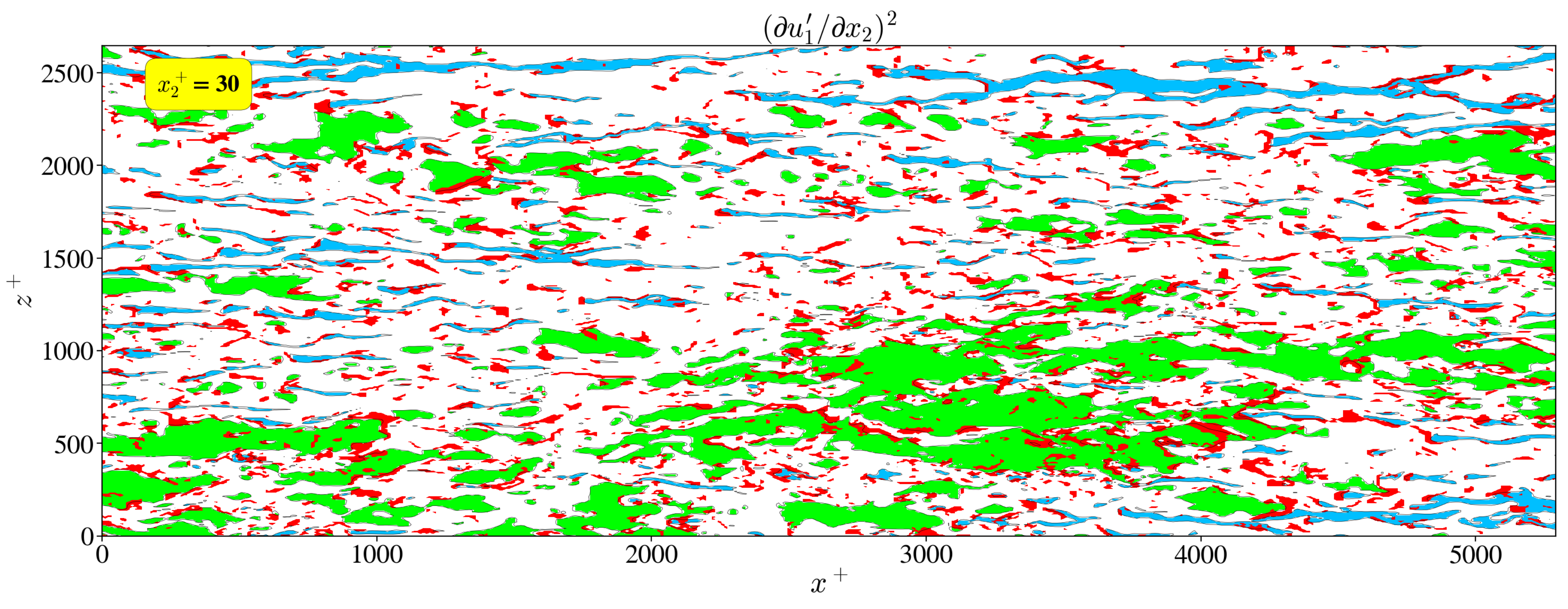}}\\
	\hspace*{3 cm}(a) \hspace{5 cm}(b)\\
	\resizebox{0.5\linewidth}{!}{\includegraphics[scale=1]{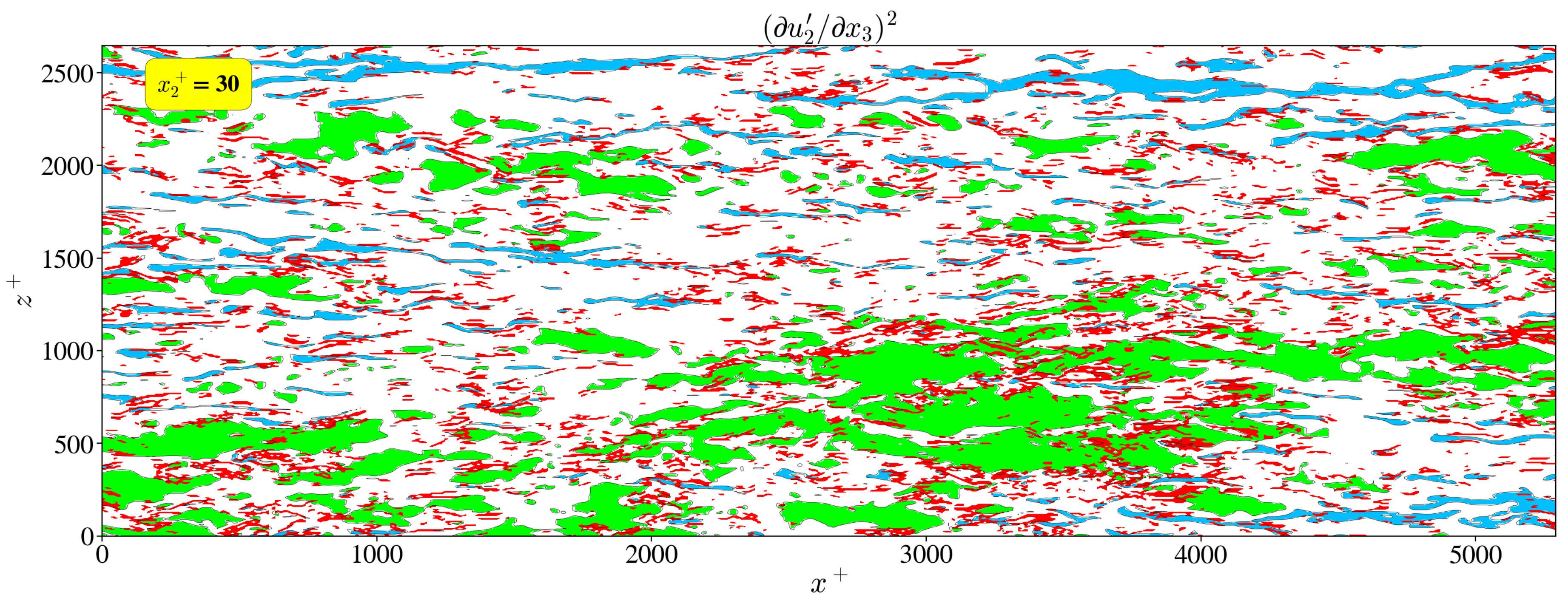}}
	\resizebox{0.5\linewidth}{!}{\includegraphics[scale=1]{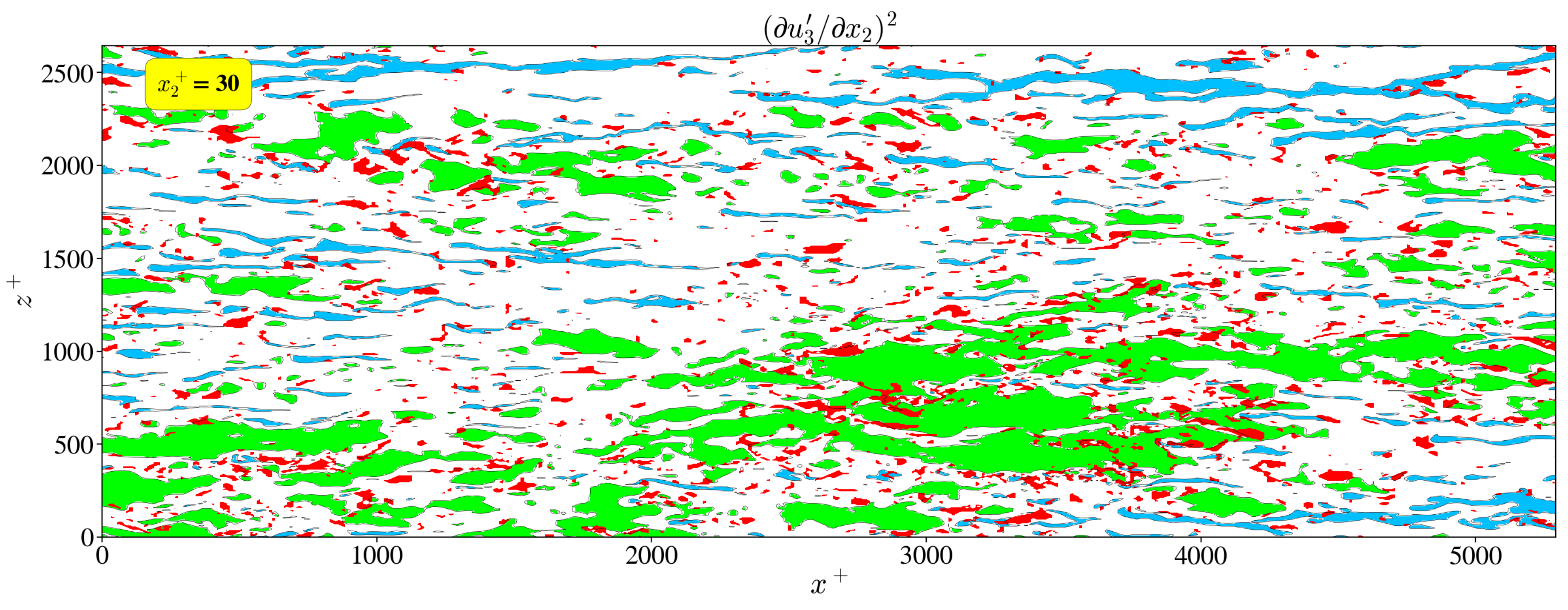}}\\
	\hspace*{3 cm}(c) \hspace{5 cm}(d)\\
	\caption{Instantaneous view of the derivative moments contributing to the main terms of the pseudo-dissipation tensor $D_{ij}$ at $x_2^+ = 30$. High speed streaks: green, low speed streaks: blue. In red: (a) Q criterion, (b) main term of $\mathcal{D}_{11}$. (c) main term of $\mathcal{D}_{22}$. (d) main term of $\mathcal{D}_{33}$. }
	\label{fig:instant_30}      
\end{figure} 

In order to get a better insight into the structures contributing the most to these dissipative terms, figures \ref{fig:instant_3} and \ref{fig:instant_30} were built from the DNS of \cite{thais11} in planes parallel to the wall at  $y^+ = 3$ and $y^+ = 30$ respectively.  The low (blue) and high (green) speed streaks are evidenced by thresholding the streamwise velocity fluctuations. The Q criterion, which is indicative of the presence of vortices, is given in red in figure (a). The dominating moments of each of the diagonal terms of $\mathcal{D}_{ij}$ are also given in red in figures (b-d). Looking first at figure \ref{fig:instant_3} for $y^+ = 3$, it appears clearly from the Q criterion (a) that the vortical activity is very limited. A striking similarity appears between this Q criterion and the term  $\langle\left[\frac{\partial u_2}{\partial x_3}\right]^2\rangle$ in (c) which is the dominant term of $\mathcal{D}_{22}$, indicating that this kind of dissipation is linked to strain rate which appears mostly on the side of the high speed streaks. The contrast is clear compared to the terms of figures (b) and (d). The red regions are much larger in scale and mostly located inside the high speed streaks. Looking now at figure \ref{fig:instant_30} for $y^+ = 30$, the first thing to notice  in (a) is that the number of vortical structures is much larger, that they are mostly streamwise oriented and elongated and that they effectively appear mostly between the streaks as sketched in figure \ref{fig:lin08}. What is also clear from figures (b)-(c) is that all the dissipative activity is now strongly linked to these vortical structures and all at a comparable scale. 

The conclusion which can be drawn from these two figures is that the dissipation very near the wall has two distinct physical origins. One dominating at the wall which are strongly linked to the high speed streaks, appear at spanwise scales comparable to them and are probably related to the sweeping motions associated to these streaks. The second, the vortical activity which is maximum around $y^+ = 20$ and which, looking back at figure \ref{fig:dnsdii}, affects clearly all the components of $\mathcal{D}_{ij}$ at different levels and through different terms. It is interesting to notice at this stage that this wall location is also the place of a peak of production of TKE. In both cases, we are very far from the classical Kolmogorov cascade and from the dissipation of TKE through the smallest eddies (These streamwise vortical structures are typically 10 Kolmogorov in diameter and about 50 in length.). Consequently, any attempt to model the near wall dissipation on the basis of the Kolmogorov theory has little chance to succeed. This probably explains the large number of near wall corrections available in the RANS literature.

\section{Discussion and Conclusions}
\label{conclusion}

For turbulence modeling it is of prime importance to characterize the dissipation rate of TKE. This key parameter is difficult to access in practice as it involves all the terms of the velocity gradient tensor and requires a very good spatial resolution. With this objective, a specific SPIV experiment allowing the derivative computation along the three directions of space was carried out. The measurement noise and the cut-off frequency of the derivative schemes were optimized in order to select the best second order centered scheme depending on the wall distance. The specificity of the SPIV set-up was also used, together with the continuty equation to remove the noise from the derivative moments. This set of data permitted computation of all the derivative moments contributing to both the full and pseudo-dissipation tensors $\varepsilon_{ij}$ and $\mathcal{D}_{ij}$. In addition, the data from a plane channel flow DNS \cite{thais11} were used to build all the same velocity derivative moments and the corresponding dissipation components.

As an introduction, the main averaged equations governing turbulence were reviewed and emphasis was placed on a precise definition of the full dissipation tensor $\varepsilon_{ij}$  and the pseudo-dissipation tensor $\mathcal{D}_{ij}$  (Note that there is considerable  confusion in the litterature about these two entities). The full and pseudo dissipations of turbulence kinetic energy were directly defined from the corresponding tensors. It should be emphasized here that, although equations (\ref{rs_dij}) and (\ref{eq:k_d}) can be deduced from equations (\ref{rs_eq})  and (\ref{eq:k}) respectively, only $\varepsilon_{ij}$ and $\varepsilon$ represent the true dissipation.

Having all the velocity derivative moments available from both experiment and DNS, made it possible to check the classical hypotheses used to simplify turbulence theory. The results show that {\it local isotropy}  is clearly a very bad assumption for this boundary layer flow within the overlap region and below. As evidenced by figures \ref{fig:spiviso} and \ref{fig:dnsiso}, several of the conditions set by equations (\ref{iso1}) to (\ref{iso3}) are violated by the derivative moments and consequently, the estimation of the full dissipation on the basis of one single moment, as done in equation (\ref{iso_simp}) is quite risky. The streamwise mean
square derivative, $\langle [\partial u_1/\partial x_1 ]\rangle$, is a particularly bad representative of the rest. 

As far as {\it local axisymmetry}  is concerned, it is observed that all conditions provided by equations (\ref{axi1}) to (\ref{axi2}) are fulfilled, except two as evidenced by figures \ref{fig:testaxi} and \ref{fig:testaxidns}. Nevertheless, {\it local axisymmetry}  provides a very good approximation above $y^+ = 100$ and a reasonable one for $ 30 < y^+ < 100$, allowing us to propose simplified equations to evaluate the full TKE dissipation $\varepsilon$ on the basis of only four derivative moments. These should be particularly useful to experimentalists using planar SPIV, since all the necessary terms can be measured in a plane.

The {\it local homogeneity}  hypothesis appeared in fact to be the most interesting one as it  implies directly that $\varepsilon_{ij} = \mathcal{D}_{ij}$ and consequently that  $\varepsilon = \mathcal{D}$. As for local axisymmetry, the data show that {\it local homogeneity}   is limited to the region above $y^+ = 100$ (Note that {\it local homogeneity}  is a requisite for either {\it local axisymmetry}  or {\it local isotropy}  to be possible.).  The data clearly show that {\it local homogeneity}  is not valid below $y^+ = 100$.

In seemingly contradiction to the breakdown of local homogeneity, the full dissipation tensor can hardly be distinguished from the pseudo one below this wall distance. The analysis performed in section \ref{sec-planehomogeneity} shows that supposing homogeneity in planes parallel to the wall is enough to validate only one of the  moments equalities demanded by equation (\ref{eq:homogderiv5}) to have $\varepsilon_{ij} = \mathcal{D}_{ij}$. Figure \ref{fig:homodiss}a \& \ref{fig:homodiss}b shows that these three equalities are nearly perfectly fulfilled over the whole wall layer in both the channel DNS and the BL, confirming the planar homogeneity of turbulence of the whole wall layer in these case. They are nevertheless not sufficent to ensure the equality between $\varepsilon_{ij}$ and $\mathcal{D}_{ij}$,  hence the difference between the full and pseudo dissipations observed in figure \ref{fig:compardissipation}. 

Looking at $\mathcal{D}_{ij}$ which, on the basis of the present results, can be considered as representative of the dissipation of the Reynolds stresses, it appears that the different moments building each component of this tensor do not have the same weight. The moments based on wall normal derivatives are mostly dominant, especially very near the wall. Looking at what is known of near wall turbulence organization and at some snapshots of the different terms in planes parallel to the wall, it appears that near wall dissipation is based on two different physical phenomenon. One dominating very near the wall is associated to the high speed streaks and to the sweeping motions embedded in them. The second one is clearly related to the quasi-streamwise vortices which have been observed by numerous authors around $y^+ = 20$ and which are at the origin also of the near wall turbulence production. Both phenomenon occur at fairly different scales. They appear in any case very different from the classical Kolmogorov cascade model. To go further, it would be of interest to extract from a sufficent number of independent realizations, length scales and intensity scales of the different dominant terms of $\mathcal{D}_{ij}$ evidenced here. This would most probably help a better modelling of the near wall dissipation which is critical to turbulence modelling. 

From the authors point of view, the main contribution of the present work is to validate quantitatively, on the basis of both experimental and DNS high quality data the hypothesis made at the early stage of turbulence modelling to replace the full dissipation tensor in the Reynolds stress equation and the full dissipation in the turbulence kinetic energy equation by the corresponding pseudo-dissipation terms. This hypothesis was done at that time for the purpose of simplification (fewer terms to model) and because very little was known about dissipation. In about half a century, both direct simulation and experimental techniques have made enough progress to allow today a detailed analysis of the numerous dissipation components and an  {\it a posteriori} validation of this well known hypothesis. It should be noted however, as pointed out in the appendix, that our theoretical conclusions are very much dependent on the assumptions that certain derivatives relations do not commute.  If it can be argued that they do, then all of these conclusions near the wall will be purely a consequence of plane homogeneity.  This will be especially problematical since ALL of the data is also homogeneous in planes.  So fully three-dimensional flows could behave very differently. 

Among the other contributions of the present paper, it is worth mentioning the clarification of the definition of dissipation and pseudo-dissipation, which has been the subject of considerable confusion and debates over the last fifty years when the early work was forgotten. Finally, the results obtained here clearly evidence that, except very near the wall ($y^+ < 100$), the {\it local homogeneity}  hypothesis is remarkably good and the {\it local axisymmetry}  can be quite helpful to measure more accurately the scalar dissipation.

\section*{Acknowledgement}
This work was supported through the International Campus on Safety and Inter modality in Transportation (CISIT). This work was carried out within the framework of the CNRS Research Federation on Ground Transports and Mobility, in articulation with the Elsat2020 project supported by the European Community, the French Ministry of Higher Education and Research, the Hauts de France Regional Council. Centrale Lille is acknowledged for providing regular financial support to the visits of Pr. George. This research has granted access to the HPC resources  of [CCRT /CINES /IDRIS] under the allocation i20142b022277 and i20162a01741 made by GENCI (Grand Equipement National de Calcul Intensif). L. Thais is  acknowledged for providing the data of his DNS of channel flow.

\bibliographystyle{plainnat}

\bibliography{biblio_wkg}

\begin{thebibliography}{34}
\providecommand{\natexlab}[1]{#1}
\providecommand{\url}[1]{\texttt{#1}}
\expandafter\ifx\csname urlstyle\endcsname\relax
  \providecommand{\doi}[1]{doi: #1}\else
  \providecommand{\doi}{doi: \begingroup \urlstyle{rm}\Url}\fi

\bibitem[Andreopoulos and Honkan({2001})]{honkan2001}
Y.~Andreopoulos and A.~Honkan.
\newblock {An experimental study of the dissipative and vortical motions in a
  turbulent boundary layer}.
\newblock \emph{J. Fluid Mech.}, {439}:\penalty0 {131--163}, {2001}.

\bibitem[Antonia et~al.({1986})Antonia, Anselmet, and Chambers]{antoniaetal86}
R.A. Antonia, F.~Anselmet, and A.~J. Chambers.
\newblock Assessment of local isotropy using measurements in a turbulent plane
  jet.
\newblock \emph{J. Fluid Mech.}, {163}:\penalty0 {365 -- 391}, {1986}.

\bibitem[Antonia et~al.({1991})Antonia, Kim, and Browne]{antonia91}
R.A. Antonia, J.~Kim, and L.W.B. Browne.
\newblock {Some characteristics of small-scale turbulence in a turbulent duct
  flow }.
\newblock \emph{J. Fluid Mech.}, {233}:\penalty0 {369--388}, {1991}.

\bibitem[Balint et~al.({1991})Balint, Wallace, and Vukolavcevic]{balint91}
J.L. Balint, J.M. Wallace, and P.~Vukolavcevic.
\newblock {The velocity and vorticity vector-fields of a turbulent
  boundary-layer .2. statistical properties}.
\newblock \emph{J. Fluid Mech.}, {228}:\penalty0 {53--86}, {1991}.

\bibitem[Batchelor(1953)]{batchelor53}
G.~K. Batchelor.
\newblock \emph{The theory of homogeneous turbulence}.
\newblock 1953.

\bibitem[Batchelor(1954)]{batchelor46}
G.~K. Batchelor.
\newblock The theory of axisymmetric turbulence.
\newblock \emph{Proc. R. Soc. Lond. A}, 186:\penalty0 480--502, 1954.

\bibitem[Chandrasekar(1950)]{chandrasekar50}
S.~Chandrasekar.
\newblock The theory of axisymmetric turbulence.
\newblock \emph{Proc. R. Soc. Lond. A}, 242:\penalty0 557--577, 1950.

\bibitem[Foucaut et~al.(2020)Foucaut, George, Stanislas, and Cuvier]{foucaut20}
J.~M. Foucaut, W.~George, M~Stanislas, and C.~Cuvier.
\newblock Velocity derivatives in a high {R}eynolds number {T}urbulent
  {B}oundary {L}ayer. {P}art 3: {M}easurement of the dissipation tensor from an
  {SPIV} experiment.
\newblock \emph{submitted to JFM}, 2020.

\bibitem[George and Castillo(1997)]{george97b}
W.~K. George and L.~Castillo.
\newblock The zero pressure-gradient turbulent boundary layer.
\newblock \emph{Applied Mechanics Reviews}, 50:\penalty0 689--729, 1997.

\bibitem[George and Hussein(1991)]{george91}
W.K. George and H.J. Hussein.
\newblock Locally axisymmetrical turbulence.
\newblock \emph{J. Fluid Mech.}, 233:\penalty0 1--23, 1991.

\bibitem[George and Tutkun(2011)]{georgetutkun2011}
W.K. George and M.~Tutkun.
\newblock The mesolayer and reynolds number dependencies of boundary layer
  turbulence.
\newblock In Marusic~I. Stanislas~M., Jimenez~J., editor, \emph{Progress in
  Wall Turbulence: Understanding and Modeling. ERCOFTAC Series}, volume~14,
  pages 183--190. {SPRINGER-VERLAG BERLIN}, 2011.

\bibitem[Gerolymos and Vallet(2016)]{gerolymos16}
G.A. Gerolymos and I.~Vallet.
\newblock The dissipation tensor $\varepsilon_{ij}$ in wall turbulence.
\newblock \emph{J. Fluid Mech.}, 807:\penalty0 386--418, 2016.

\bibitem[Honkan and Andreopoulos({1997})]{honkan97}
A.~Honkan and Y.~Andreopoulos.
\newblock {Vorticity, strain-rate and dissipation characteristics in the
  near-wall region of turbulent boundary layers}.
\newblock \emph{J. Fluid Mech.}, {350}:\penalty0 {29--96}, {1997}.

\bibitem[Jakirli\'c and Hanjali\'c(2002)]{jakirlic02}
S.~Jakirli\'c and K.~Hanjali\'c.
\newblock A new approach to modelling near-wall turbulence energy and stress
  dissipation.
\newblock \emph{Journal of Fluid Mechanics}, 459:\penalty0 139--166, 2002.

\bibitem[Jim\'enez and Pinelli(1999)]{jimenez99b}
J.~Jim\'enez and A.~Pinelli.
\newblock The autonomous cycle of near wall turbulence.
\newblock \emph{J. Fluid Mech.}, 389:\penalty0 335--359, 1999.

\bibitem[Kline and Portela(1997)]{Kline1997}
S.~J. Kline and L.~M. Portela.
\newblock \emph{A view of the structure of turbulent boundary layers}.
\newblock Computational Mechanics Publications, R. Panton Editor, 1997.

\bibitem[Kline et~al.(1967)Kline, Reynolds, Schraub, and Runstadler]{Kline67}
S.~J. Kline, W.~C. Reynolds, F.~A. Schraub, and P.~W. Runstadler.
\newblock The structure of turbulent boundary layers.
\newblock \emph{Journal of Fluid Mechanics}, 30:\penalty0 741--773, 1967.

\bibitem[Kolmogorov(1941)]{Kolmogorov41a}
A.~N. Kolmogorov.
\newblock The local structure of turbulence in incompressible viscous fluid for
  very large reynolds numbers.
\newblock \emph{C. R. Acad. Sci. URSS}, 30:\penalty0 301--305, 1941.

\bibitem[Leschziner(2015)]{leschziner2015}
M.~Leschziner.
\newblock \emph{Statistical Turbulence Modelling for Fluid Dynamics
  Demystified}.
\newblock Imperial College Press, 2015.

\bibitem[Lin et~al.(2008)Lin, Laval, Foucaut, and Stanislas]{lin08}
J.~Lin, J.-P. Laval, J.-M. Foucaut, and M.~Stanislas.
\newblock Quantitative characterization of coherent structures in the buffer
  layer of near-wall turbulence. part 1: streaks.
\newblock \emph{Experiments in Fluids}, 45\penalty0 (6):\penalty0 999--1013,
  2008.

\bibitem[Manceau et~al.(2002)Manceau, Carlson, and Gatski]{manceau02a}
R.~Manceau, J.~R. Carlson, and T.~B. Gatski.
\newblock A rescaled elliptic relaxation approach: neutralizing the effect on
  the log layer.
\newblock \emph{Phys. Fluids}, 14\penalty0 (11):\penalty0 3868--3879, 2002.

\bibitem[Panton(1997)]{panton97}
R.~Panton.
\newblock Self-sustaining mechanisms of wall turbulence.
\newblock \emph{Computational Mechanics}, 1997.

\bibitem[Pope(2000)]{pope00}
S.~B. Pope.
\newblock \emph{Turbulent Flows}.
\newblock Cambridge Univ. Press, 2000.

\bibitem[Reynolds(1976)]{reynolds76}
W.~C. Reynolds.
\newblock Computation of turbulence flows.
\newblock \emph{Ann. Rev. of Fluid Mechanics}, 8:\penalty0 183--208, 1976.

\bibitem[Saddoughi and Veeravalli(1994)]{saddoughi94}
S.~G. Saddoughi and S.~V. Veeravalli.
\newblock Local isotropy in turbulent boundary layers at high {R}eynolds
  number.
\newblock \emph{J. Fluid Mech.}, 268:\penalty0 333--372, 1994.

\bibitem[Sillero et~al.(2013)Sillero, Jimenez, and Moser]{sillero13}
J.A. Sillero, J.~Jimenez, and R.D. Moser.
\newblock One-point statistics for turbulent wall-bounded flows at reynolds
  numbers up to $\delta^+\approx2000$.
\newblock \emph{Phys. Fluids}, 25, 2013.

\bibitem[Smits and Marusic(2011)]{smits11}
A.~J. Smits and I.~Marusic.
\newblock High~reynolds number wall turbulence.
\newblock \emph{Annu. Rev. Fluid Mech.}, 43:\penalty0 353--375, 2011.

\bibitem[Stanislas et~al.(2020)Stanislas, Foucaut, George, Cuvier, and
  Laval]{stanislas20}
M.~Stanislas, J.~M. Foucaut, W.~George, C.~Cuvier, and J.~P. Laval.
\newblock Velocity derivatives in a high {R}eynolds number {T}urbulent
  {B}oundary {L}ayer. {P}art 1: {D}issipation and {E}nergy {B}alance.
\newblock \emph{submitted to JFM}, 2020.

\bibitem[Taylor(1935)]{Taylor1935}
G.~I. Taylor.
\newblock Statistical theory of turbulence.
\newblock \emph{Proc. R. Soc. London A}, 151:\penalty0 421 -- 478, 1935.

\bibitem[Thais et~al.(2011)Thais, Tejada-Mart{\'i}nez, Gatski, and
  Mompean]{thais11}
L.~Thais, A.~E. Tejada-Mart{\'i}nez, T.~B. Gatski, and G.~Mompean.
\newblock A massively parallel hybrid scheme for direct numerical simulation of
  turbulent viscoelastic channel flow.
\newblock \emph{Comp. \& Fluid}, 43:\penalty0 134--142, 2011.

\bibitem[Theodorsen(1952)]{theodorsen52}
T.~Theodorsen.
\newblock Mecanism of turbulence.
\newblock In \emph{Proc. Midwest Conf. Fluid Mech. $2^{nd}$ Edn}, pages 1--18,
  Columbus, Ohio, 1952.

\bibitem[Wosnik et~al.(2000)Wosnik, Castillo, and George]{wosnik00}
M.~Wosnik, L.~Castillo, and W.K. George.
\newblock Theory for turbulent pipe and channel flows.
\newblock \emph{J. Fluid Mech.}, 421:\penalty0 115--145, 2000.

\bibitem[Zhao et~al.(2015)Zhao, George, and van Wachem]{Fanetal2015}
F.~Zhao, W.K. George, and B.~G.~M. van Wachem.
\newblock Four-way coupled simulations of small particles in turbulent channel
  flow: The effects of particle shape and stokes number.
\newblock \emph{Physics of Fluids.}, 27\penalty0 (doi:
  http://dx.doi.org/10.1063/1.4927277):\penalty0 083301, 2015.

\bibitem[Zhou et~al.(1999)Zhou, Adrian, Balachander, and Kendall]{Zhou1999}
J.~Zhou, R.~J. Adrian, S.~Balachander, and T.~M. Kendall.
\newblock Mechanisms for generating coherent packets of hairpin vortices in
  channel flow.
\newblock \emph{J.~Fluid Mech.}, 387:\penalty0 353--396, 1999.

\end{thebibliography}

\end{document}